\documentclass[fleqn,usenatbib,onecolumn]{mnras}

\usepackage{newtxtext,newtxmath}

\usepackage[T1]{fontenc}

\DeclareRobustCommand{\VAN}[3]{#2}
\let\VANthebibliography\thebibliography
\def\thebibliography{\DeclareRobustCommand{\VAN}[3]{##3}\VANthebibliography}

\usepackage{graphicx}	
\usepackage{amsmath}	

\usepackage{amssymb}
\usepackage{natbib}
\usepackage{natbib}
\usepackage{threeparttable}
\usepackage{hyperref}
\usepackage{subfigure}
\usepackage{amsmath}
\usepackage{bm}

\title[Mutiple-color study of eleven contact binaries]{The study of eleven contact binaries with mass ratios less than 0.1}

\author
[Xin-Yi Liu et al.]{
Xin-Yi Liu,$^{1}$
Kai Li,$^{1}$\thanks{E-mail: kaili@sdu.edu.cn}
Raul Michel,$^{2}$
Xiang Gao,$^{1}$
Xing Gao,$^{3}$
Fei Liu,$^{1}$
Shi-Peng Yin,$^{1}$
Xi Wang,$^{1}$
Guo-You Sun$^{4}$
\\
$^{1}$ Shandong Key Laboratory of Optical Astronomy and Solar-Terrestrial Environment, School of Space Science and Physics, Institute of Space Sciences,\\ 
Shandong University, Weihai, Shandong 264209, China\\
$^{2}$ Observatorio Astron\'{o}mico Nacional, Instituto de Astronom\'{ı}a, Universidad Nacional Aut\'{o}noma de M\'{e}xico, Apartado Postal 877, Ensenada, B.C. 22830, M\'{e}xico\\
$^{3}$ Xinjiang Astronomical Observatory, 150 Science 1-Street, Urumqi 830011, China\\
$^{4}$ Wenzhou Astronomical Association, Beibaixiang Town, Yueqing, Zhejiang, 325603, China
}

\date{Accepted XXX. Received YYY; in original form ZZZ}

\pubyear{2022}

\begin{document}
\bibliographystyle{mnras}
\label{firstpage}
\pagerange{\pageref{firstpage}--\pageref{lastpage}}
\maketitle

\begin{abstract}	
Multi-band photometric observations of eleven totally eclipsing contact binaries were carried out. Applying the Wilson-Devinney program, photometric solutions were obtained. There are two W-subtype systems, which are CRTS J133031.1+161202 and CRTS J154254.0+324652, and the rest systems are A-subtype systems. CRTS J154254.0+324652 has the highest fill-out factor with 94.3$\%$, and the lowest object is CRTS J155009.2+493639 with only 18.9$\%$. The mass ratios of the eleven systems are all less than 0.1, which means that they are extremely low mass ratio binary systems. We performed period variation investigation and found that the orbital periods of three systems decrease slowly, which may be caused by the angular momentum loss, and of six systems increase slowly, which indicates that the materials may transfer from the secondary component to the primary component. LAMOST low$-$resolution spectra of four objects were analyzed, and using the spectral subtraction technique, H$\alpha$ emission line was detected, which means that the four objects exhibit chromospheric activity. In order to understand their evolutionary status, the mass-luminosity and mass-radius diagrams were plotted. The two diagrams indicate that the primary component is in the main sequence evolution stage, and the secondary component is above TAMS, indicating that they are over-luminous. To determine whether the eleven systems are in stable state, the ratio of spin angular momentum to orbital angular momentum ($J_{s}/J_{o}$) and the instability parameters were calculated, and we argued that CRTS J234634.7+222824 is on the verge of a merger.
\end{abstract}



\section{INTRODUCTION}  
In the field, open and globular clusters, contact binaries are usually observed \citep{52,53,54,64,55}. \citet{1} estimated that there is one contact binary in every 500 main sequence stars. The large number of contact binaries provides abundant samples for studying the process of merging and the related physical mechanism, the phenomenon of materials transfer, and the presence of a third body \citep{82}. The two components are surrounded by the common envelope and hence have almost similar surface temperatures \citep{56}. There are two subtypes of contact binaries.
The massive component of A-subtype binary is hotter, while the one of W-subtype binary is cooler \citep{100}. The common convection envelope \citep{29} is located between the inner and outer critical equipotential surfaces, where the materials transfer occurs frequently. Short seperation of two components results in short periods and most systems are W UMa-type contact binaries with periods between 0.25 and 0.5 days \citep{12}. The most common explanation for the formation of W UMa-type contact binaries is that they are formed by the angular momentum loss (AML) of the initially detached systems through the magnetic wind \citep{39,40,41,42,43,44}. The ultimate fate of the W UMa-type contact binaries may be to merge into a fast-rotating single object like FK Com-type star or blue straggler when the mass ratio is less than the instability mass ratio ($q_{inst}$) \citep{5,45,4,46}. In open and globular clusters, a large number of W UMa-type contact binaries were discovered among blue stragglers, which confirmed this conjecture \citep{57,53,59,54}. 

\citet{47} proposed that deep ($f>50\%$), low mass ratio ($q<0.25$) overcontact binaries may be the progenitors of blue straggler/ FK Com–type stars. If the the fill-out factor of overcontact binaries is quite high, exceeding $70\%$ \citep{83} or $86\%$ \citep{7}, dynamic instability will occur, and overcontact binaries will merge into a fast-rotating single star. The uncertainty of the value is due to the consideration of loss of angular momentum via gravitational wave radiation or magnetic stellar winds. So far, only one merged contact binary has been observed, that of V1309 Sco $-$ Nova Sco 2008 (V1309 Sco) \citep{2}. Unfortunately, V1309 Sco was observed after the merger event, so it is not possible to conduct targeted observations on it to study the properties and the merger mechanism of contact binaries. It is estimated that the Galactic rate of contact binaries mergers such as V1309 Sco is every 10 yr \citep{49}. Therefore, the identification of pre-merger candidates is of great significance for the monitoring of merger events of contact binaries.

\citet{4} proposed that the instability (Darwin's instability) of contact binaries occurs when $J_o \leqslant 3J_s$, where $J_o$ is the orbital angular momentum and $J_s$ is the common spin angular momentum. Neglecting the spin angular momentum of the secondary, \citet{5} calculated that the $q_{inst}$ (instability mass ratio) of the contact binary containing two unevolved main-sequence stars is 0.09. Considering the rotation of the secondary and using the dimensionless rotation radius of $k_1^2=k_2^2=0.06$, \citet{7} calculated the $q_{inst}$ of the contact binary to be 0.071-0.078. \citet{8} proposed a new criterion for calculating $q_{inst}$, that is, when $J_T=J_o+J_s$ reaches a critical value \textbf{$dJ_{T}/dA = 0$}, Darwin's instability will occur. Assuming $k_1^2 \neq k_2^2$ for the component’s dimensionless gyration radii, \citet{8} obtained the theoretical $q_{inst}=0.094 - 0.109$ and when considering the effects of rotation, \citet{86} obtained the theoretical $q_{inst}=0.070 - 0.074$. \citet{9} proposed that $q_{inst}$ depends on the mass and structure of the primary, and it can be as low as 0.05. Many systems with extremely low mass ratios have been observed: M4 V53, $q \sim 0.078$ \citep{61}; SX Crv, $q \sim 0.066$ \citep{62}, $q \sim 0.072$ \citep{63}; V857 Her, $q \sim 0.065$ \citep{69}; V1187 Her, $q \sim 0.044$ \citep{72}; J082700, $q \sim 0.055$ and J132829, $q \sim 0.089$ \citep{70}. The discovery of these systems challenges the existing theories about the instability mass ratios of contact binaries. More observations and studies of contact binaries with mass ratios less than 0.1 are essential to refine the current theories.

\citet{12} performed a photometric analysis of 2335 late-type contact binaries extracted from the Catalina Sky Survey \citep[CSS,][]{73} and obtained their physical parameters, among which 112 contact binaries have mass ratios below 0.1. We selected some contact binaries with mass ratios lower than 0.1, performed photometric observations, and finally identified eleven totally eclipsing contact binaries with mass ratios less than 0.1. Table 1 listed the basic informations of the eleven contact binaries, including name, initial minimum moment, period, maximum magnitude of V-band, and amplitude. The V-band maximum magnitudes and amplitudes are from \citet{73}. In this paper, we porformed the photometric analysis of multiple-color light curves and investigations of orbital period changes of the eleven binary systems.

\begin{table*}
\caption{The basic information of the eleven targets.}
\label{table1}
\begin{tabular}{ccccccccccc} 
\hline
Target                & Here after name & HJD$_{0}$      & Period(days) & Vmax(mag) & Amplitude \\
\hline
CRTS J133031.1+161202 & J133031         & 2459323.10154  & 0.3026661    & 13.17     &    0.27   \\
CRTS J154254.0+324652 & J154254         & 2459319.97154  & 0.3549879    & 14.90     &    0.25   \\
CRTS J155009.2+493639 &	J155009         & 2459365.86122  & 0.4609100    & 13.61     &    0.20   \\
CRTS J155106.5+303534 & J155106         & 2459367.89252  & 0.3809891    & 14.88     &    0.22   \\
CRTS J160755.2+332342 & J160755         & 2459378.11819  & 0.3572878    & 15.13     &    0.26   \\	
CRTS J162327.1+031900 & J162327         & 2459353.30508  & 0.4745615    & 13.74     &    0.22   \\
CRTS J164000.2+491335 & J164000         & 2459417.37225  & 0.3907817    & 13.47     &    0.25   \\
CRTS J170307.9+020101 & J170307         & 2459390.94490  & 0.2908836    & 13.29     &    0.21   \\
CRTS J223837.9+321932 & J223837         & 2459463.28323  & 0.4441697    & 13.82     &    0.23   \\
CRTS J224827.6+341351 & J224827         & 2459463.78173  & 0.3210262    & 14.87     &    0.21   \\
CRTS J234634.7+222824 & J234634         & 2459464.42501  & 0.2906934    & 13.82     &    0.23   \\
\hline
\end{tabular}
\end{table*}

\section{OBSERVATIONS}
We conducted multi-band photometric observations on the eleven targets in six months (from April 2021 to September 2021), using 60cm Ningbo Bureau of Education and Xinjiang Observatory Telescope (NEXT), Weihai Observatory 1.0-m telescope of Shandong University (WHOT, \citet{71}), 85 cm telescope at the Xinglong Station of National Astronomical Observatories (NAOs85cm), the 84 cm telescope at the Observatorio Astron\'{o}mico Nacional San Pedro M\'{a}rtir (OAN-SPM). A back-illuminated FLI 230–42 CCD camera is used on NEXT. The CCD camera have $2048 \times 2048$ pixels, resulting in a field of view is 22$^{'} \times 22^{'}$. A PIXIS 2048B CCD camera is equipped on WHOT. The scale of each pixel is approximately 0.35$^{''}$, which provides a field of view of about 12$^{'} \times 12^{'}$. NAOs85cm uses an Andor DZ936 CCD camera. The scale of each pixel is approximately 0.94$^{''}$, which provides a field of view of about 32$^{'} \times 32^{'}$. OAN-SPM equips an E2V back-illuminated CCD detector. The scale of each pixel is 0.26$^{''}$, which gives the field of view is 9$^{'} \times 9^{'}$.

The observation time, observation band, exposure time and telescope used for each target are listed in Table 2. The observational error is about $10^{-3}-10^{-2}$ mag. For data reduction, the  data of WHOT, NAOs85cm and NEXT were processed by C-Munipack\footnote{\href{http://c-munipack.sourceforge.net/}{http://c-munipack.sourceforge.net/}}, and the OAN-SPM data were done with IRAF. The photometric images were performed bias, dark and flat corrections and processed with aperture photometry and differential photometry method. The differential magnitudes between the variable stars and the comparison stars and those between the comparison stars and the check stars were determined. The observation time and the differential magnitudes between the variable stars and the comparison stars are shown in Table A1. The selected comparison stars and check stars are also listed in Table 2. 

LAMOST (The Large Sky Area Multi-Object Fiber Spectroscopic Telescope, also known as the Goushoujing Telescope) is a special reflecting Schmidt telescope. Its field of view is 5$^{\circ}$ and it equipped 4000 fibers, which improve the rate of spectra acquisition \citep{78,79,87}. The resolution is around 1800 for low resolution mode and the wavelength ranges from 3700 to 9000 \AA \citep{80}. We searched for objects in LAMOST Data Release 8 (DR8) for the low resolution mode using a matching radius of 2 arcsec and obtained the spectra of J133031, J154254, J155009, J160755, J162327. The information including the observational date, the effective temperature (T), surface gravity (log $g$), radial velocity (RV), metallicity abundance ([Fe/H]), signal noise ratio of r filter (SNR$_{r}$) of the five objects' spectra are shown in Table 3.

\begin{table*}
\begin{center}
\caption{The observation details of the eleven contact binaries.}
\label{table2}
\begin{tabular}{clllccc} 
\hline
Target  & Observaing Date              & Exposure Time (s)     & Comparsion Star        & Check Star             & Telescope\\
\hline
J133031 & 2021 Apr 18                  & V100 R45  I35         & 2MASS 13301646+1608087 & 2MASS 13300988+1606279 & WHOT     \\
J154254 & 2021 Apr 15                  & R40  I30              & 2MASS 15425099+3245591 & 2MASS 15430105+3251410 & OAN-SPM  \\
        & 2021 May 02                  & R80  I130             & 2MASS 15425099+3245591 & 2MASS 15430105+3251410 & NAOs85cm \\
J155009 & 2021 May 31  June 03         & B60  V25  R15 I15     & 2MASS 15503806+4937136 & 2MASS 15502157+4940514 & OAN-SPM  \\
        & 2021 June 28                 & B120 V60  R40 I35     & 2MASS 15503806+4937136 & 2MASS 15502157+4940514 & WHOT     \\
J155106 & 2021 June 02                 & R60  I50              & 2MASS 15505151+3036225 & 2MASS 15505887+3034427 & OAN-SPM  \\ 
        & 2021 July 10                 & R130 I120             & 2MASS 15505151+3036225 & 2MASS 15505887+3034427 & WHOT     \\
J160755 & 2021 July 01                 & R60  I50              & 2MASS 16080618+3322554 & 2MASS 16073601+3319151 & OAN-SPM  \\
        & 2021 July 12                 & R150 I130             & 2MASS 16080618+3322554 & 2MASS 16073601+3319151 & WHOT     \\ 
J162327 & 2021 May 09 18 22 23         & g80  r70  i90         & 2MASS 16231038+0323098 & 2MASS 16234372+0319288 & NEXT     \\
        & 2021 June 15 17              & g80  r70  i90         & 2MASS 16231038+0323098 & 2MASS 16234372+0319288 & NEXT     \\            
J164000 & 2021 July 15 18 21 25        & g65  r55  i75         & 2MASS 16400156+4916155 & 2MASS 16073601+3319151 & NEXT     \\
        & 2021 Aug 01 11               & g65  r55  i75         & 2MASS 16400156+4916155 & 2MASS 16073601+3319151 & NEXT     \\
J170307 & 2021 June 24 25              & V80  R40  I30         & 2MASS 17030255+0200346 & 2MASS 17031099+0200020 & OAN-SPM  \\
J223837 & 2021 Aug 11 18 25 Sep 05     & g90  r75  i100        & 2MASS 22385725+3217007 & 2MASS 22384257+3215508 & NEXT     \\
J224827 & 2021 Sep 07 09               & V80  R40  I40         & 2MASS 22484442+3416547 & 2MASS 22483392+3415584 & OAN-SPM  \\
J234634 & 2021 Sep 02 04 06            & g80  r60              & 2MASS 23465840+2230443 & 2MASS 23463566+2222211 & NEXT     \\
\hline
\end{tabular}
\end{center}
\end{table*}

\begin{table*}
\begin{center}
\caption{The information of spectra released by LAMOST and the equivalent width (EW) of H$\alpha$ emission line.}
\label{table3}
\begin{tabular}{ccccccccccccc} 
\hline
Spectra   & Date           & $T$(K) & log $g$ & RV(km s$^{-1}$) & [Fe/H] & SNR$_{r}$ & EW(\AA) \\
\hline
J133031   & Apr.10 2013(1) & 5946   & 4.050   & 34.52 & -0.459 & 88  & 3.23(0.02)\\
          & Apr.10 2013(2) & 6004   & 4.127   & 20.20 & -0.396 & 132 & 3.41(0.04)\\
		  & Feb.06 2016    & 5831   & 4.045   & 12.31 & -0.417 & 92  & 3.80(0.13)\\
J154254   & Mar.30 2019    & 5862   & 4.196   & -35.05& -0.127 & 96  & 3.85(0.03)\\
J155009   & Apr.24 2013    &  -     &   -     &   -   &   -    & 6   & -         \\
          & Mar.25 2016    & 6907   & 3.948   & -45.38& -0.099 & 14  & -         \\
J160755   & Apr.22 2015    &  -     &   -     &   -   &   -    & 28  & 1.71(0.01)\\ 
J162327   & Apr.24 2013    & 6942   & 4.205   &-27.26 & -0.442 & 37  & 3.72(0.11)\\
          & Mar.25 2016    & 6899   & 4.248   &-32.08 & -0.402 & 71  & 5.71(0.13)\\
\hline
\end{tabular}
\end{center}
\end{table*}

\section{PHOTOMETRIC SOLUTIONS}
In order to obtain the basic physical parameters of the eleven contact binaries, we used the 2013 version of the Wilson–Devinney (W-D) program \citep{13,14,15} to analyze their light curves. The effective surface temperature of the primary component is estimated by the colour indices and LAMOST data, considering the 3D extinction \citep{16}. We calculated $g-r$, $B-V$, $J-K$ based on the magnitudes of B, V, g, r bands from AAVSO Photometric All Sky Survey \citep{89} and J and K bands from 2MASS \citep{90}. All the above mentioned temperatures and color indices are listed in Table 4. We took the average of all temperatures as the initial temperature of the primary component. Limb-darkening coefficients were obtained from the tables given by \citet{19}. When the system temperature is lower than 7200K, the gravity darkening and bolometric albedo coefficients were set to be $g_{1}=g_{2}=0.32$, $A_{1}=A_{2}=0.5$, otherwise, $g_{1}=g_{2}=1$, $A_{1}=A_{2}=1$ \citep{91}. For the eleven targets, the values of semi-major axis were set to be the values calculated by Equation (4) given in \citet{115}. J133031, J154254, J155009 and J162327 have non-solar chemical compositions and the values are found in LAMOST. For the four targets, the parameter of $[M/H]$ was set as the value determined by LAMOST in W-D program. In the process, the contact model (model 3) was adopted. The adjustable parameters are: the orbital inclination of the binary ($i$), the effective surface temperature of the secondary component ($T_{2}$), the dimensionless luminosity of each filter of the primary component ($L_{1}$), and the dimensionless surface potentials ($\Omega_{1}=\Omega_{2}$). 

In many cases, due to lack of radial velocity measurements, the q-search method was applied to determine the mass ratio. However, this may be inaccurate. But for the totally eclipsing contact binaries, the mass ratios are quite reliable. \citet{17} analysed the spectroscopic and photometric mass ratios of 80 contact binaries and found that they show some deviations in partially eclipsing binaries’ data, and they are basically same in totally eclipsing binaries’ data. \citet{18} and \citet{92} proved this conclusion. Iterations were carried out using 0.1 step length between q=1 and q=5, 0.05 step length between q=0.2 and q=1, and 0.01 step length between q=0.01 and q=0.2, and during the process, the mass ratio was fixed. Then, we obtained a series of convergent solutions. The obtained convergent solutions of J154254 are only between q=0.07 and q=0.16. The sum square of residuals versus mass ratios are plotted and shown in Fig. 1. The mass ratio with the smallest residuals was selected as the initial mass ratio. Then we set mass ratio as an adjustable parameter. By running the W-D program, we got the photometric solutions. Due to the light curves of J133031, J154254, J155009, J170307, J224827, J234634 existing obvious O'Connell effect, we used the dark spot model to fit the distortion in the light curve. When using the dark spot model, the convergent solution is not unique. According to \citet{107}, the reasonable accuracy in spot parameters can be obtained when the photometric accuracy is 0.0001 mag or better. The photometric accuracy can't be so high, so the convergent solutions with dark spot we got are possible solutions. The convergent solutions determined by the above steps are preliminary solutions. Considering the spread of $T_{1}$ estimates leads to significant temperature and luminosity uncertainties, we used the following equations \citep{108,109} to further determine the temperatures of two components,
\begin{equation}
\begin{aligned}
&T_{1}' = (((1+k^{2})T_{1}^{4})/(1+k^{2}(T_{2}/T_{1})^{4}))^{0.25}\\
&T_{2}' = T_{1}'(T_{2}/T_{1})\\
\end{aligned}
\end{equation}		
where $T_{1}$ is the initial temperature of the primary component, $T_{2}/T_{1}$ and k are the temperature and the radius ratio, $T_{1}'$ and $T_{2}'$ are more accurate temperatures derived by Equation (1). We set more accurate temperatures derived by Equation (1) as initial values in W–D program, while the primary temperature was fixed and the secondary temperature was adjustable. By running the W–D program, convergent solutions were obtained after one iteration. Then, we set the values of semi-major axes obtained in Section 6 as initial values, and they were fixed in W–D program. By running the W–D program again, convergent solutions were obtained after one iteration. After obtaining convergent solutions, for eleven targets, the third light was set as an adjustable parameter, and running W–D program, only J223837 and J234634 obtained convergent solutions, while the other targets can’t. The observed and theoretical light curves are shown in Fig. 2. The photometric solutions of the eleven targets are shown in Table 5. The relative errors of mass ratios we provided in Table 5 is about 0.1$\%$, which is underestimated \citep{110,111}. According to \citet{110}, if there is no third light, the relative errors of mass ratios of totally eclipsing contact binaries are about 1.0$\%$, and when the third light exists, the relative errors will become larger, but less than 10.0$\%$.

J133031, J154254, J155009, 155106, J164000 were observed by the Transiting Exoplanet Survey Satellite \citep[TESS,][]{105}. However, the datas of J154254 and 155106 are too poor, which may be caused by the target being too far from the center of the field of view, so we didn't use the data of the two targets for analysis. For J133031, J155009 and J164000, we set the obtained photometric solutions using our observations as initial parameters and ran the W-D program to get the photometric solutions and theoretical light curves of the TESS observation data. According to the 30 minutes observation cadence of TESS data, phase smearing effect was taken into account. After obtaining convergent solutions, the third light was set as an adjustable parameter, and running W–D program, J133031, J155009 and J164000 all obtained convergent solutions. However, the three targets of our data can't get convergent solutions when the third light parameter was adjustable in W-D program. This is because the scale of our data is very small and the scale of TESS data is 21 arcsec, when observing targets, the nearby companion stars will be included. Therefore, we searched J133031, J155009 and J164000 in Gaia DR3 \citep{112,113,114} using matching radius of 10 arcsec and 42 arcsec. When using a matching radius of 10 arcsec, no companion objects have been found, so it is correct that the three targets of our data can't get convergent solutions when the third light parameter was adjustable in W-D program. When using a matching radius of 42 arcsec, we found that J133031 exists three companion objects, J155009 exists one companion object and J164000 exists two companion objects. According to the magnitude given by Gaia DR3, we calculated flux ratios of third light flux to total flux using the following equation.
\begin{equation}
m_{2}-m_{1}=-2.5lg(E_{2}/E_{1})
\end{equation}           
The observed and theoretical light curves are shown in Fig. 3. The photometric solutions of TESS data and flux ratios calculated by Equation (2) are shown in Table 6. It is found that flux ratios obtained from the light curve analysis using W-D program are close to the ratios calculated by Equation (2). It should be noted that we use solutions obtained from our data as final solutions for the discussion in Section 6.

\begin{table*}
\begin{center}
\caption{The effective surface temperature of the eleven targets' primary components.}
\label{table4}
\normalsize
\begin{tabular}{ccccccccccccc} 
\hline
Target  & $(g-r)_{0}$  & $(B-V)_{0}$  & $(J-K)_{0}$  & T$_{g-r}$(K) & T$_{B-V}$(K) & T$_{J-K}$(K) & T$_{LAMOST}$(K)  & T$_{eff}$(K)\\
\hline
J133031 & 0.40 & 0.58 & 0.41 & 5990         & 5990         & 5550         & 5831/5946/6004 & 5880\\   
J154254 & 0.40 & 0.61 & 0.35 & 5990         & 5930         & 5860         & 5862           & 5910\\
J155009 & 0.27 & 0.48 & 0.24 & 6550         & 6350         & 6550         & 6907           & 6580\\
J155106 & 0.09 & 0.22 & 0.27 & 7220         & 7760         & 6350         & -              & 7110\\
J160755 & 0.07 & 0.24 & 0.02 & 7400         & 7590         & 8800         & -              & 7930\\
J162327 & 0.15 & 0.14 & 0.18 & 7020         & 6640         & 6810         & 6942/6899      & 6860\\
J164000 & 0.15 & 0.35 & 0.13 & 7020         & 6820         & 7400         & -              & 7080\\
J170307 & 0.58 & 0.72 & 0.48 & 5480         & 5480         & 5270         & -              & 5410\\
J223837 & 0.15 & 0.32 & 0.22 & 7020         & 7020         & 6670         & -              & 6900\\
J224827 & 0.29 & 0.56 & 0.38 & 6350         & 6050         & 5660         & -              & 6020\\
J234634 & 0.43 & 0.59 & 0.38 & 5930         & 5930         & 5680         & -              & 5840\\
\hline
\end{tabular}
\end{center}
\end{table*}

\begin{figure*}
\centering
\subfigure{\label{fig:subfig:a}
\includegraphics[width=0.32\linewidth]{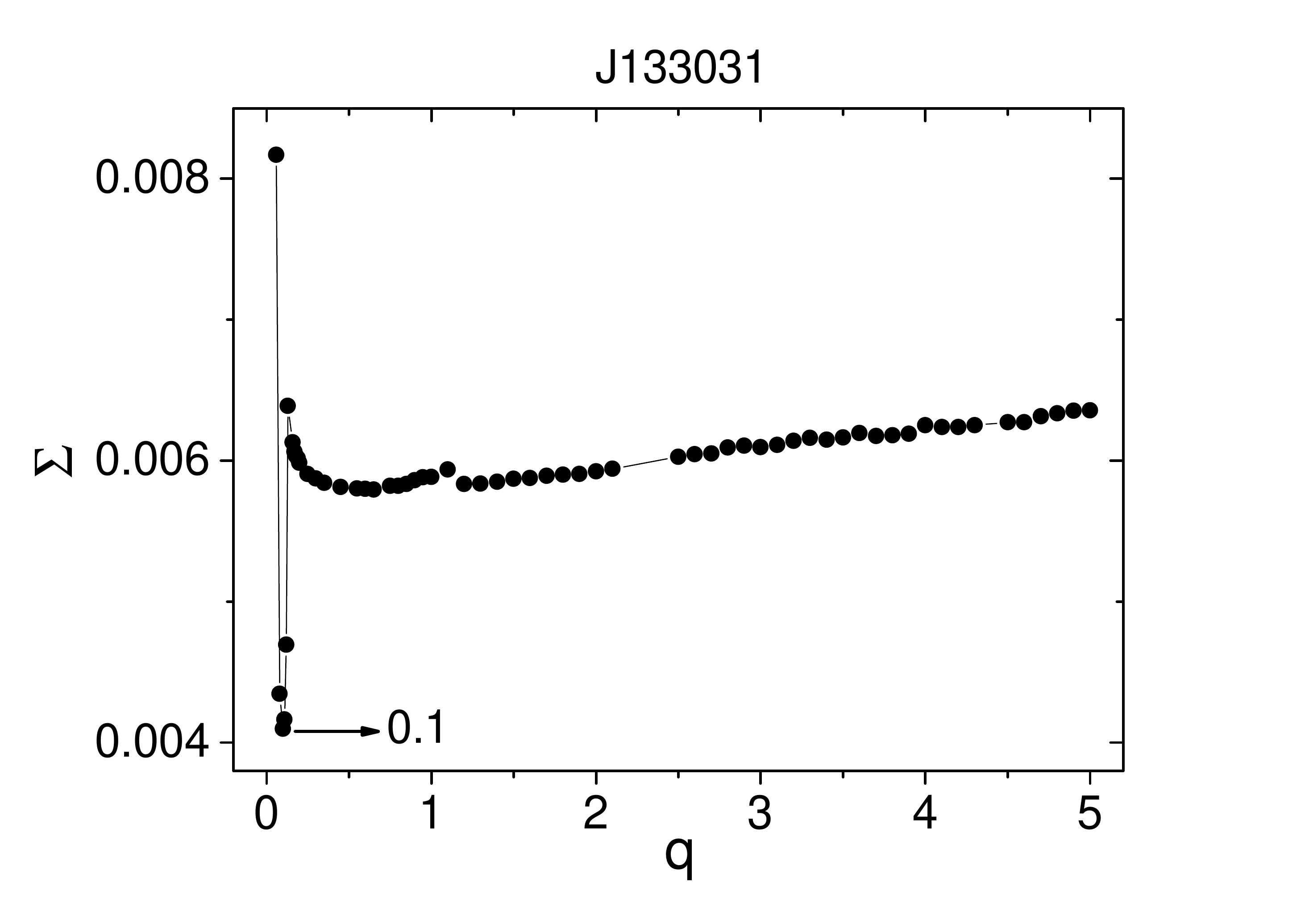}}
\subfigure{\label{fig:subfig:b}
\includegraphics[width=0.32\linewidth]{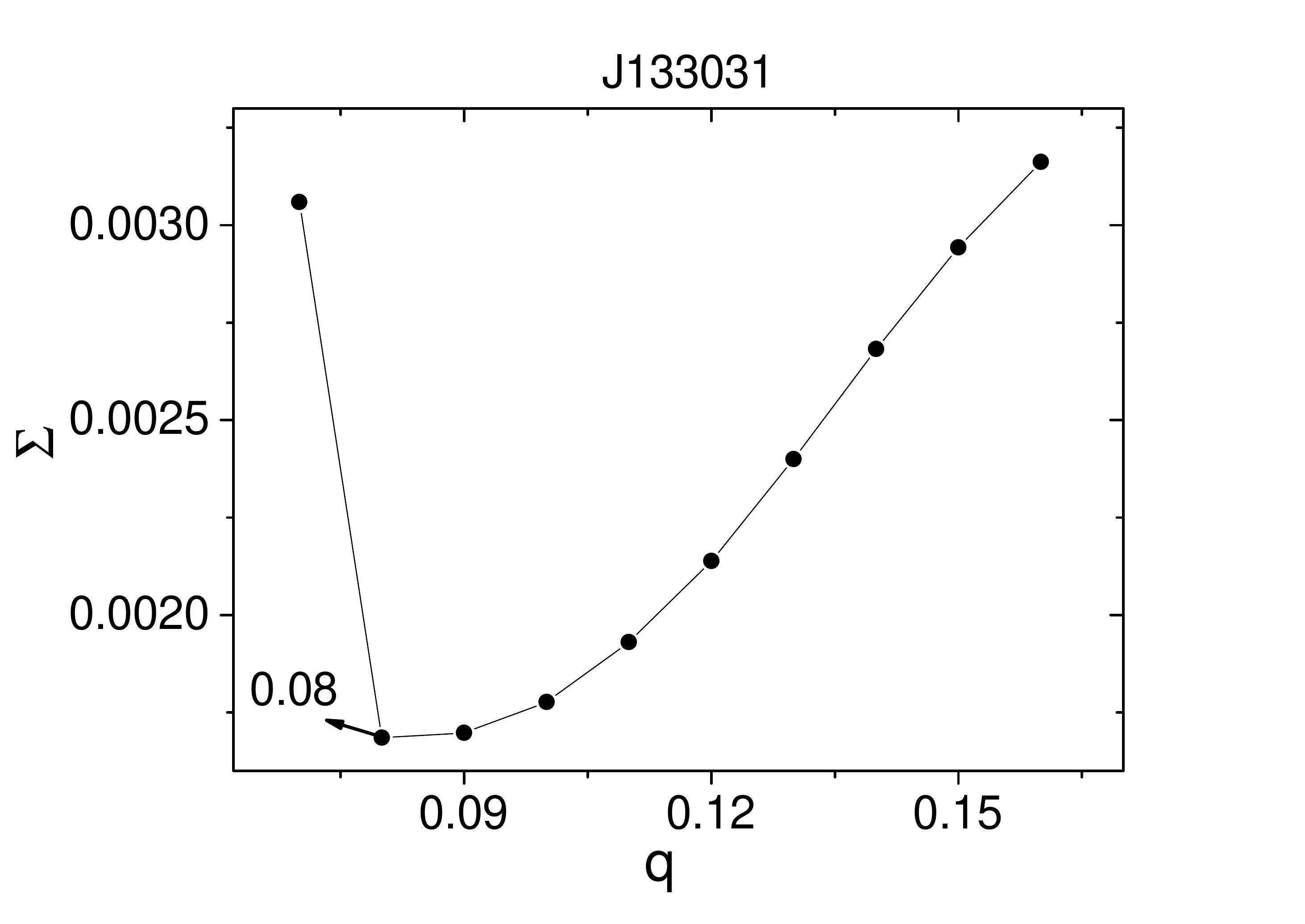}}
\subfigure{\label{fig:subfig:c}
\includegraphics[width=0.32\linewidth]{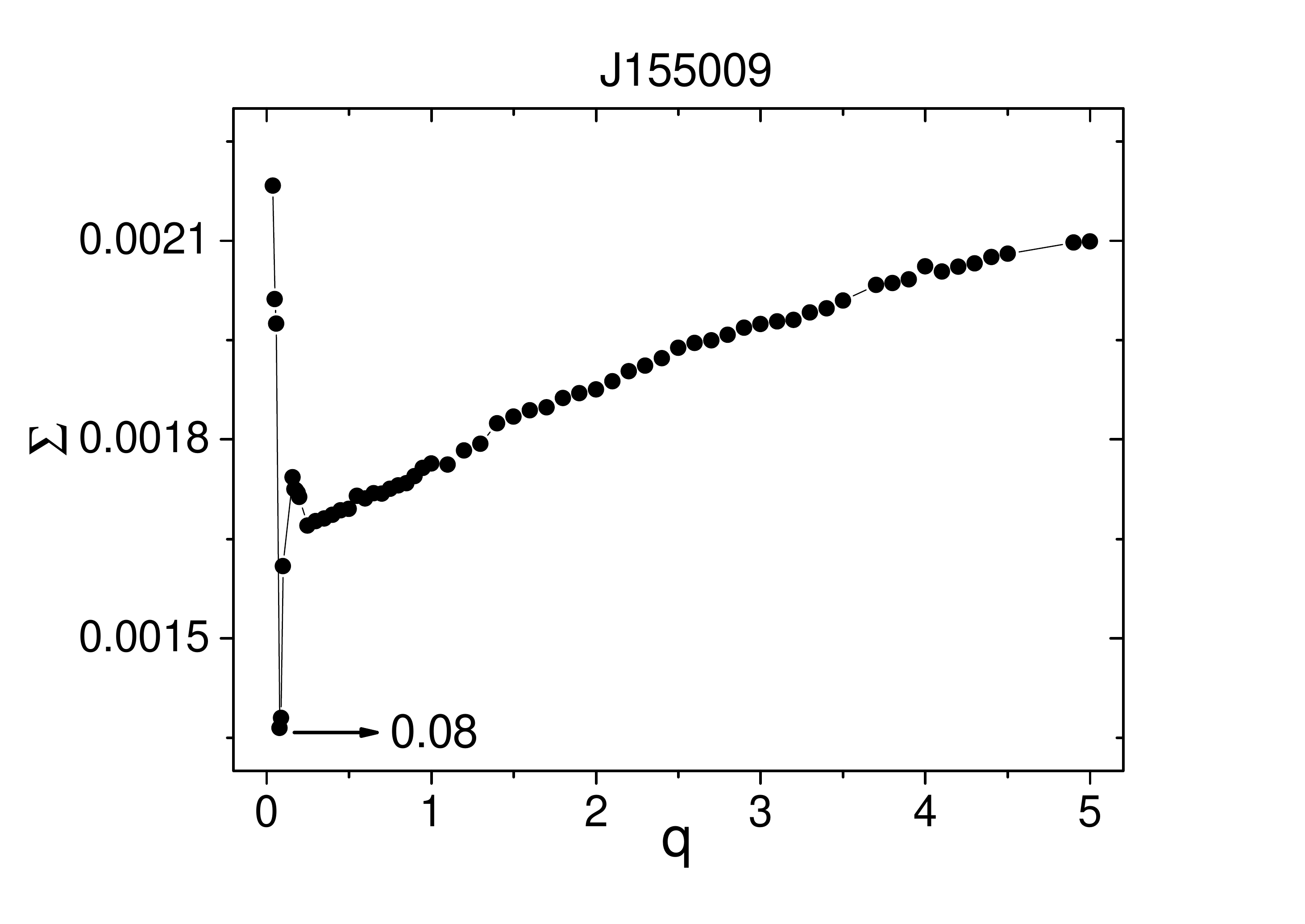}}
\subfigure{\label{fig:subfig:a}
\includegraphics[width=0.32\linewidth]{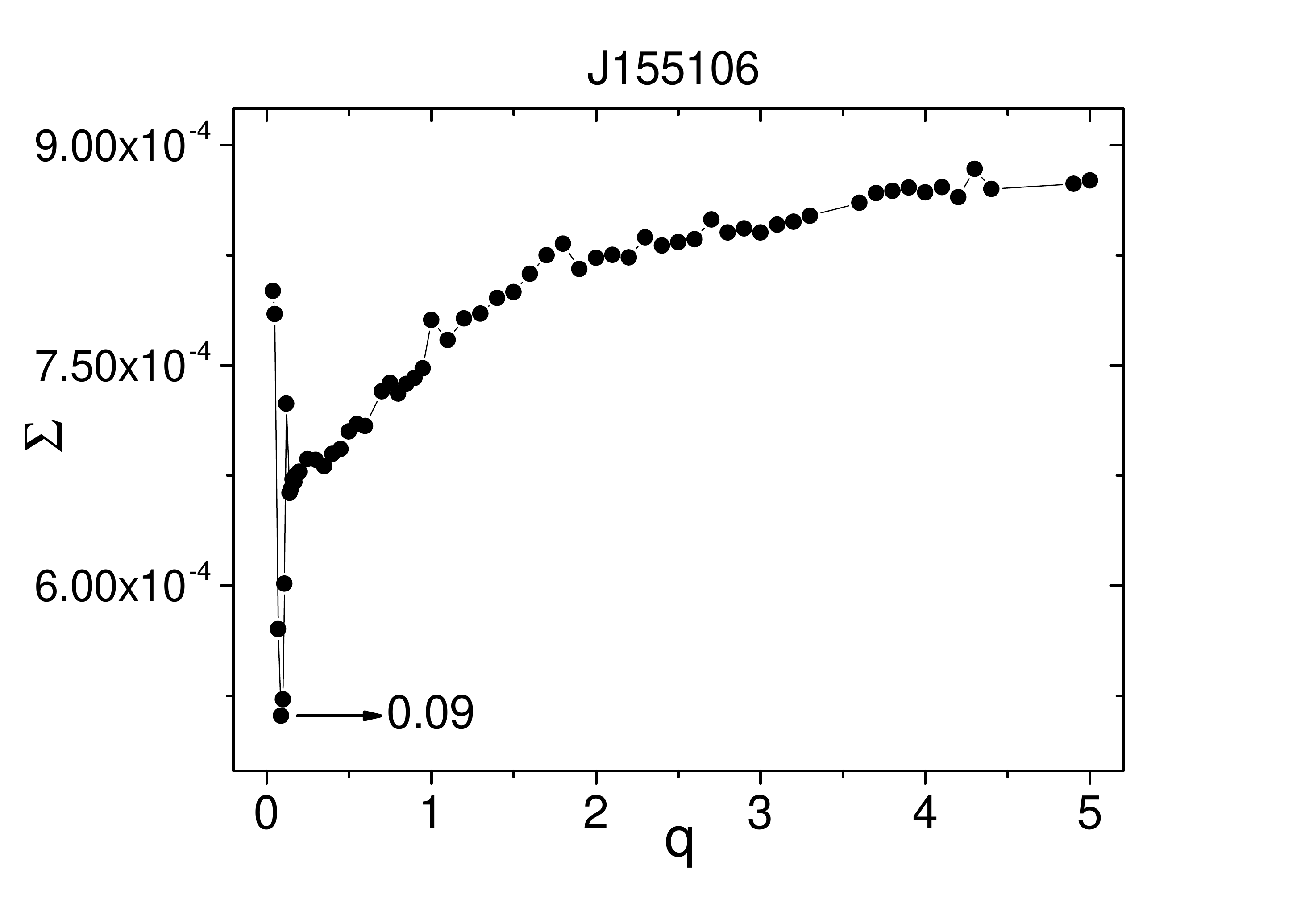}}
\subfigure{\label{fig:subfig:b}
\includegraphics[width=0.32\linewidth]{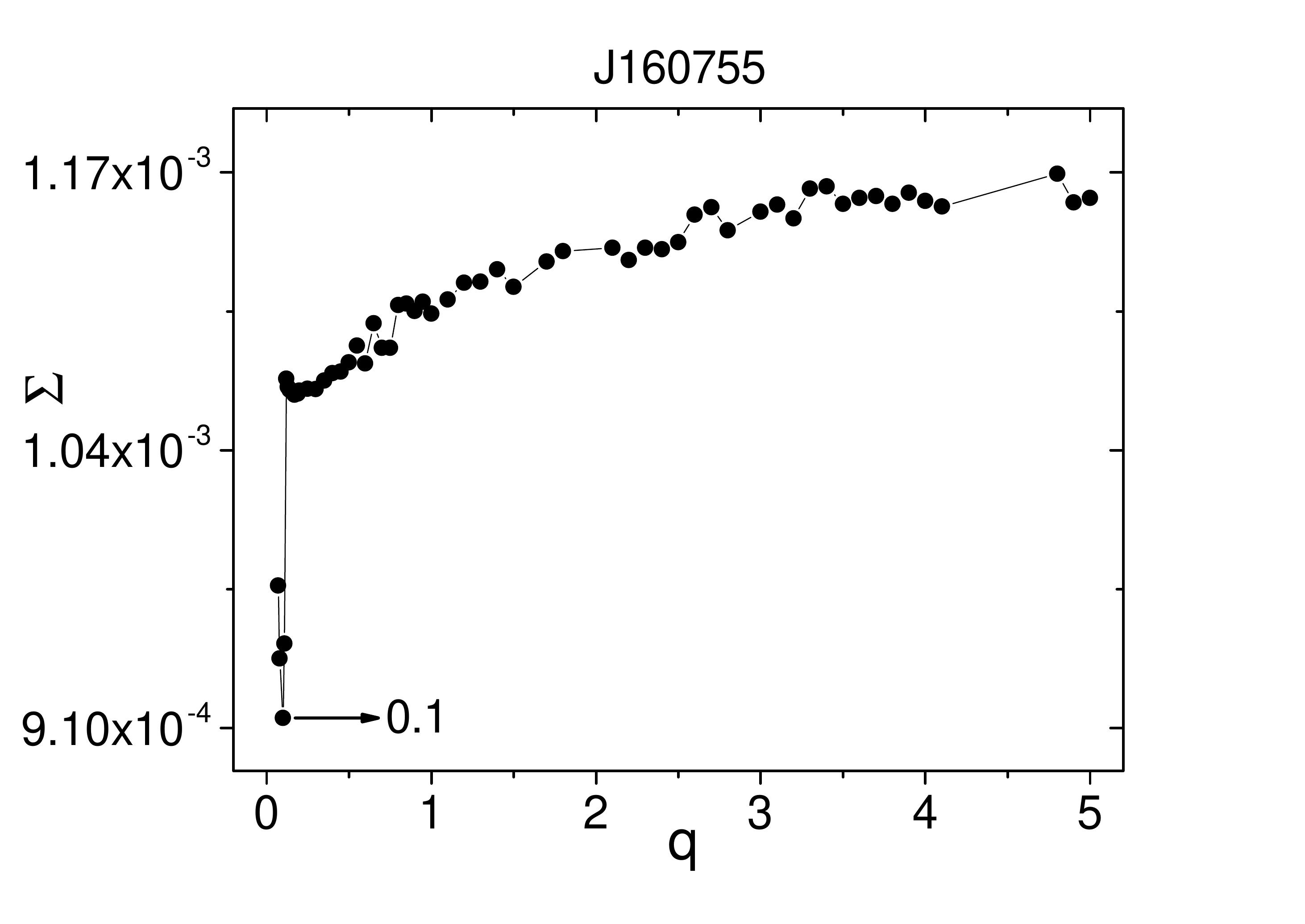}}
\subfigure{\label{fig:subfig:c}
\includegraphics[width=0.32\linewidth]{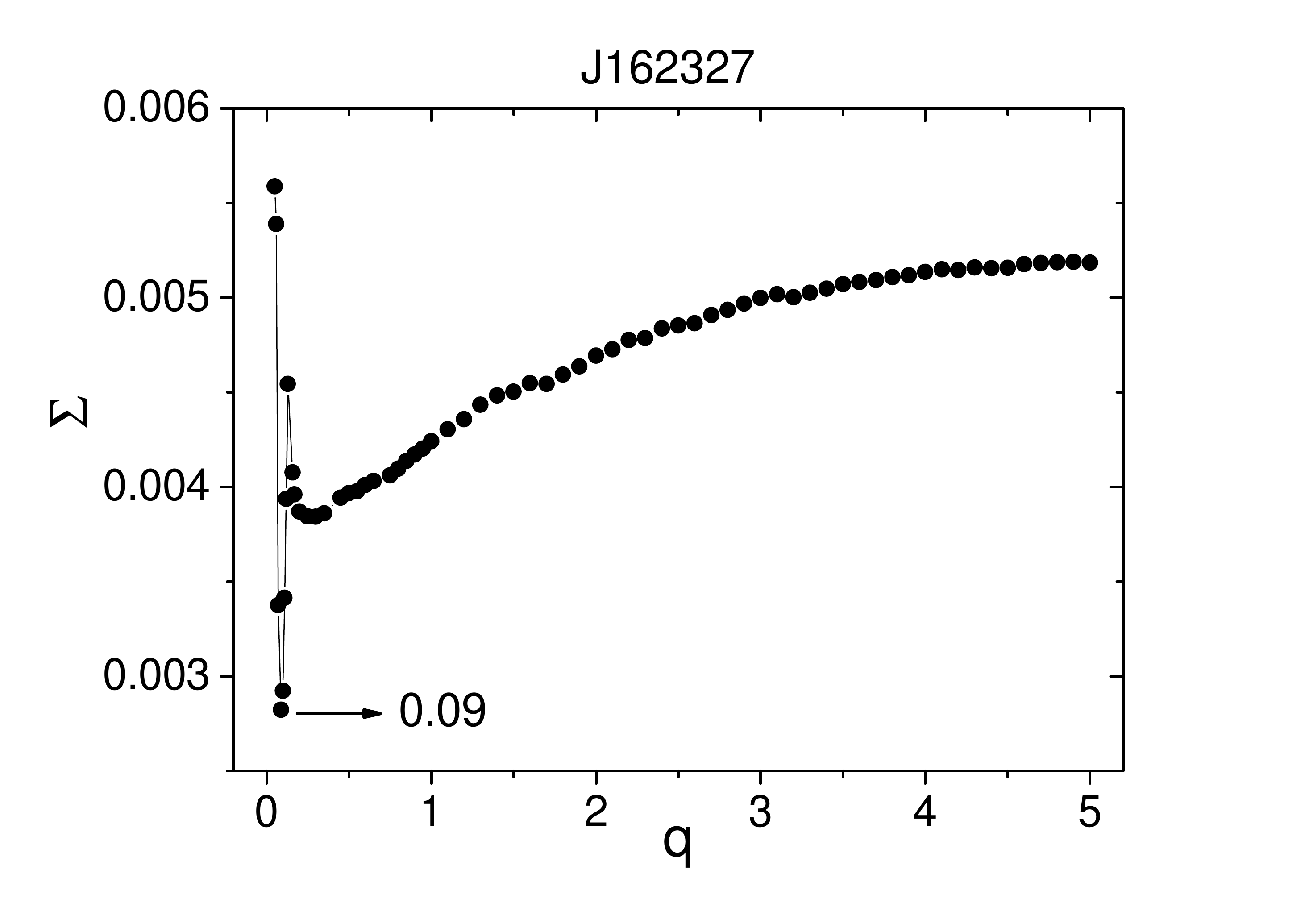}}
\subfigure{\label{fig:subfig:a}
\includegraphics[width=0.32\linewidth]{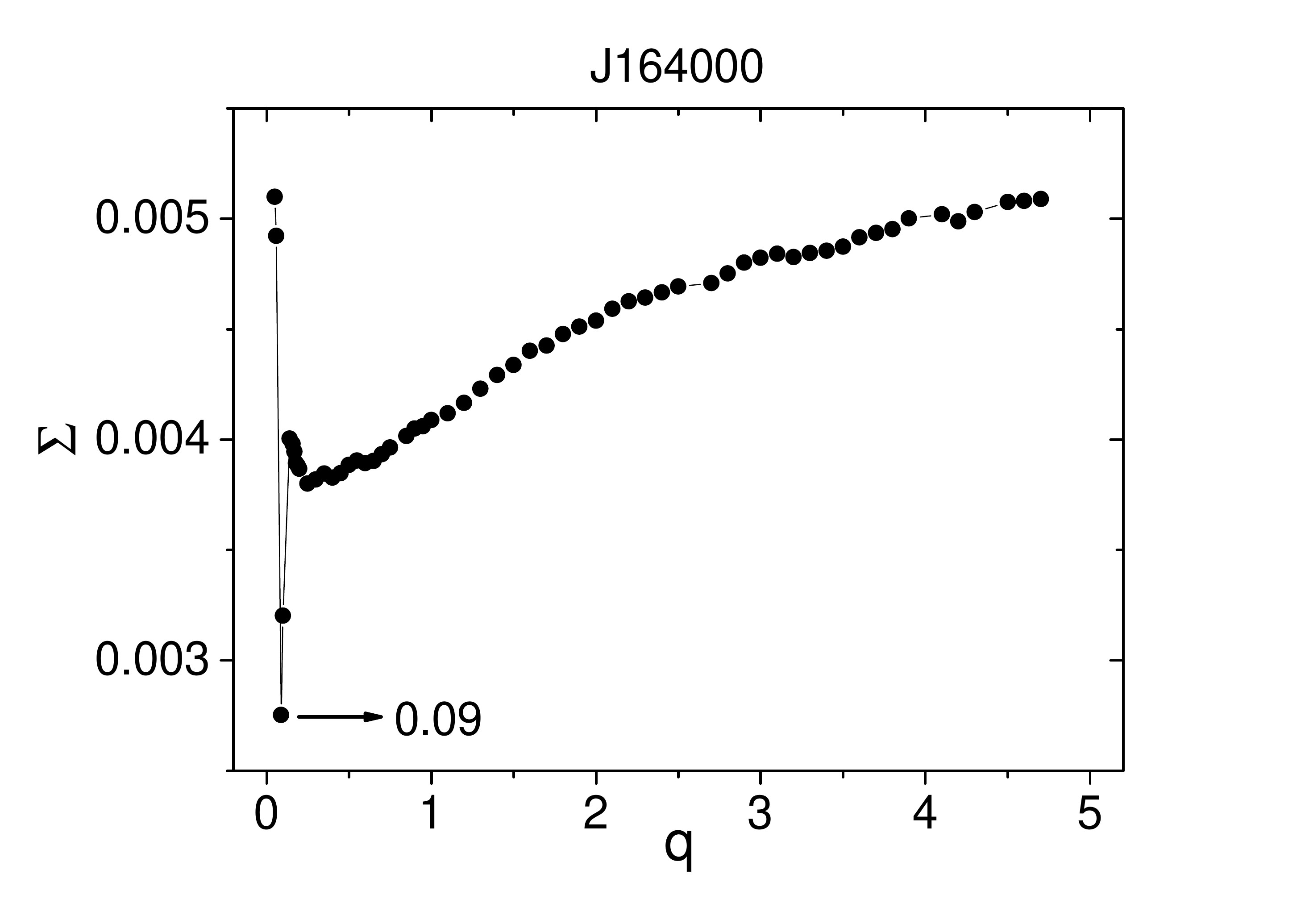}}
\subfigure{\label{fig:subfig:b}
\includegraphics[width=0.32\linewidth]{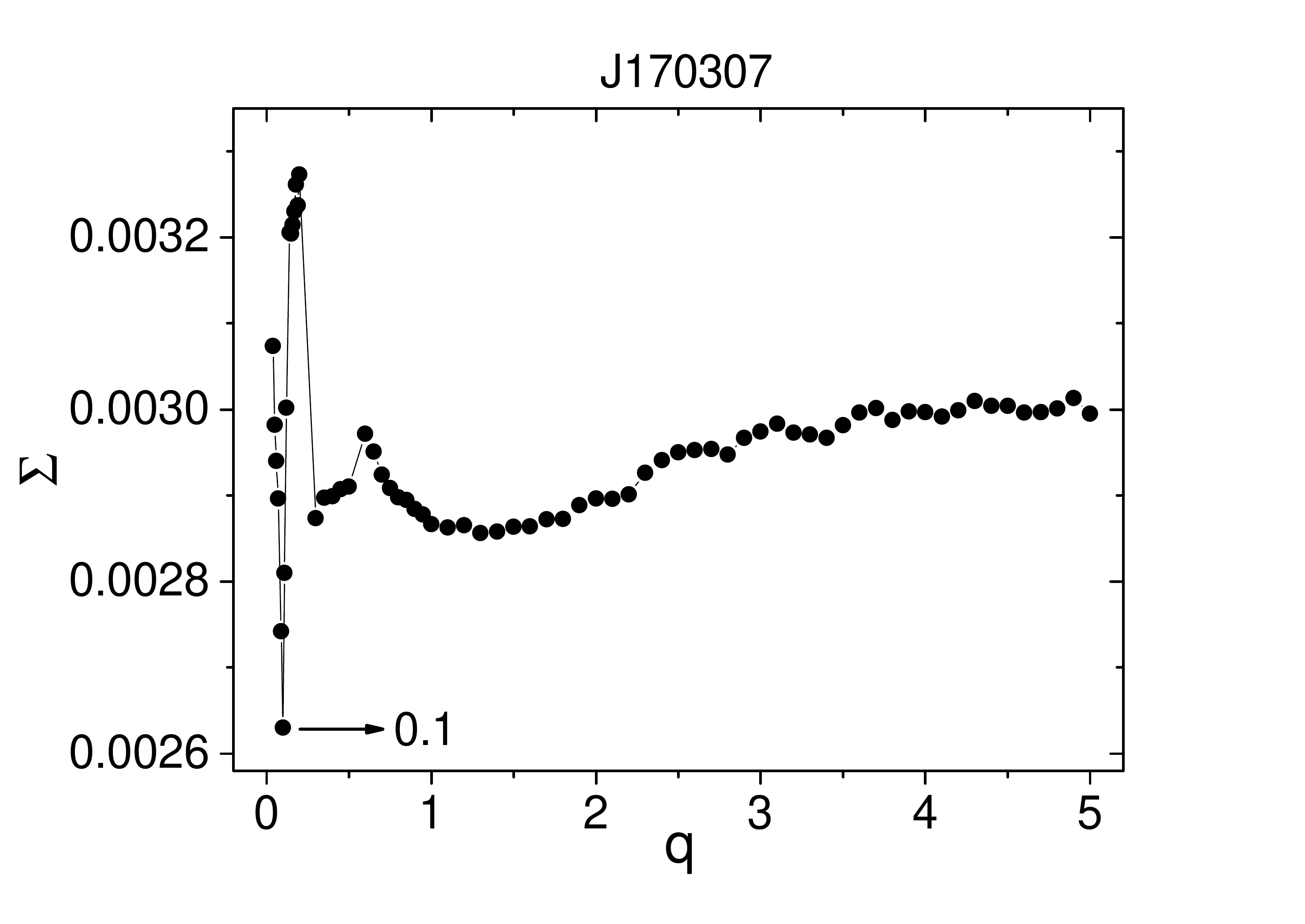}}
\subfigure{\label{fig:subfig:c}
\includegraphics[width=0.32\linewidth]{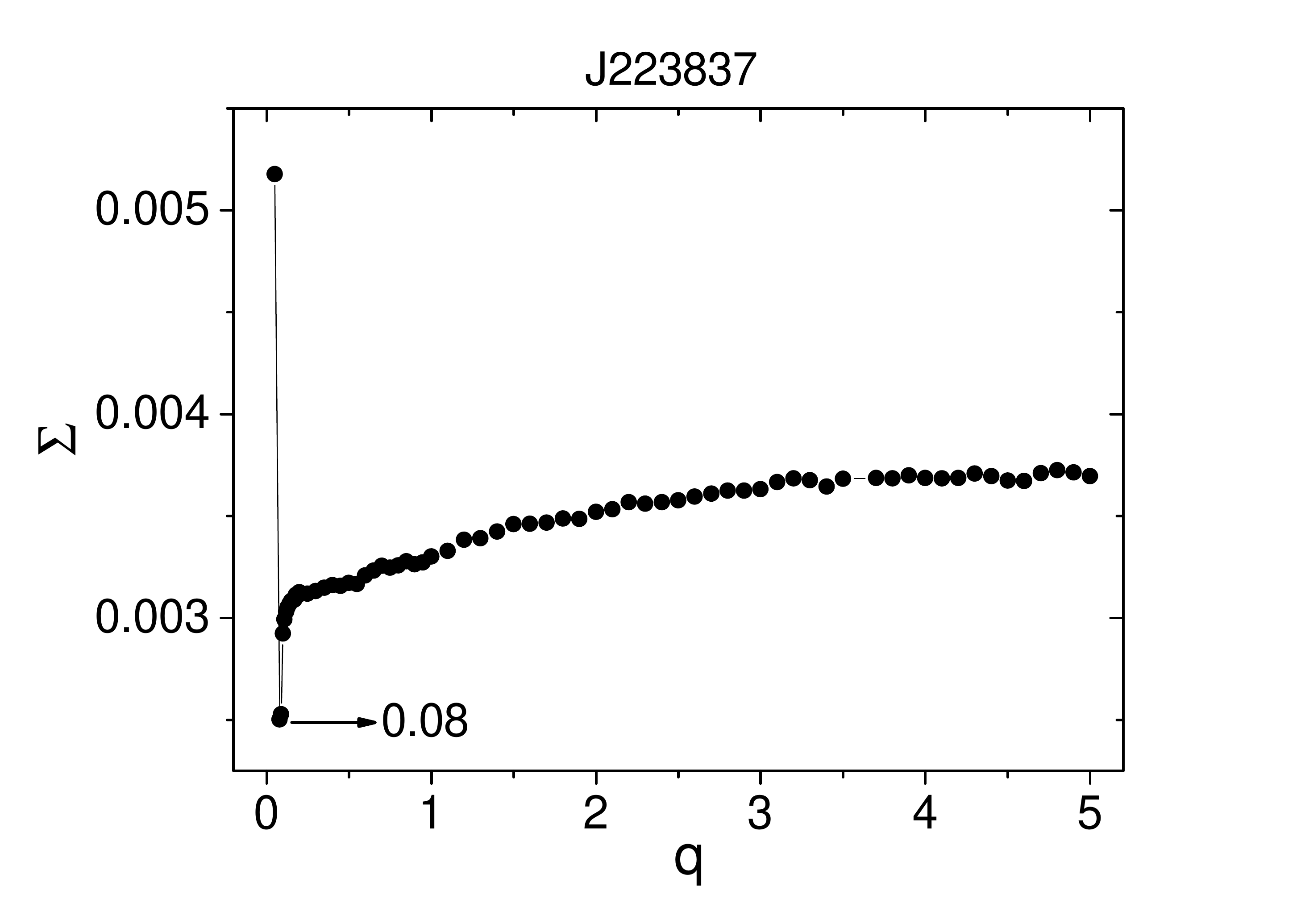}}
\subfigure{\label{fig:subfig:a}
\includegraphics[width=0.35\linewidth]{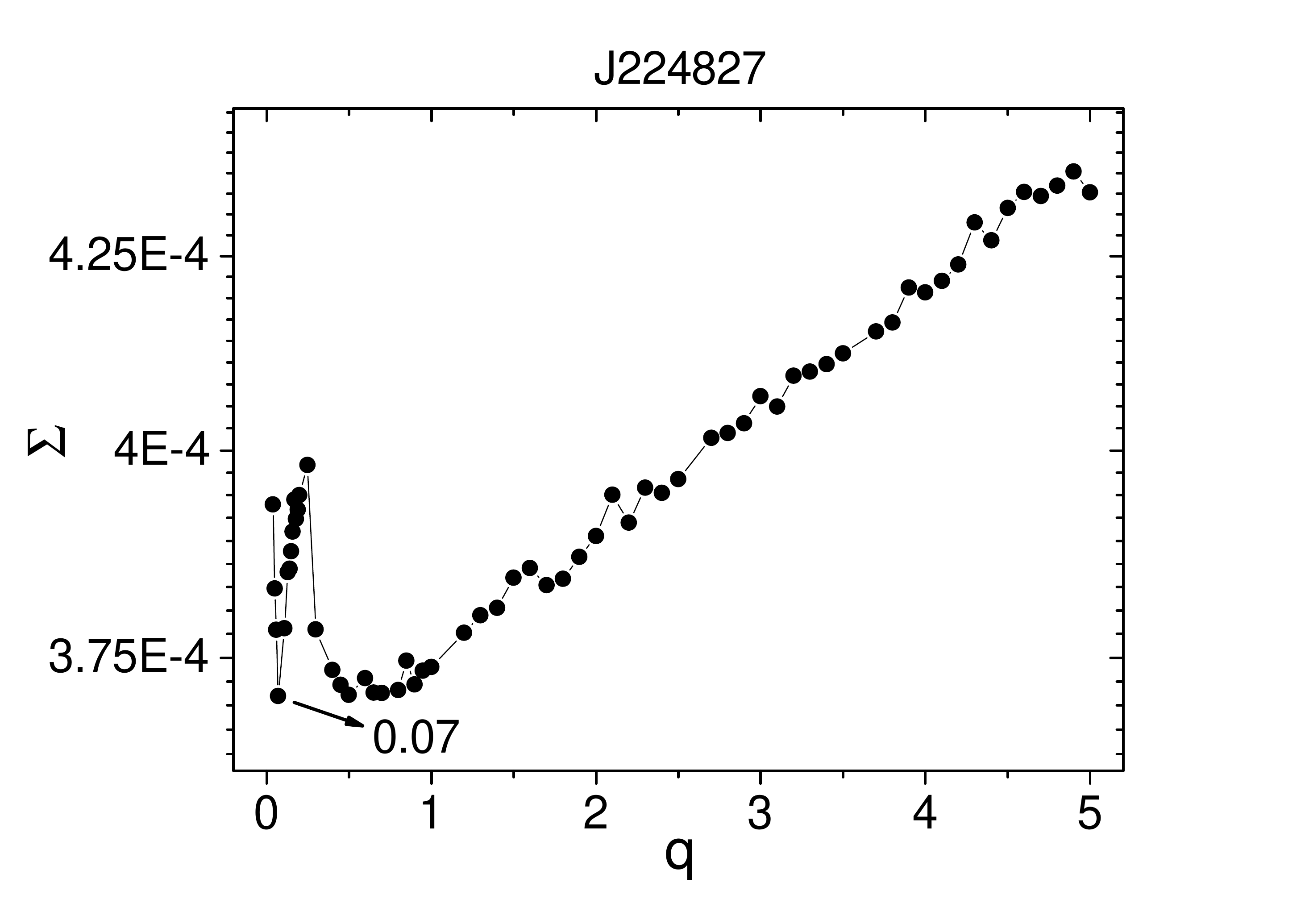}}
\subfigure{\label{fig:subfig:b}
\includegraphics[width=0.35\linewidth]{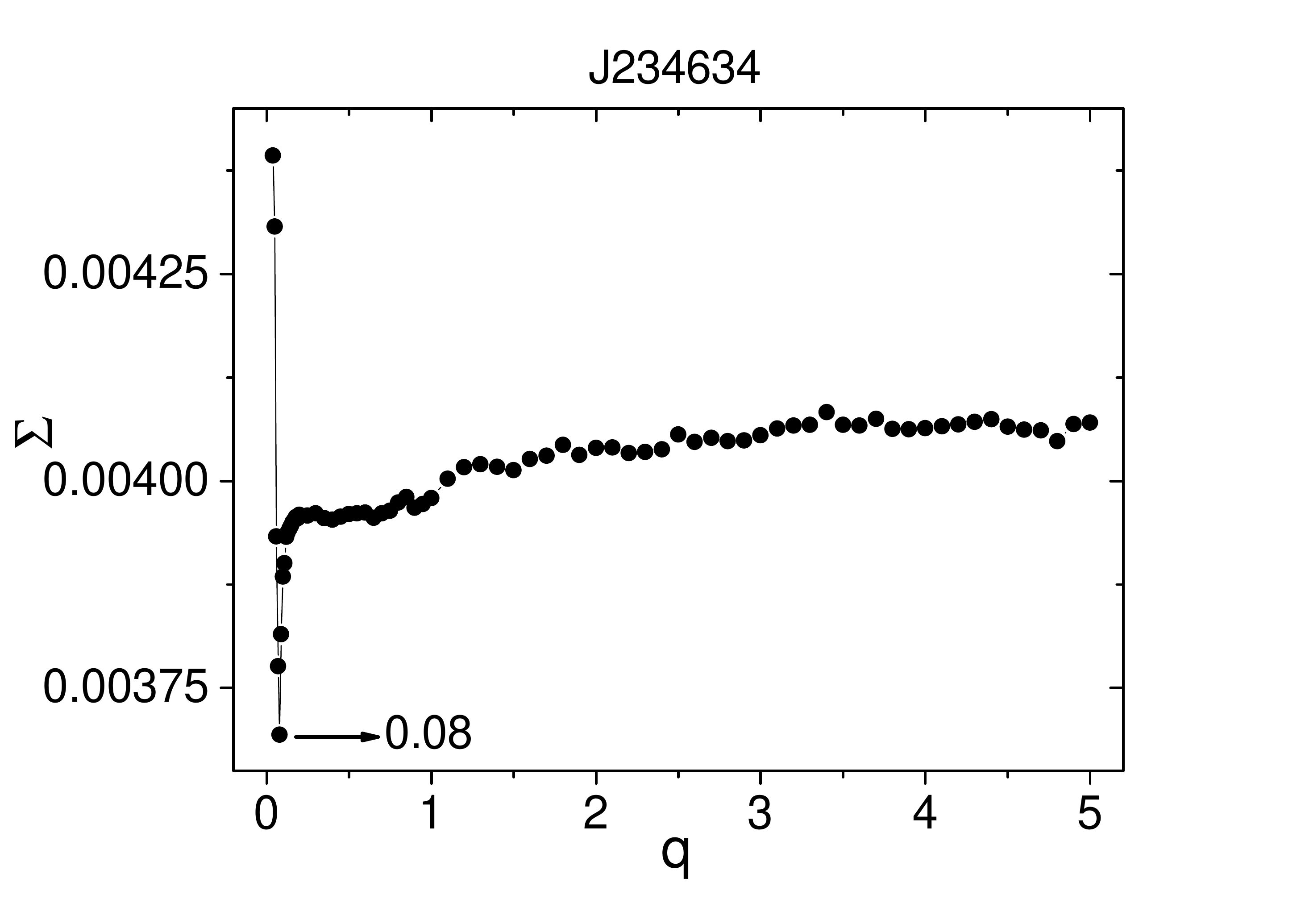}}
\caption{This figure displays the sum square of residuals $\Sigma$ versus mass ratio q.}
\end{figure*}

\begin{figure*}
\centering
\subfigure{\label{fig:subfig:a}
\includegraphics[width=0.32\linewidth]{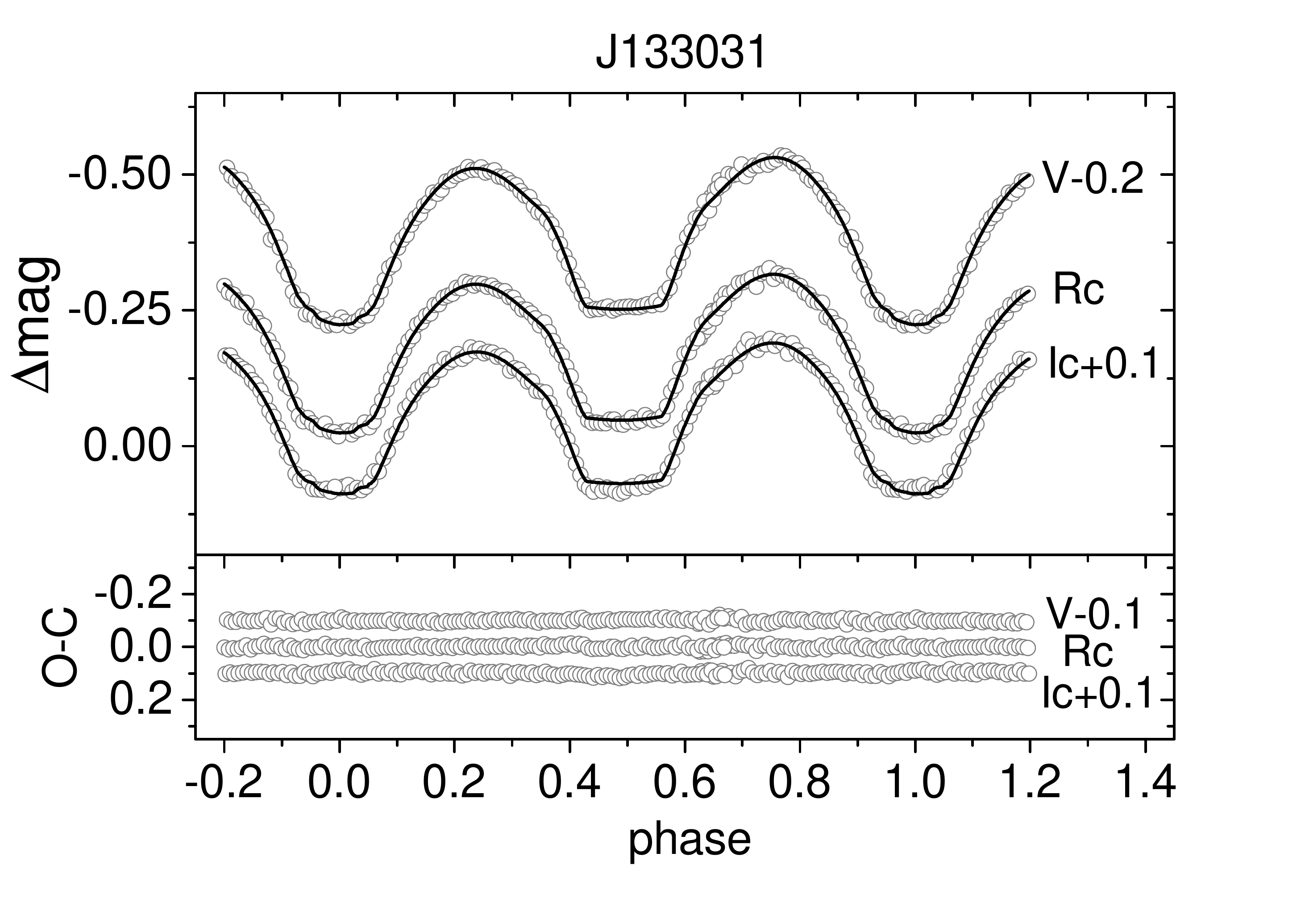}}
\subfigure{\label{fig:subfig:b}
\includegraphics[width=0.32\linewidth]{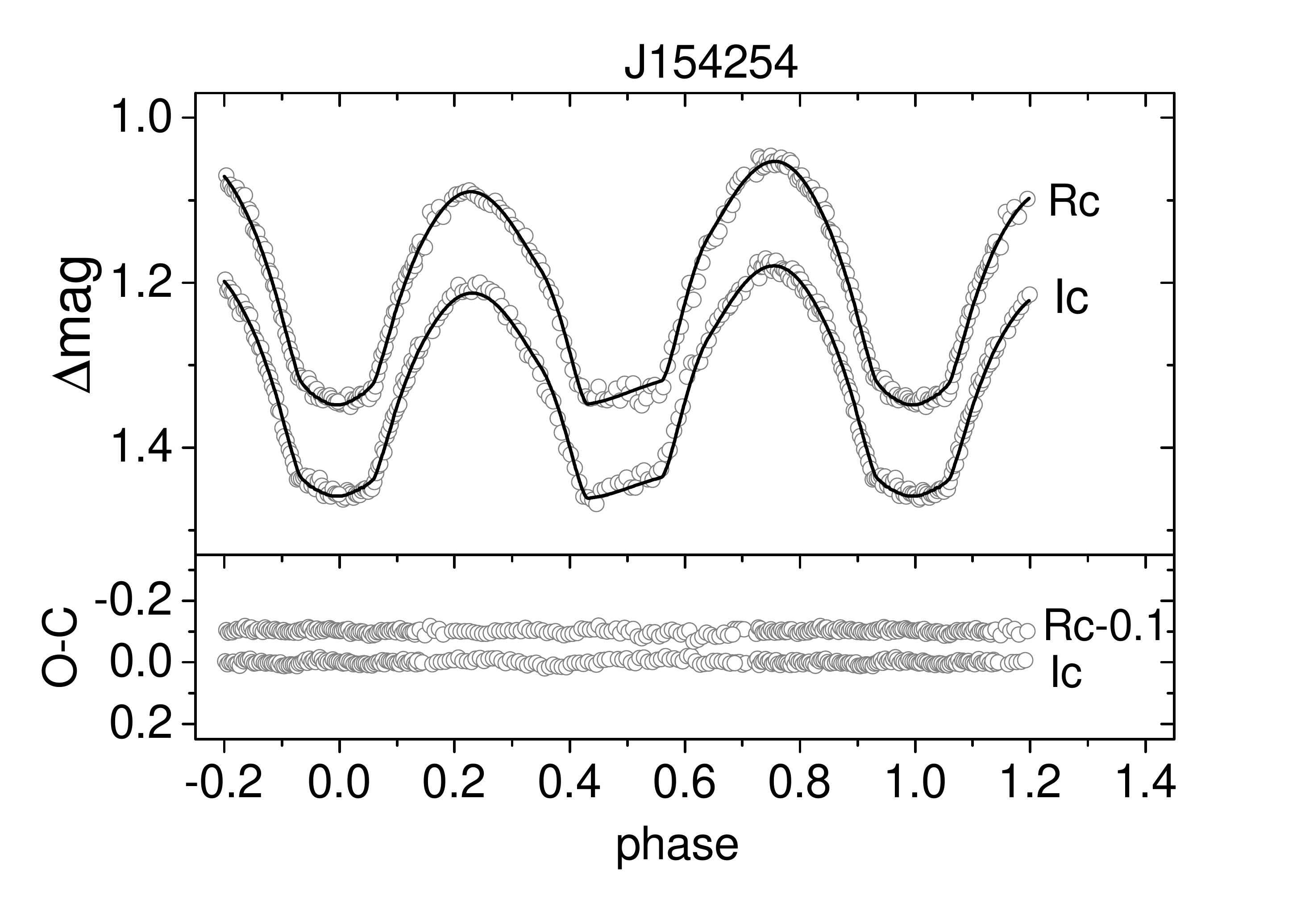}}
\subfigure{\label{fig:subfig:c}
\includegraphics[width=0.32\linewidth]{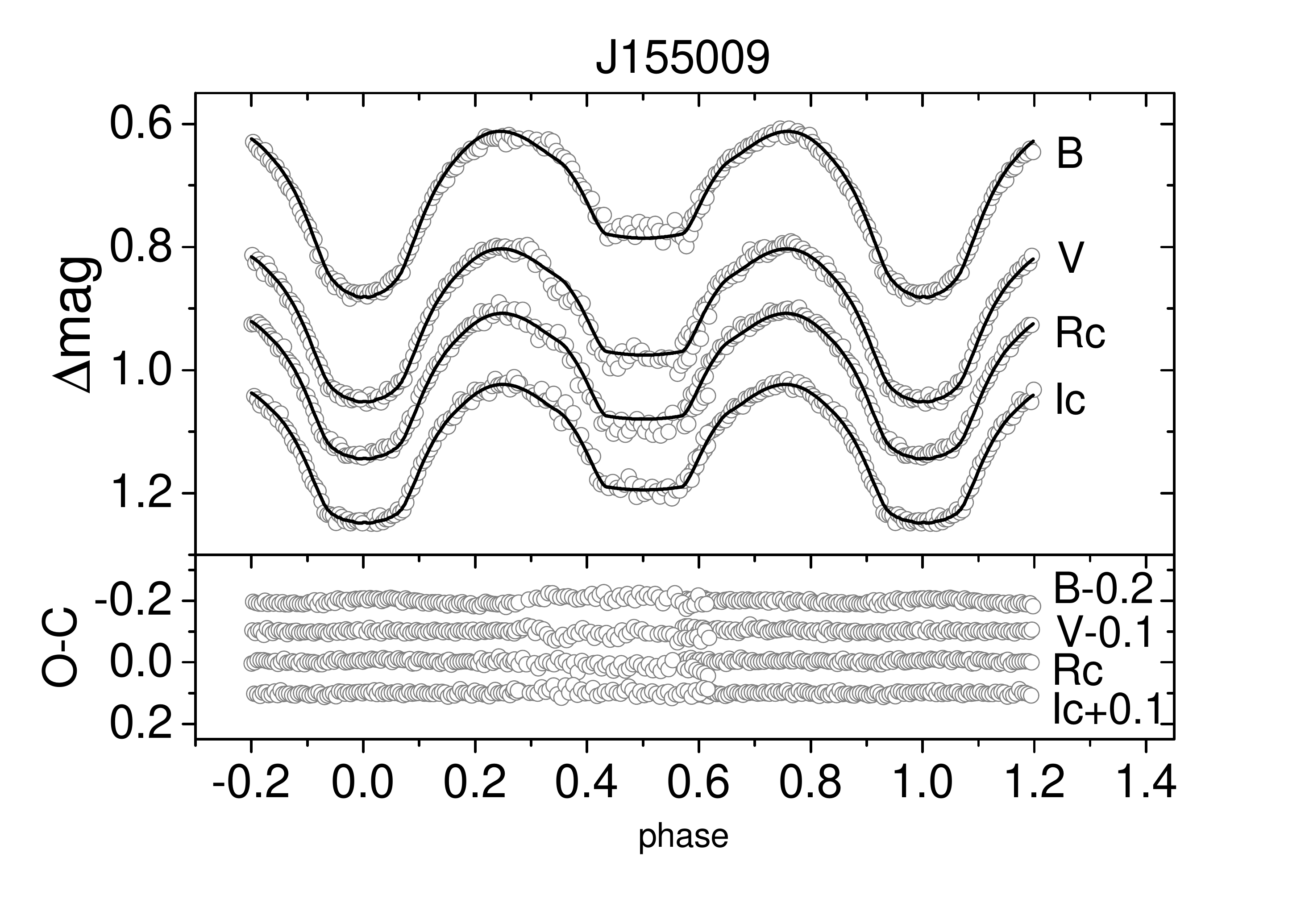}}
\subfigure{\label{fig:subfig:a}
\includegraphics[width=0.32\linewidth]{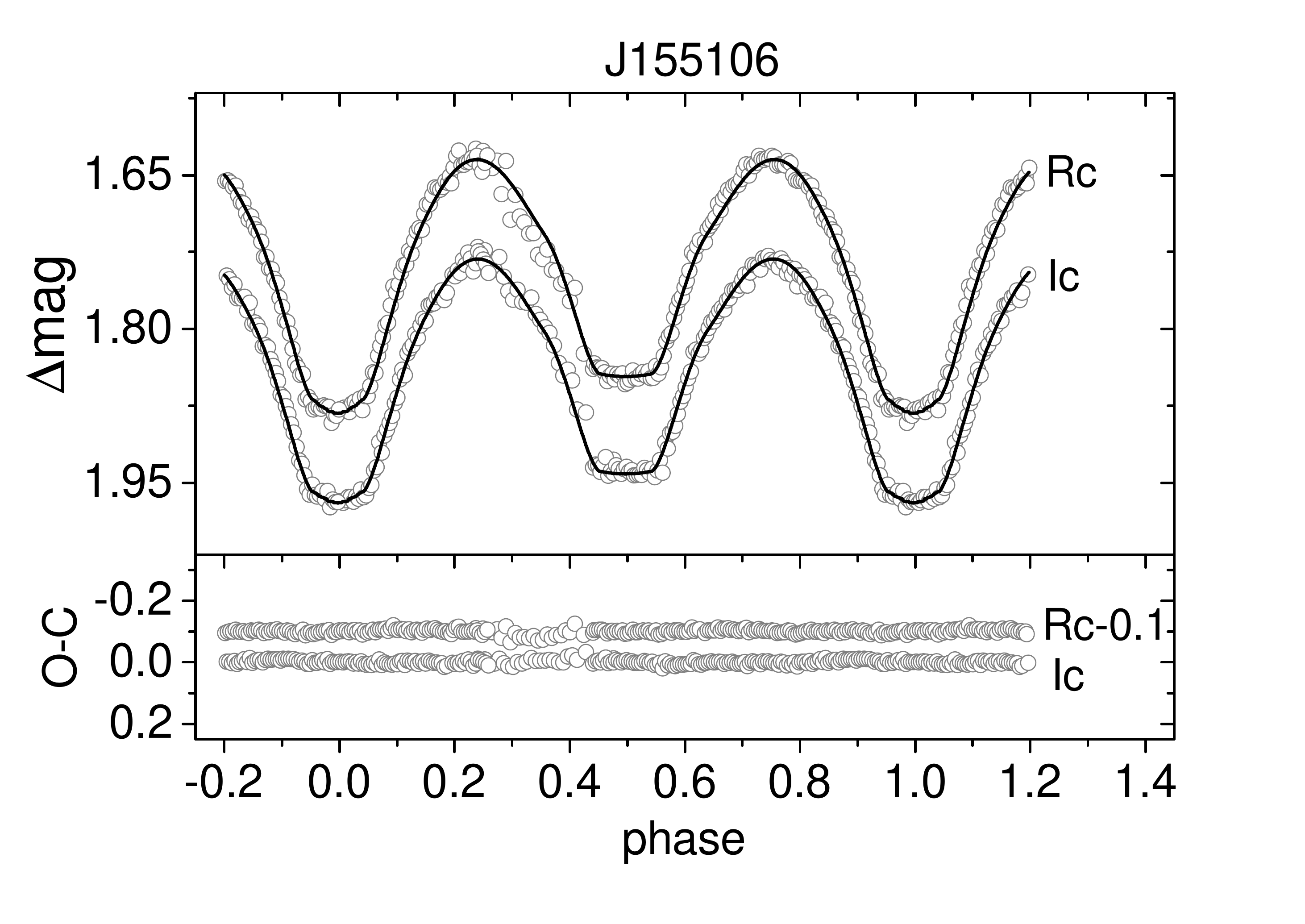}}
\subfigure{\label{fig:subfig:b}
\includegraphics[width=0.32\linewidth]{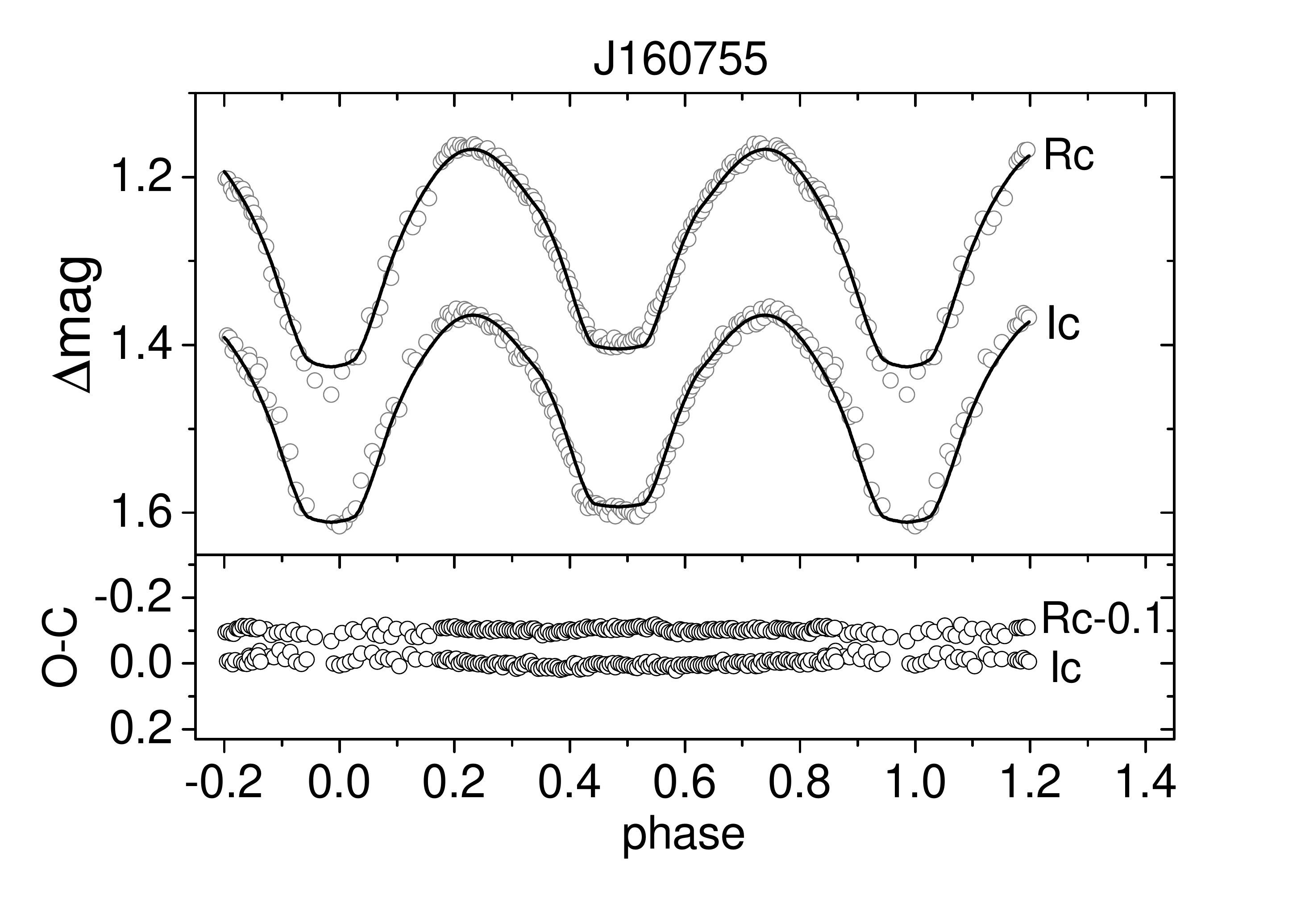}}
\subfigure{\label{fig:subfig:c}
\includegraphics[width=0.32\linewidth]{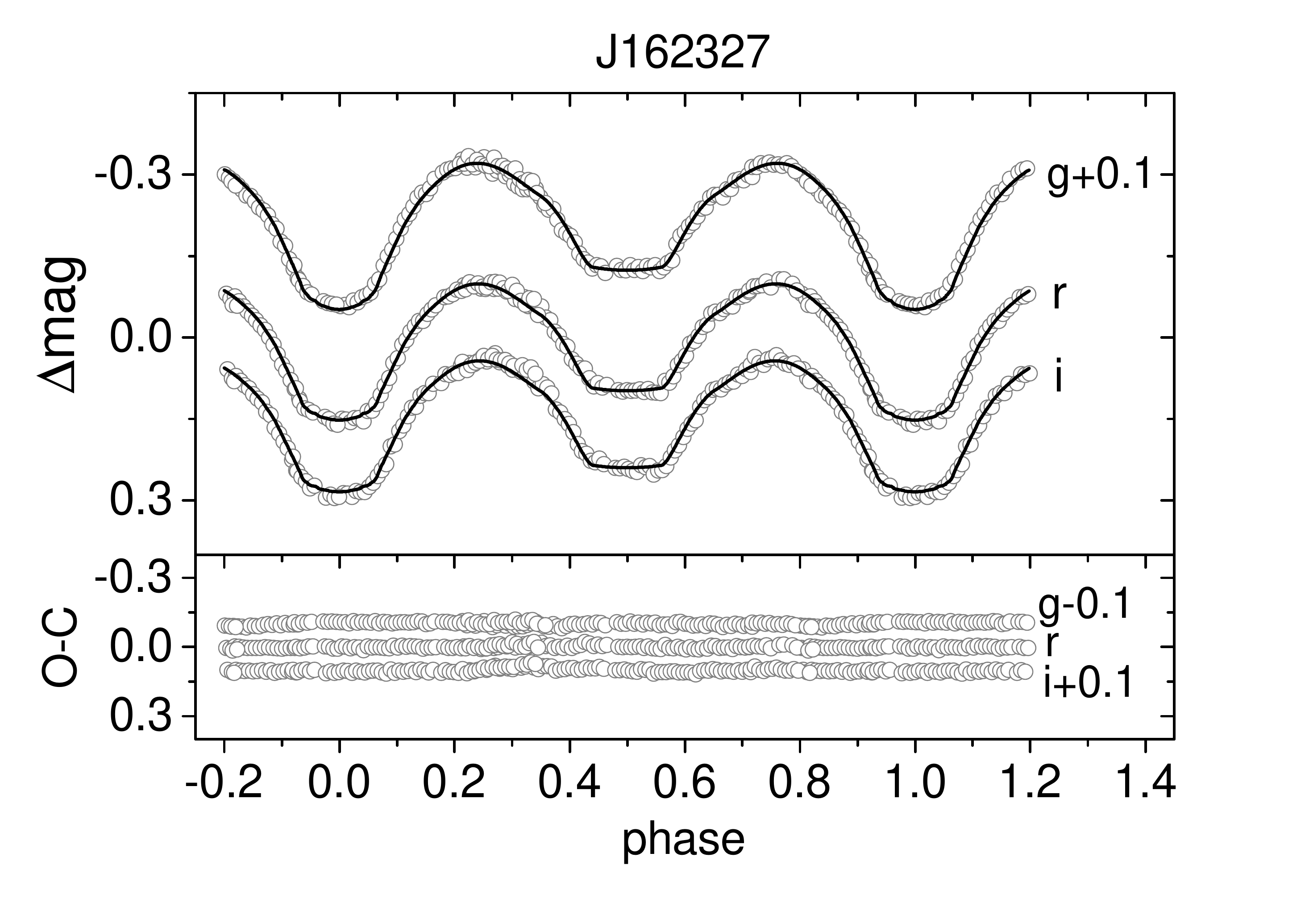}}
\subfigure{\label{fig:subfig:a}
\includegraphics[width=0.32\linewidth]{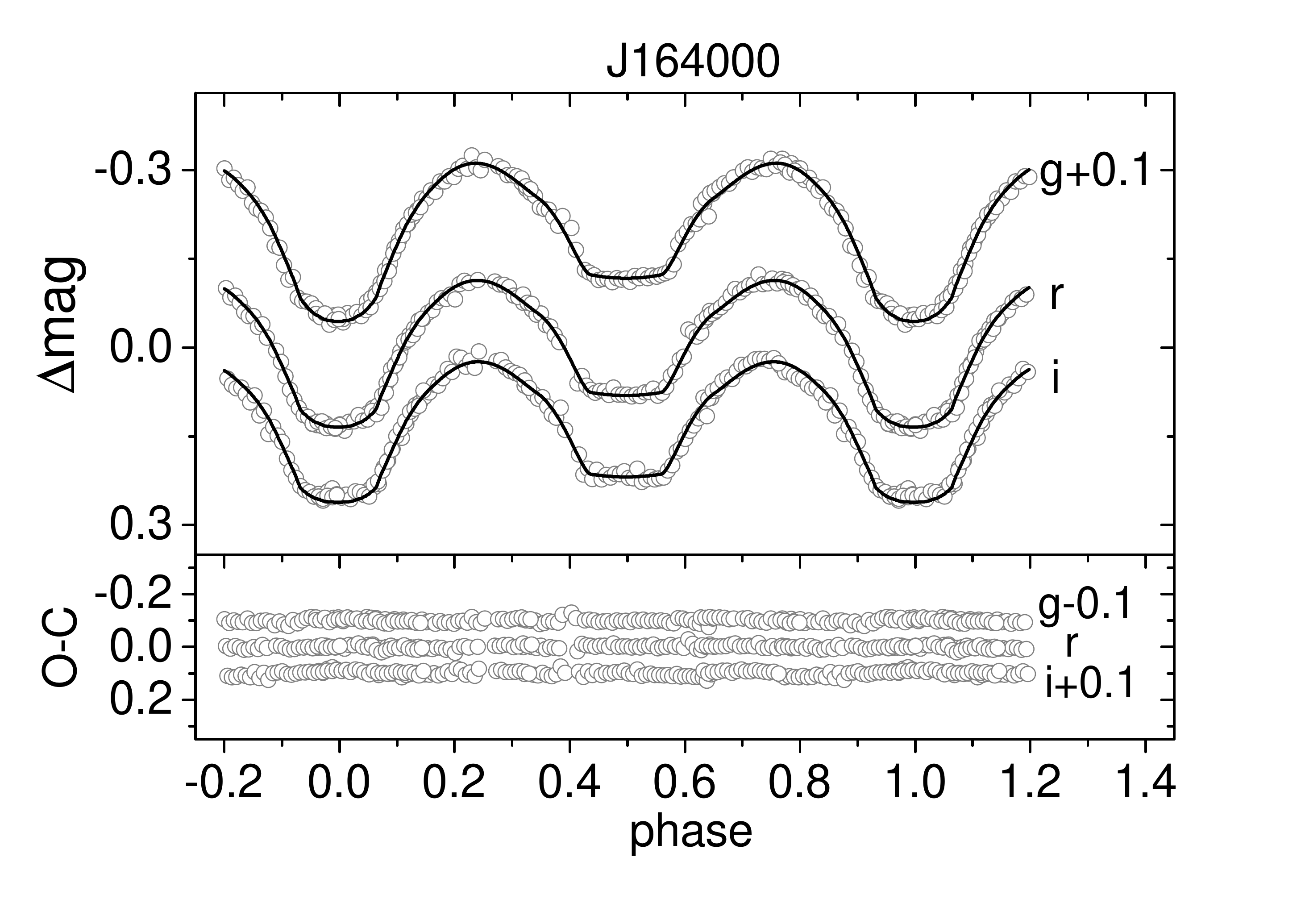}}
\subfigure{\label{fig:subfig:b}
\includegraphics[width=0.32\linewidth]{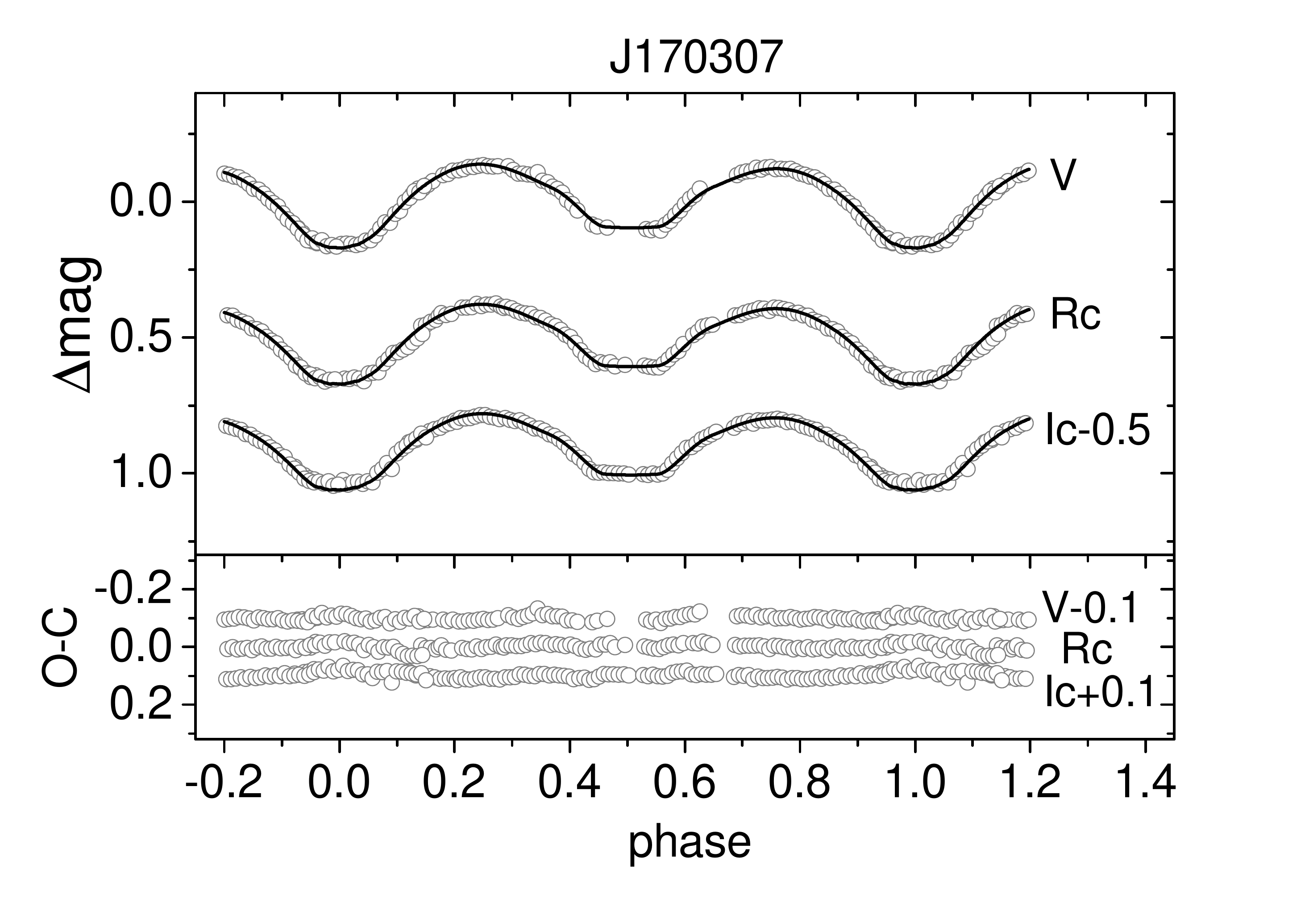}}
\subfigure{\label{fig:subfig:c}
\includegraphics[width=0.32\linewidth]{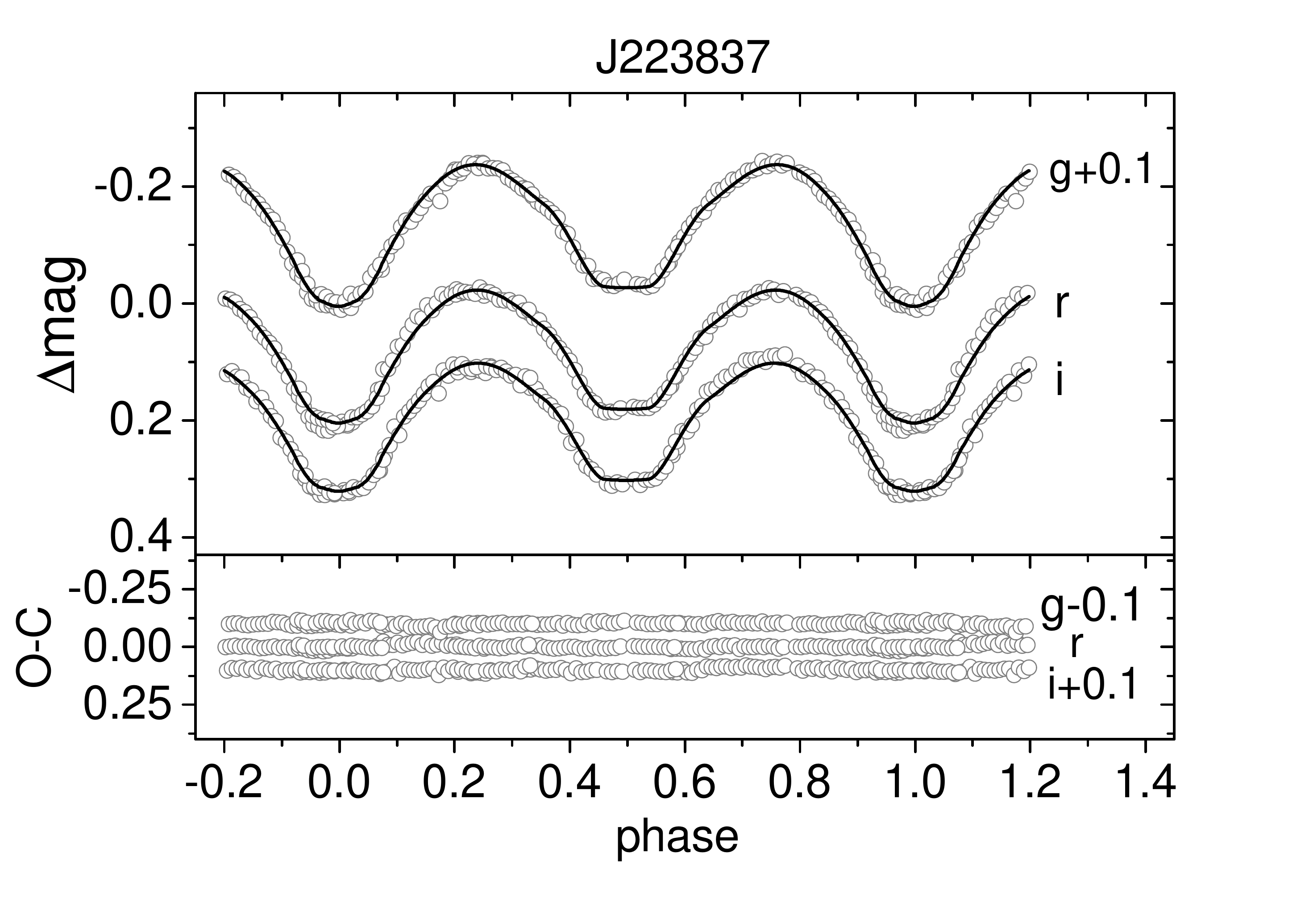}}
\subfigure{\label{fig:subfig:a}
\includegraphics[width=0.32\linewidth]{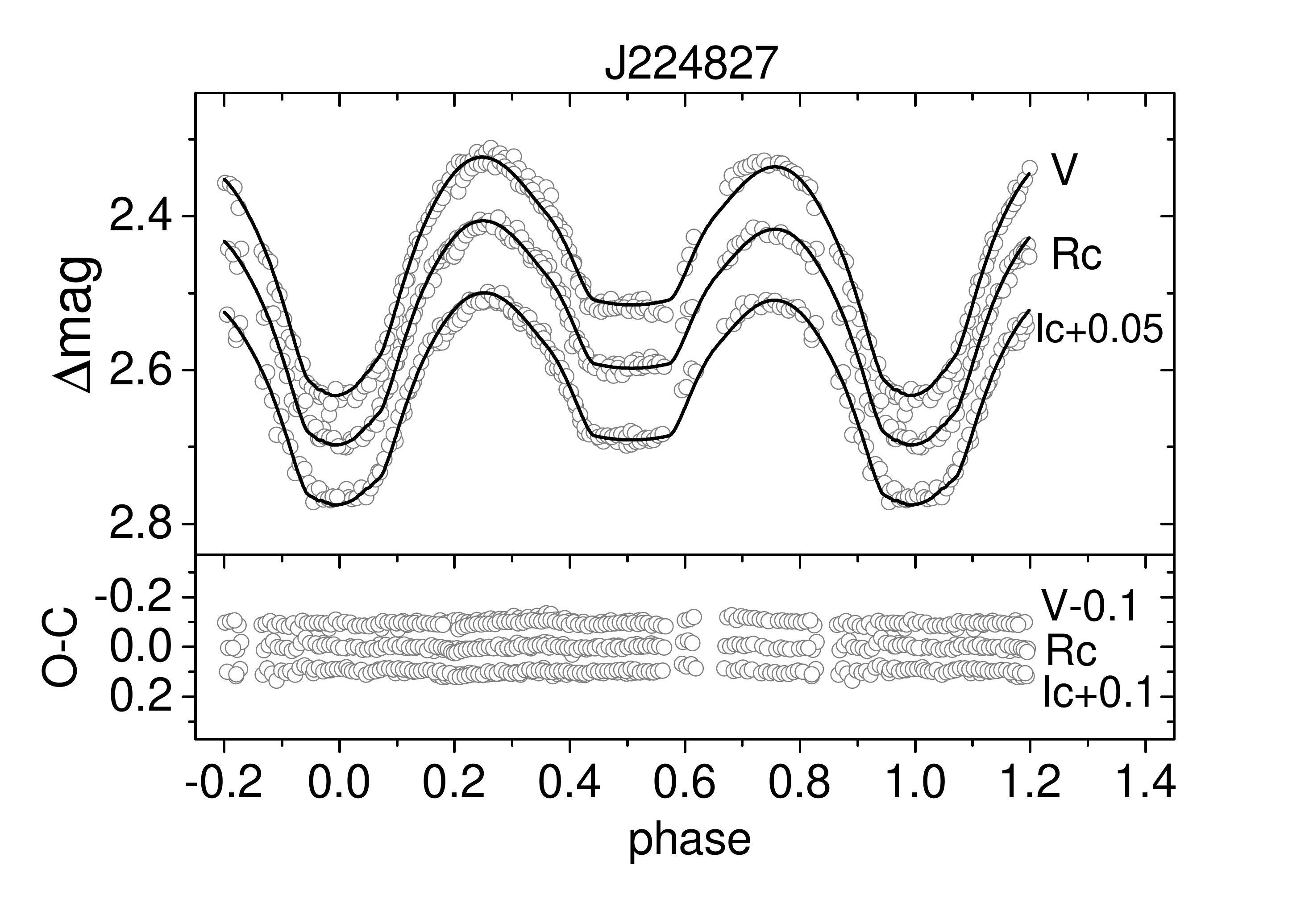}}
\subfigure{\label{fig:subfig:b}
\includegraphics[width=0.32\linewidth]{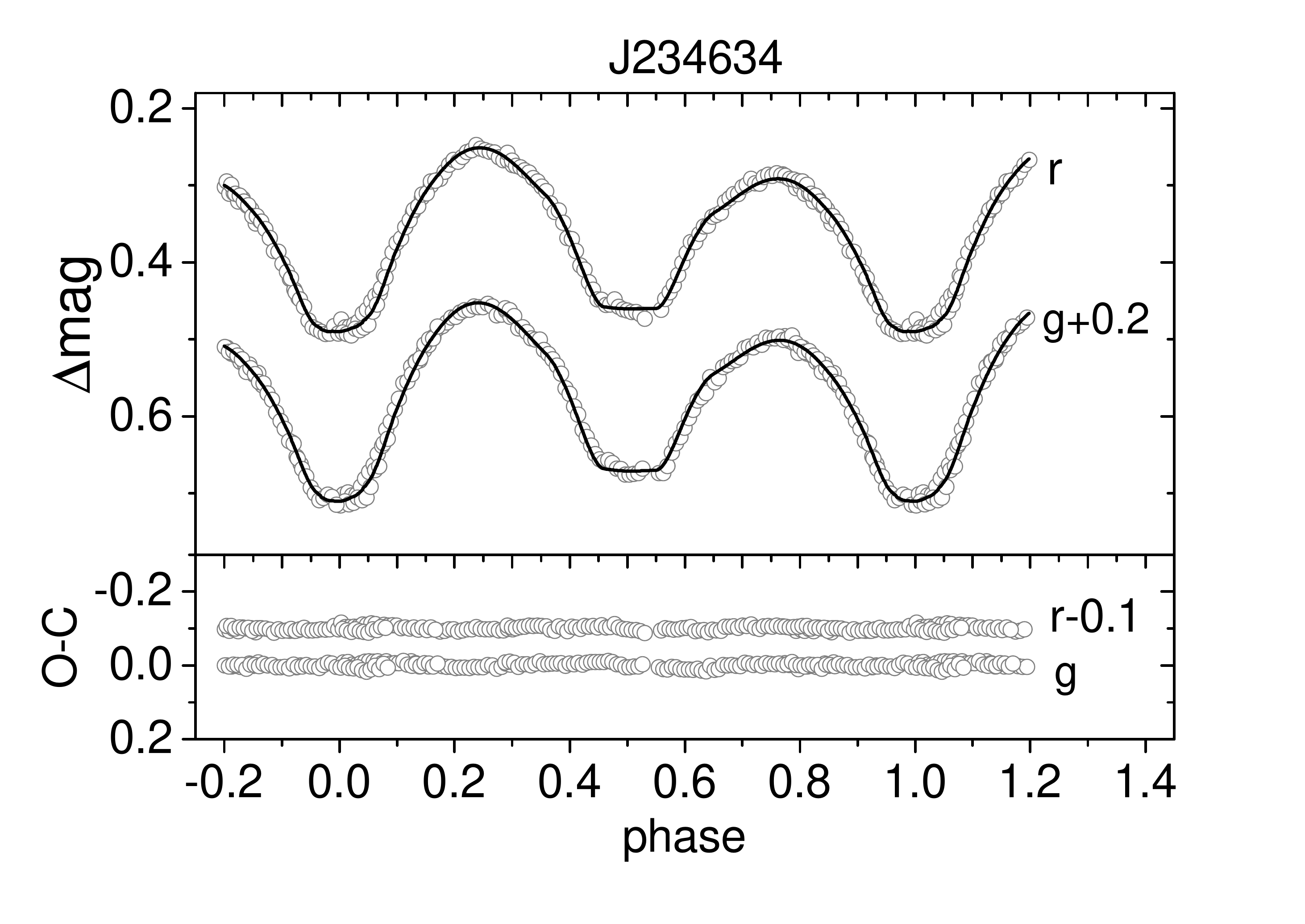}}
\caption{This figure displays observed (symbols) and theoretical (solid lines) light curves.}
\centering
\end{figure*}

\begin{figure*}
\centering
\subfigure{\label{fig:subfig:a}
\includegraphics[width=0.45\linewidth]{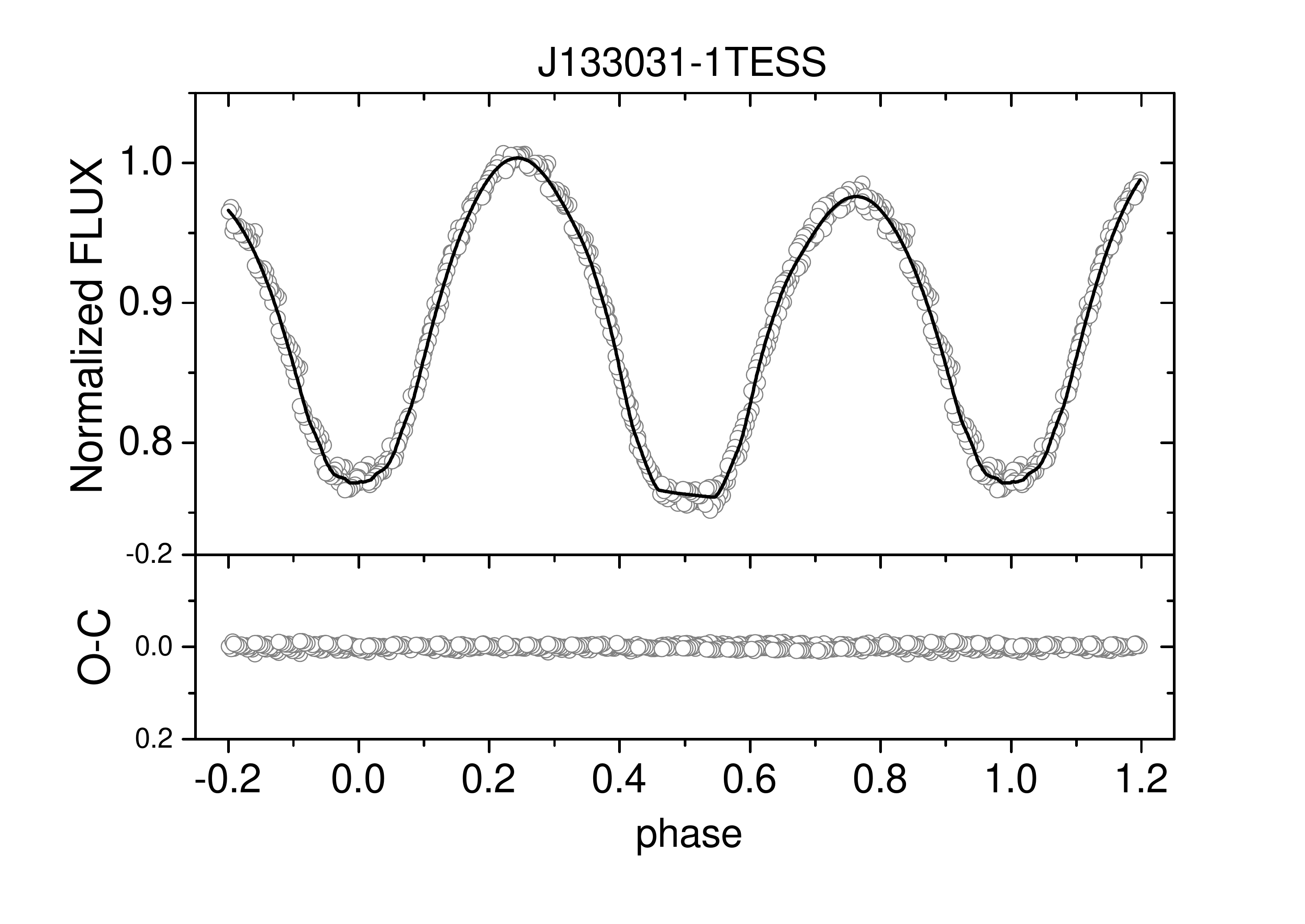}}
\subfigure{\label{fig:subfig:b}
\includegraphics[width=0.45\linewidth]{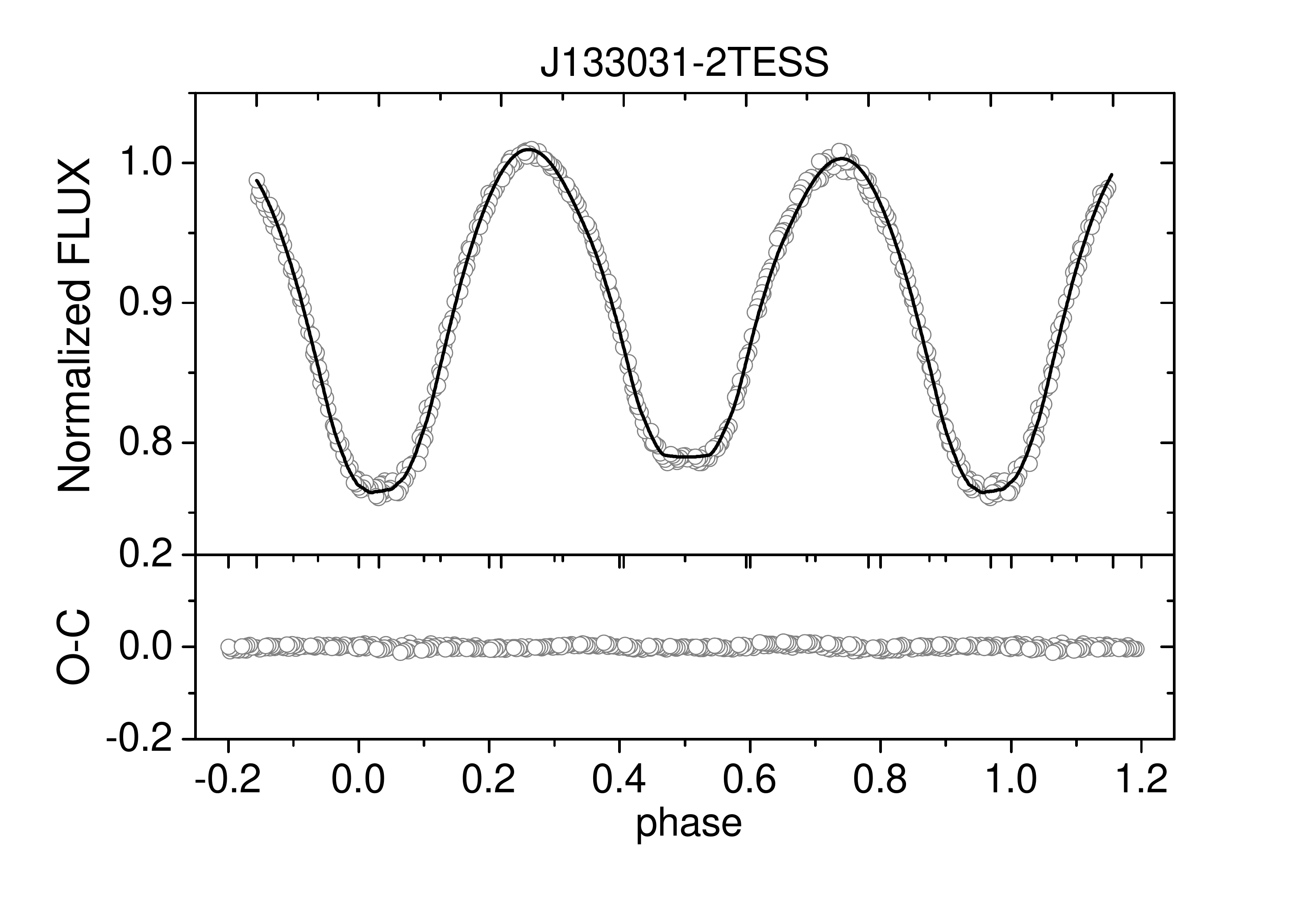}}
\subfigure{\label{fig:subfig:a}
\includegraphics[width=0.45\linewidth]{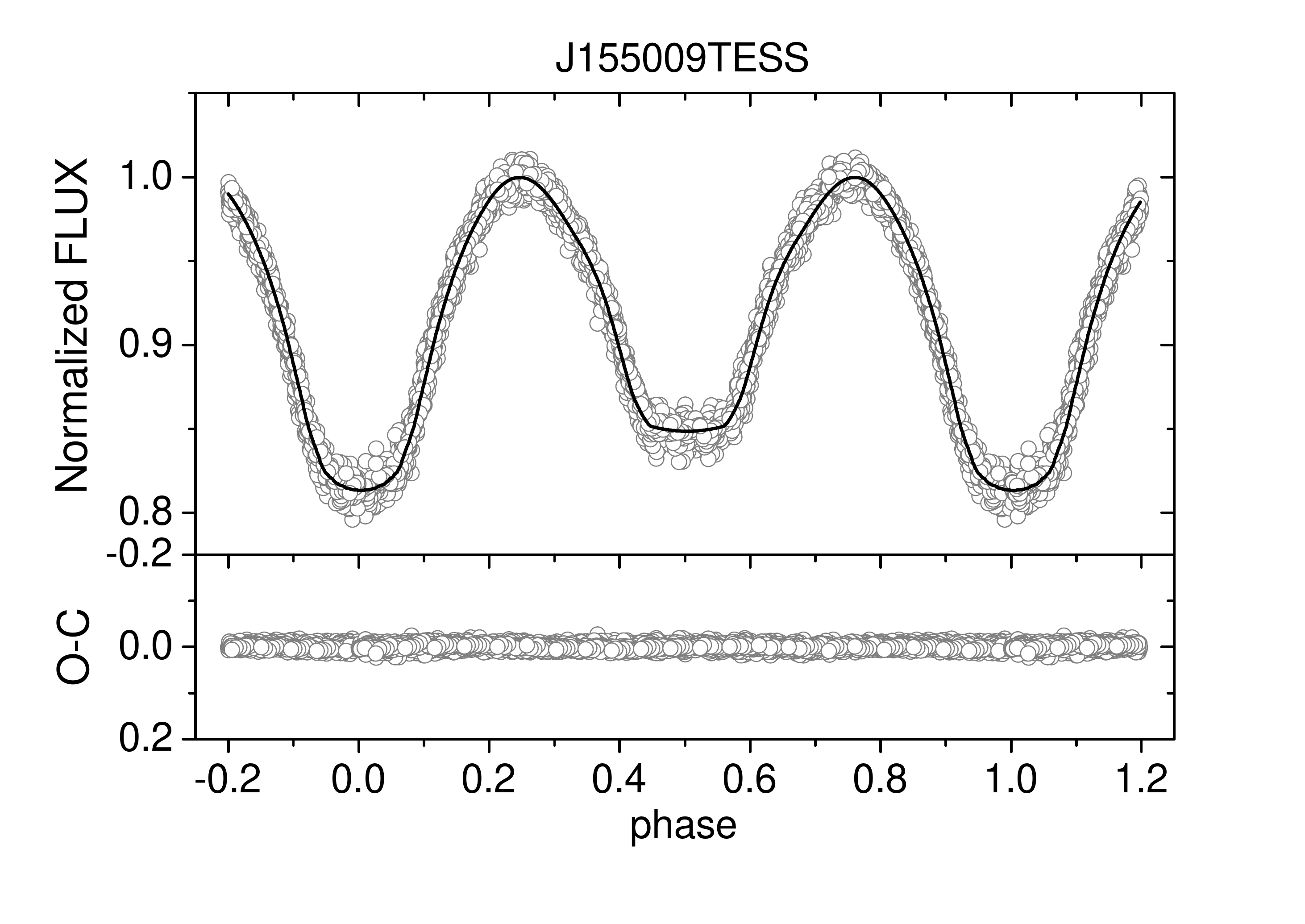}}
\subfigure{\label{fig:subfig:b}
\includegraphics[width=0.45\linewidth]{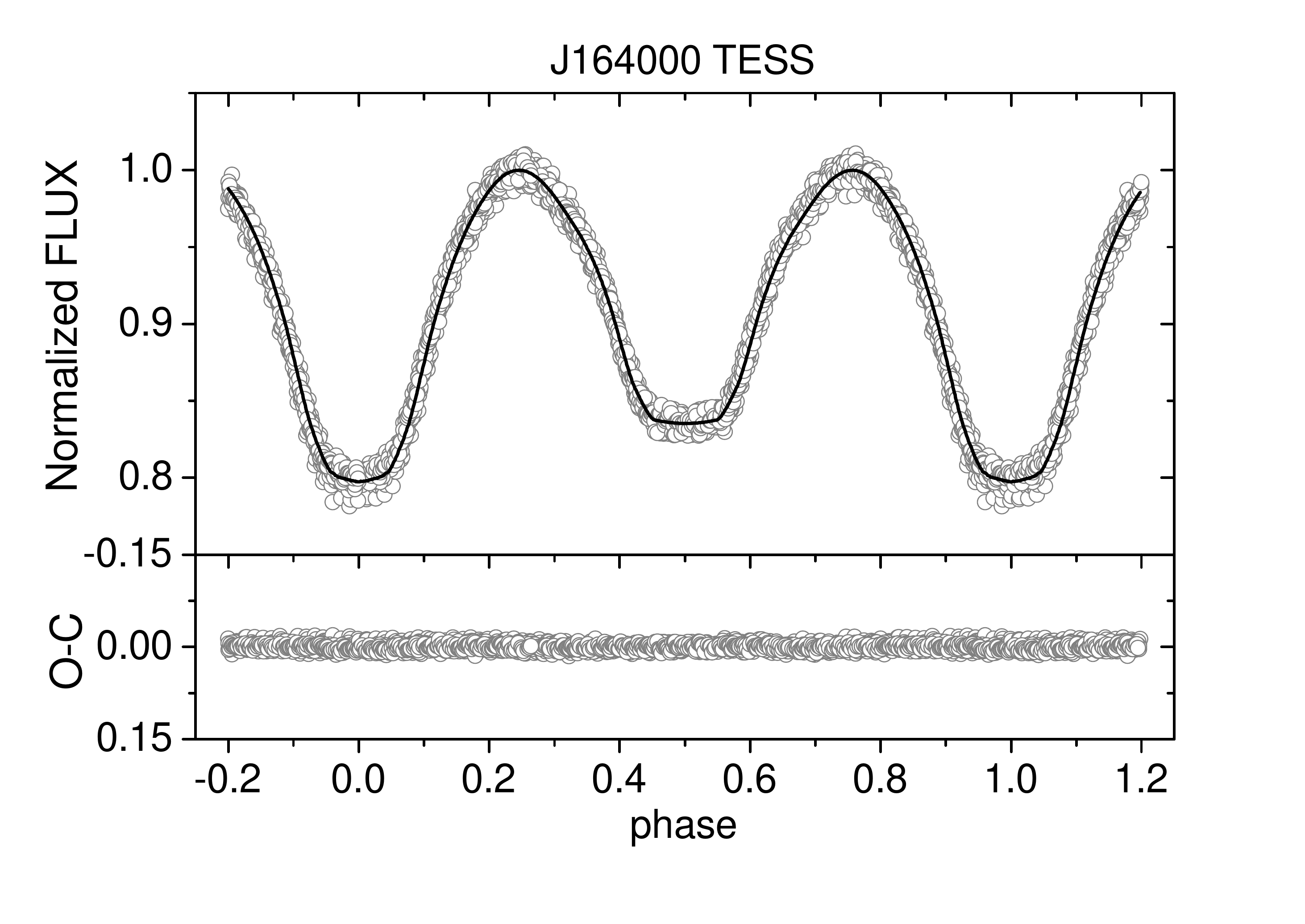}}
\caption{This figure displays observed (symbols) and theoretical (solid lines) light curves of TESS observation data.}
\centering
\end{figure*}

\begin{table*}
\begin{center}
\renewcommand\tabcolsep{1.2pt} 
\caption{Photometric Solutions}
\label{table5}
\small
\begin{tabular}{lllllllllllllllllll}
\hline
Target          &J133031       &J154254       & J155009       &J155106       &J160755       &J162327       &J164000       &J170307       &J223837       &J224827       &J234634        \\
\hline
$T_{1}'$(K)     &5860          &5885          &6619           &7147          &7987          &6914          &7137          &5433          &6919          &6077          &5851           \\
$T_{2}'$(K)     &6019(12)      &6067(16)      &6199(16)       &6819(21)      &7476(62)      &6380(18)      &6574(18)      &5237(15)      &6756(25)      &5541(27)      &5746(13)       \\
$i$             &88.3(2)       &84.4(3)       &85.7(2)        &74.8(5)       &74.9(6)       &82.5(4)       &83.5(2)       &75.7(4)       &74.0(3)       &83.2(6)       &74.6(3)        \\
$q(M_{2}/M_{1})$&0.0980(4)     &0.0867(2)     &0.0815(7)      &0.0885(15)    &0.0991(23)    &0.0968(8)     &0.0951(7)     &0.0930(12)    &0.0933(11)    &0.0791(4)     &0.0855(6)      \\
$\Omega$        &1.917(2)      &1.863(1)      &1.890(4)       &1.883(7)      &1.933(8)      &1.932(4)      &1.926(4)      &1.895(5)      &1.911(5)      &1.845(1)      &1.893(3)       \\
$(L_{1}/L_{t})_{B}$&---        &---           &0.930(1)       &---           &---           &---           &---           &---           &---           &---           &---            \\
$(L_{1}/L_{t})_{V}$&0.867(1)   &---           &0.924(1)       &---           &---           &---           &---           &0.901(1)      &---           &0.928(1)      &---            \\
$(L_{1}/L_{t})_{R}$&0.869(1)   &0.871(1)      &0.921(1)       &0.902(1)      &0.903(1)      &---           &---           &0.898(1)      &---           &0.923(1)      &---            \\
$(L_{1}/L_{t})_{I}$&0.871(1)   &0.873(1)      &0.918(1)       &0.899(1)      &0.899(1)      &---           &---           &0.896(1)      &---           &0.919(1)      &---            \\
$(L_{1}/L_{t})_{g}$&---        &---           &---            &---           &---           &0.920(1)      &0.922(1)      &---           &0.820(1)      &---           &0.896(1)       \\
$(L_{1}/L_{t})_{r}$&---        &---           &---            &---           &---           &0.913(1)      &0.914(1)      &---           &0.852(1)      &---           &0.895(1)       \\
$(L_{1}/L_{t})_{i}$&---        &---           &---            &---           &---           &0.909(1)      &0.910(1)      &---           &0.863(1)      &---           &---            \\
$(L_{3}/L_{t})_{g}$&---        &---           &---            &---           &---           &---           &---           &---           &0.086(15)     &---           &0.009(11)      \\
$(L_{3}/L_{t})_{r}$&---        &---           &---            &---           &---           &---           &---           &---           &0.048(16)     &---           &0.008(10)      \\
$(L_{3}/L_{t})_{i}$&---        &---           &---            &---           &---           &---           &---           &---           &0.034(17)     &---           &---            \\
$r_{1}$         &0.599(1)      &0.618(1)      &0.603(1)       &0.608(3)      &0.592(2)      &0.592(1)      &0.593(1)      &0.606(2)      &0.599(2)      &0.623(1)      &0.604(1)       \\
$r_{2}$         &0.225(4)      &0.239(6)      &0.199(8)       &0.226(23)     &0.220(20)     &0.213(7)      &0.212(7)      &0.229(17)     &0.218(11)     &0.229(8)      &0.209(7)       \\
$f$             &0.560(27)     &0.943(12)     &0.189(71)      &0.693(124)    &0.361(36)     &0.274(64)     &0.287(60)     &0.698(86)     &0.449(78)     &0.916(22)     &0.373(56)      \\
$Spot$          & star1        &star1         & star1         &---           & ---          & ---          & ---          &star1         & ---          & star1        & star1         \\
$\theta (deg)$  &29.14(76)     &104.90(154)   &165.08(48)     &---           &---           &---           &---           &133.99(83)    &---           &113.34(181)   &120.08(39)     \\
$\lambda (deg)$ &295.66(190)   &251.78(111)   &0.35(219)      &---           &---           &---           &---           &48.30(194)    &---           &24.24(263)    &81.30(81)      \\
$r_{s} (deg)$   &17.22(25)     &14.87(17)     &31.21(29)      &---           &---           &---           &---           &16.69(30)     &---           &19.86(42)     &19.86(13)      \\
$T_{f}$         &0.81(1)       &0.81(1)       &0.85(1)        &---           &---           &---           &---           &0.71(2)       &---           &0.93(1)       &0.83(1)        \\
\hline
\end{tabular}
\end{center}
\end{table*}

\begin{table*}
\begin{center}
\caption{Photometric solutions of TESS datas}
\label{table6}
\begin{tabular}{lllll}
\hline
Target           & J133031-1    & J133031-2    & J155009    & J164000 \\
\hline 
$T_{1}$(K)       & 5860         & 5860         & 6619       & 7137\\
$T_{2}$(K)       & 6256(9)      & 6018(11)     & 6079(7)    & 6495(8)\\
$i$              & 87.9(3)      & 79.4(4)      & 86.2(1)    & 87.9(1)\\
$q(M_{2}/M_{1})$ & 0.1082(8)    & 0.0910(2)    & 0.0824(4)  & 0.0995(2)\\
$\Omega$         & 1.946(4)     & 1.872(1)     & 1.884(2)   & 1.938(1)\\
$(L_{1}/L_{t})$  & 0.831(1)     & 0.868(2)     & 0.885(1)   & 0.895(1)\\
$(L_{3}/L_{t})$  & 0.015(13)    & ---          & 0.035(5)   & 0.015(5)\\
$(L_{3}/L_{t})_{Gaia}$&0.046    & ---          & 0.014      & 0.007   \\
$r_{1}$          & 0.591(1)     & 0.616(1)     & 0.606(1)   & 0.590(1)\\
$r_{2}$          & 0.232(8)     & 0.250(11)    & 0.205(4)   & 0.215(2)\\
$f$              & 0.540(52)    & 0.985(11)    & 0.350(29)  & 0.287(21)\\
$Spot$           & star1        & star1        & ---        & ---\\
$\theta (deg)$   & 115.93(104)  & 152.77(51)   & ---        & ---\\
$\lambda (deg)$  & 89.76(109)   & 27.89(176)   & ---        & ---\\
$r_{s} (deg)$    & 15.97(14)    & 25.85(31)    & ---        & ---\\
$T_{f}$          & 0.84(1)      & 0.81(1)      & ---        & ---\\
\hline
\end{tabular}
\end{center}
\end{table*}

\section{ORBITAL PERIOD INVESTIGATIONS}
The study of orbital period variation is of great significance for discussing the dynamic interaction of the binary systems, the existence of a third body and the materials transfer between the two components of binary systems \citep{95,96,106,98,99}. Therefore, the research of the orbital period variation has always been an important part of the binary systems. We collected minimum moments of the eleven targets as many as possible from our observation data and sky survey data bases including the All-Sky Automated Survey for SuperNovae \citep[ASAS-SN,][]{101,102}, the Zwicky Transient Facility (ZTF) survey \citep{103,104}, Wide Angle Search for Planets \citep[WASP,][]{81}, CSS and TESS. Data from ASAS-SN, ZTF and CSS is discrete and the observation cadence of TESS is 30 minutes, hence their minimum moments can't be obtained directly. In order to derive as many minimum moments as possible, we used the method proposed by \citet{33} to obtain the minimum moments. All minimum moments, errors and their sources are listed in Table 7. 

We calculated the values of O–C employing the following equation,
\begin{equation}
\begin{aligned}
BJD = BJD_{0} + P \times E\\
\end{aligned}
\end{equation}
where BJD is the minimum moment, BJD$_{0}$ is the initial minimum moment, $P$ is period, and $E$ is cycle number. All initial minimum moments and periods are listed in Table 1. All initial minimum moments listed in Table 1 are HJD, and we transformed them from HJD to BJD on the website\footnote{\href{http://astroutils.astronomy.ohio-state.edu/time/hjd2bjd.html}{http://astroutils.astronomy.ohio-state.edu/time/hjd2bjd.html}}. The general trends of O-C curves are obtained and shown in Fig. 4. According to the least squares method, we made linear fitting and quadratic fitting to the O-C curves. In order to determine which fitting is better, we calculated AIC and BIC values based on the two fittings. The criteria are defined as follows, 
\begin{equation}
\begin{aligned}
&AIC=nlog(RSS/n)+2k\\
&BIC=nlog(RSS/n)+klogn\\
\end{aligned}
\end{equation}
where $n$ is the number of data points and $k$ is the number of parameters estimated by the model. The smaller of the AIC and BIC values, the better of the fitting result. The corrected values of initial minimum moments and period are shown in Table 8. We found that the general trend of the O-C curves of J133031, J160755 and J162327 is straight line, of J155109, J170307, J223837, J224827 and J234634 is an upward parabola, which means that the period is slowly increasing, and of J154254, J155106 and J164000 is a downward parabola, which means that the period is slowly decreasing.

\begin{table*}
\begin{center}
\caption{The minimum moments of eleven contact binaries.}
\label{table7}
\begin{tabular}{lccccccc} 
\hline
Target      & BJD            & Error    & Min & Epoch    & O-C      & Residuals & Ref\\	
\hline	       
J133031	    & 2456856.98783  & 0.00061	& p	  & -8148	 &  0.00884 & -0.00152  & (1)\\ 
			& 2456856.83570  & 0.00073	& s	  & -8148.5	 &	0.00804 & -0.00232  & (1)\\
			& 2458269.83219  & 0.00061	& p	  & -3480	 &  0.00784 &	0.00419 & (2)\\
			& 2458269.98242  & 0.00041	& s	  & -3479.5	 &	0.00674 &	0.00309 & (2)\\
			& 2458561.90067  & 0.00076	& p	  & -2515	 &  0.00354 &	0.00127 & (2)\\
			& 2458562.05258  & 0.00122	& s	  & -2514.5	 &	0.00411 &	0.00185 & (2)\\
			& 2458962.93344  & 0.00118	& p	  & -1190	 &  0.00372 &	0.00336 & (2)\\
			& 2458963.07890  & 0.00113	& s	  & -1189.5	 &	-0.00215& 	-0.00251& (2)\\
			& 2459269.07260  & 0.00101	& s	  & -178.5	 &	-0.00387& 	-0.00278& (2)\\
			& 2459269.22821  & 0.00083	& p	  & -178	 &   0.00040& 	0.00150 & (2)\\
			& 2458283.75483  & 0.00061	& p	  & -3434	 &   0.00784&	0.00426 & (2)\\
			& 2458283.90540  & 0.00033	& s	  & -3433.5	 &	 0.00708&	0.00350 & (2)\\
			& 2458573.85709  & 0.00057	& s	  & -2475.5	 &	 0.00465&	0.00244 & (2)\\
			& 2458574.00739  & 0.00072	& p	  & -2475	 &   0.00361&	0.00141 & (2)\\
			& 2458899.07456  & 0.00080	& p   & -1401	 &   0.00740&	0.00673 & (2)\\
			& 2458899.21834  & 0.00116	& s	  & -1400.5	 &	-0.00016& 	-0.00082& (2)\\
			& 2459291.92714  & 0.00064	& p	  & -103	 &  -0.00062& 	0.00058 & (2)\\
			& 2459292.07438  & 0.00073	& s   & -102.5	 &	-0.00472& 	-0.00351& (2)\\
			& 2458547.97803  & 0.00096	& p	  & -2561	 &   0.00354& 	0.00121 & (2)\\
			& 2453827.89995  & 0.00209	& p	  & -18156	 &	 0.00329& 	-0.02146& (3)\\
			& 2453828.06695  & 0.00213	& s	  & -18155.5 &	 0.01896& 	-0.00579& (3)\\
			& 2454949.93652  & 0.00394	& p	  & -14449	 &	 0.05662& 	0.03720 & (3)\\
			& 2454950.09233  & 0.00270	& s	  & -14448.5 &	 0.06110& 	0.04168 & (3)\\
			& 2456062.90590  & 0.00217	& s	  & -10771.5 &	-0.02857& 	-0.04271& (3)\\
			& 2456063.04713  & 0.00324	& p	  & -10771	 &	-0.03868& 	-0.05281& (3)\\
			& 2459323.10237  & 0.00075  & p   &  0	     &   0.00000&  0.00135  & (4)\\
			& 2459323.25146  & 0.00037  & s   &  0.5	 &  -0.00224& 	-0.00089& (4)\\
            & 2458932.51436  & 0.00021  & s	  & -1290.5	 &	0.00259 &  0.00213  & (5)\\
            & 2458932.66292  & 0.00023  & p	  & -1290    &  -0.00018& -0.00064  & (5)\\
            & 2458938.56630  & 0.00021  & s   & -1270.5	 &	0.00121 & 0.00078   & (5)\\
            & 2458938.71520  & 0.00013  & p	  & -1270    &  -0.00123& -0.00165  & (5)\\
            & 2458948.25108  & 0.00025  & s	  & -1238.5	 &	0.00067 & 0.00029   & (5)\\
            & 2458948.40102  & 0.00018  & p	  & -1238    &  -0.00072& -0.00110  & (5)\\
            & 2458952.63938  & 0.00012  & p	  & -1224  	 &	0.00031 & -0.00004  & (5)\\
            & 2458952.79101  & 0.00020  & s   & -1223.5  &  0.00061 & 0.00025   & (5)\\              
\hline
\end{tabular}
\begin{tablenotes}
\footnotesize
\item[1] (1)ASAS-SN; (2)ZTF; (3)CSS; (4)WHOT; (5)TESS; (6)WASP (7) OAN-SPM (8)NAOs85cm (9)NEXT
\item[2] This table is available in its entirety in machine-readable form in the online version of this article.
\end{tablenotes}	
\end{center}
\end{table*} 

\begin{figure*}
\centering
\subfigure{\label{fig:subfig:a}
\includegraphics[width=0.32\linewidth]{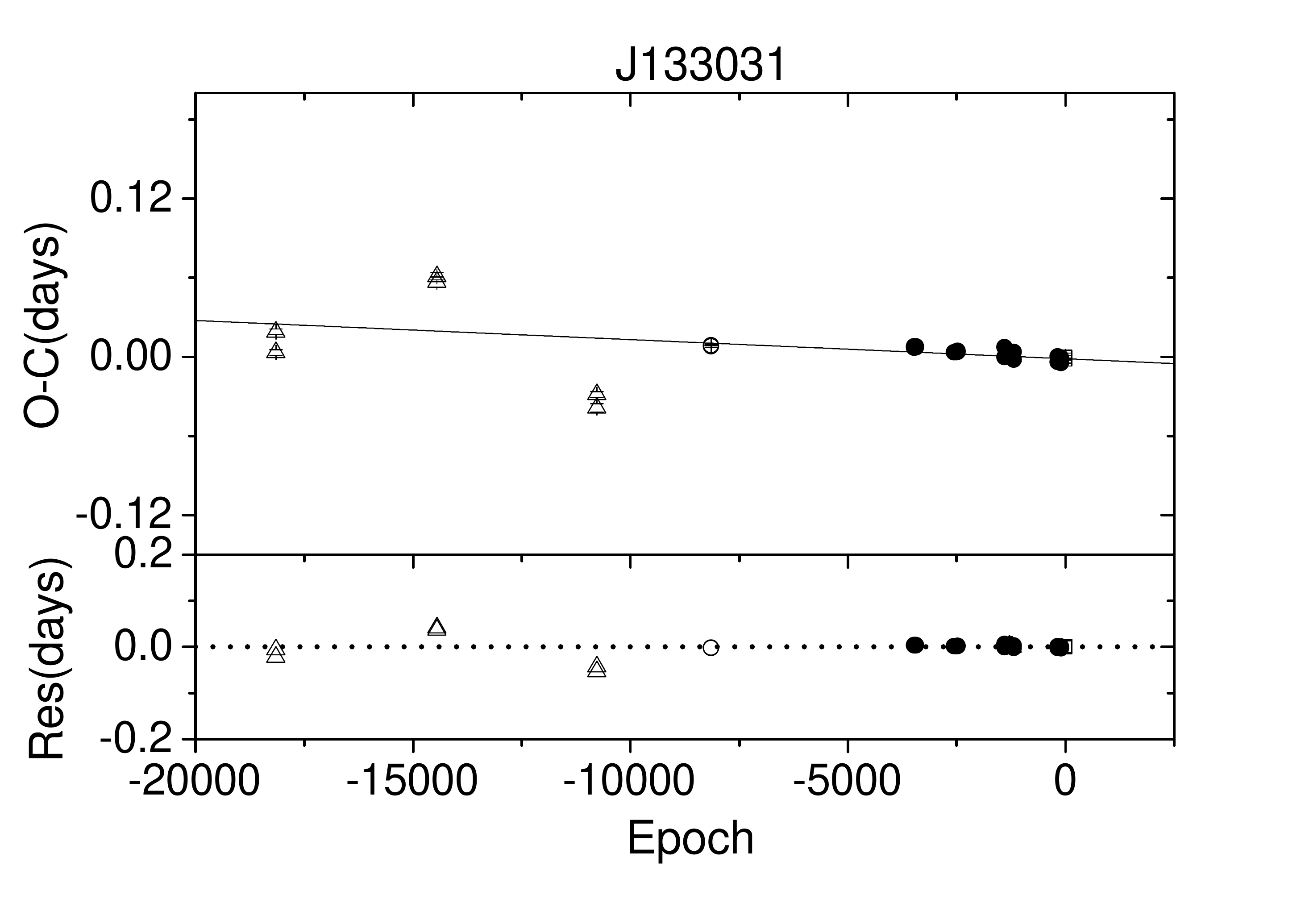}}
\subfigure{\label{fig:subfig:b}
\includegraphics[width=0.32\linewidth]{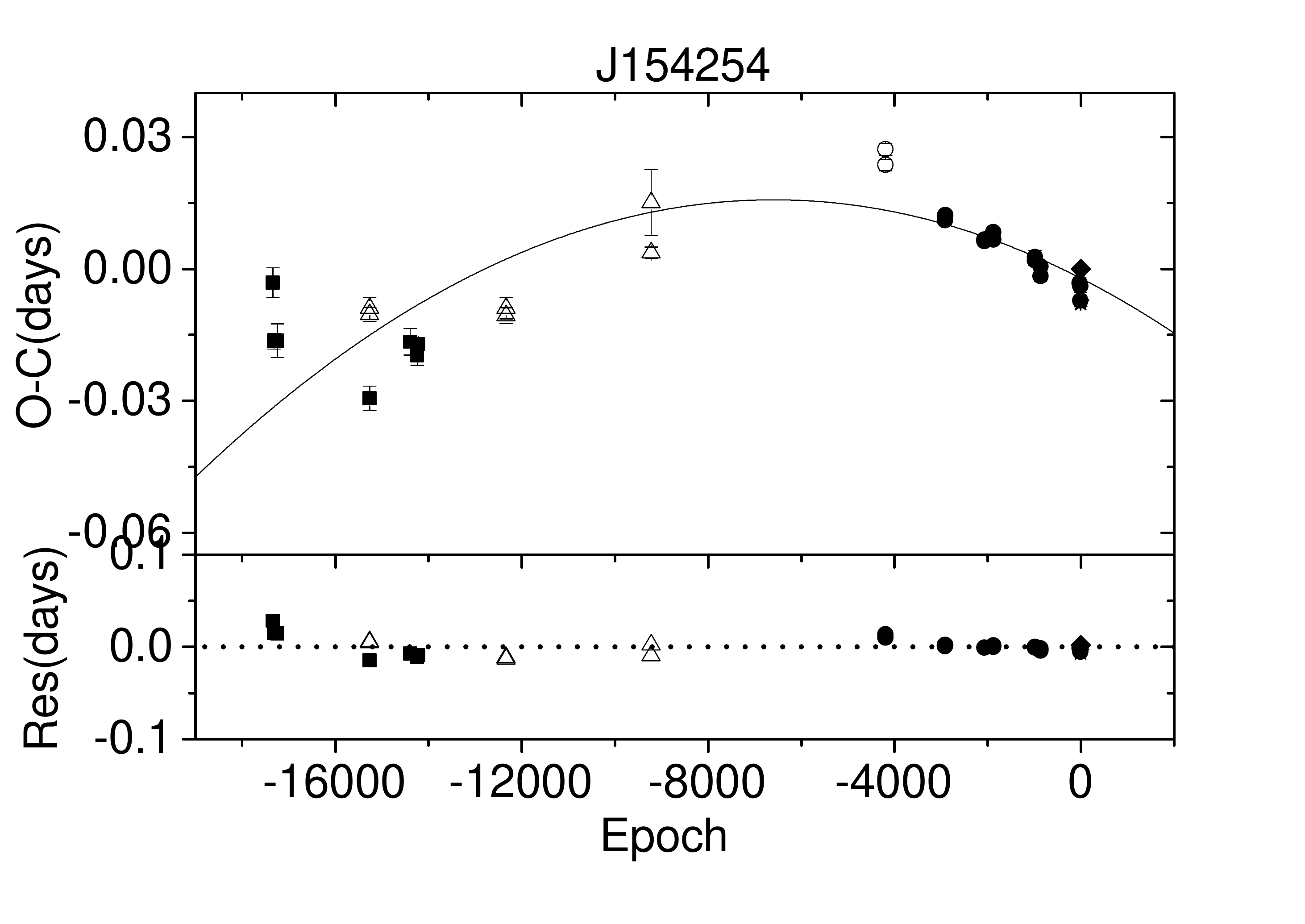}}
\subfigure{\label{fig:subfig:c}
\includegraphics[width=0.32\linewidth]{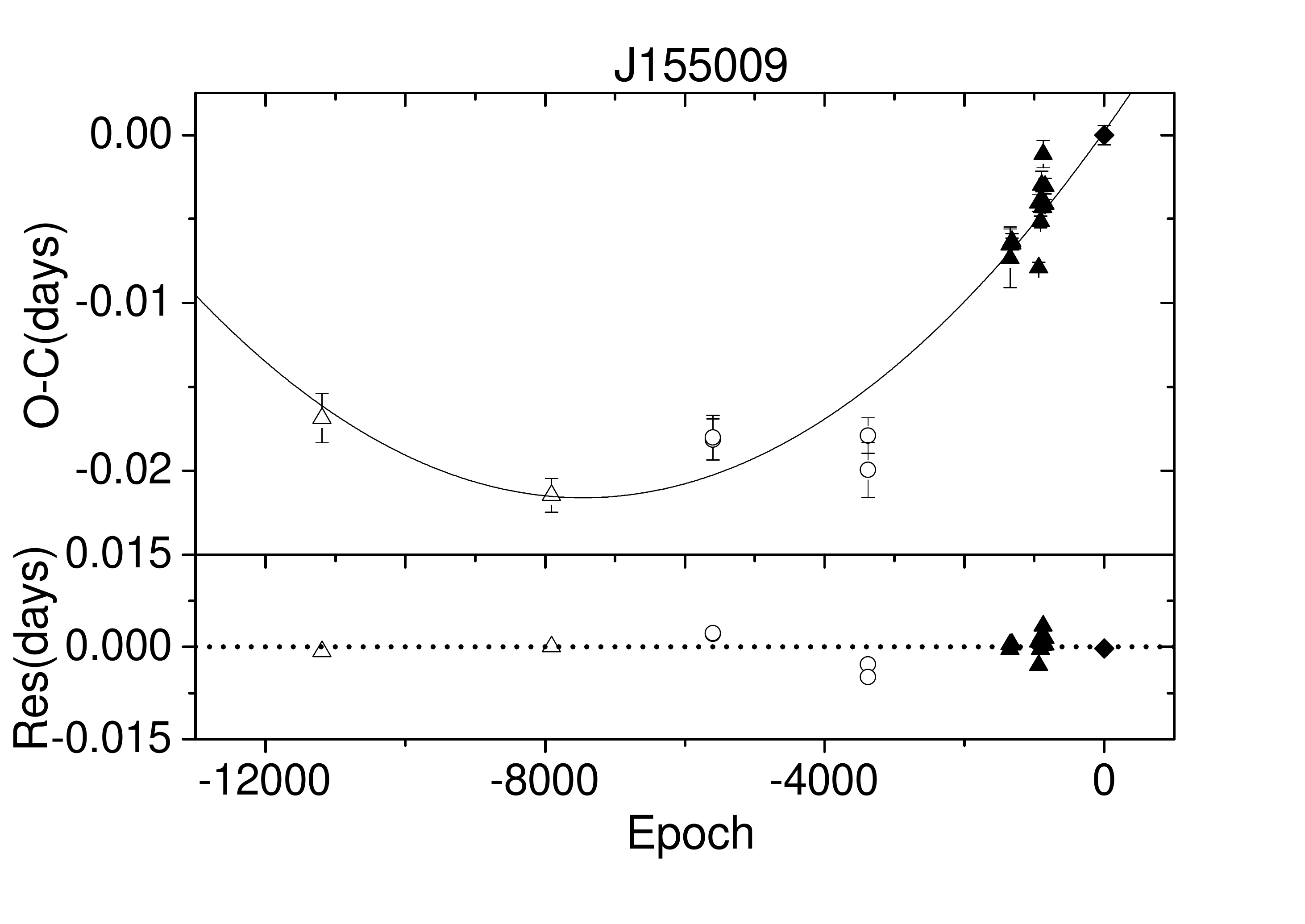}}
\subfigure{\label{fig:subfig:a}
\includegraphics[width=0.32\linewidth]{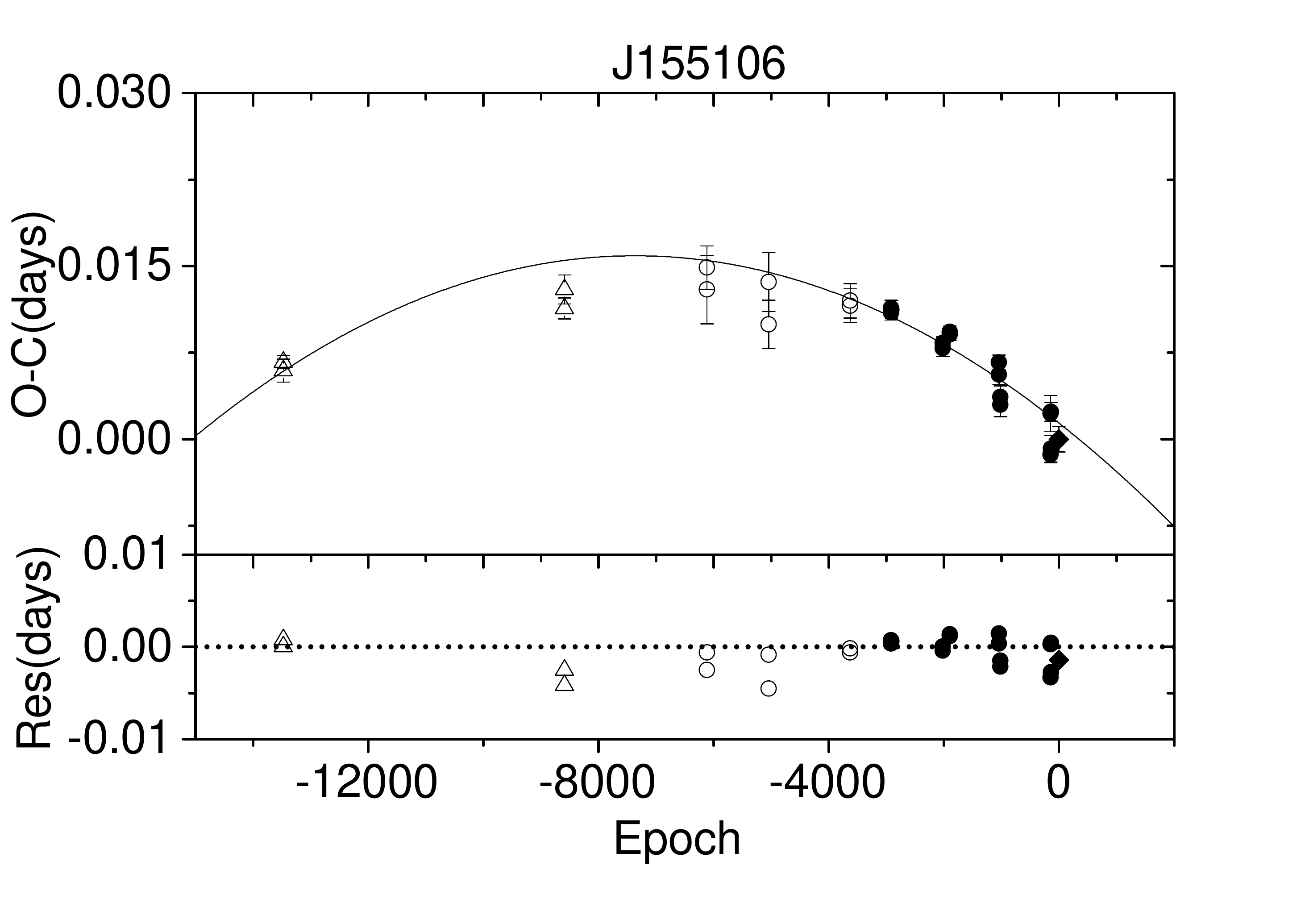}}
\subfigure{\label{fig:subfig:b}
\includegraphics[width=0.32\linewidth]{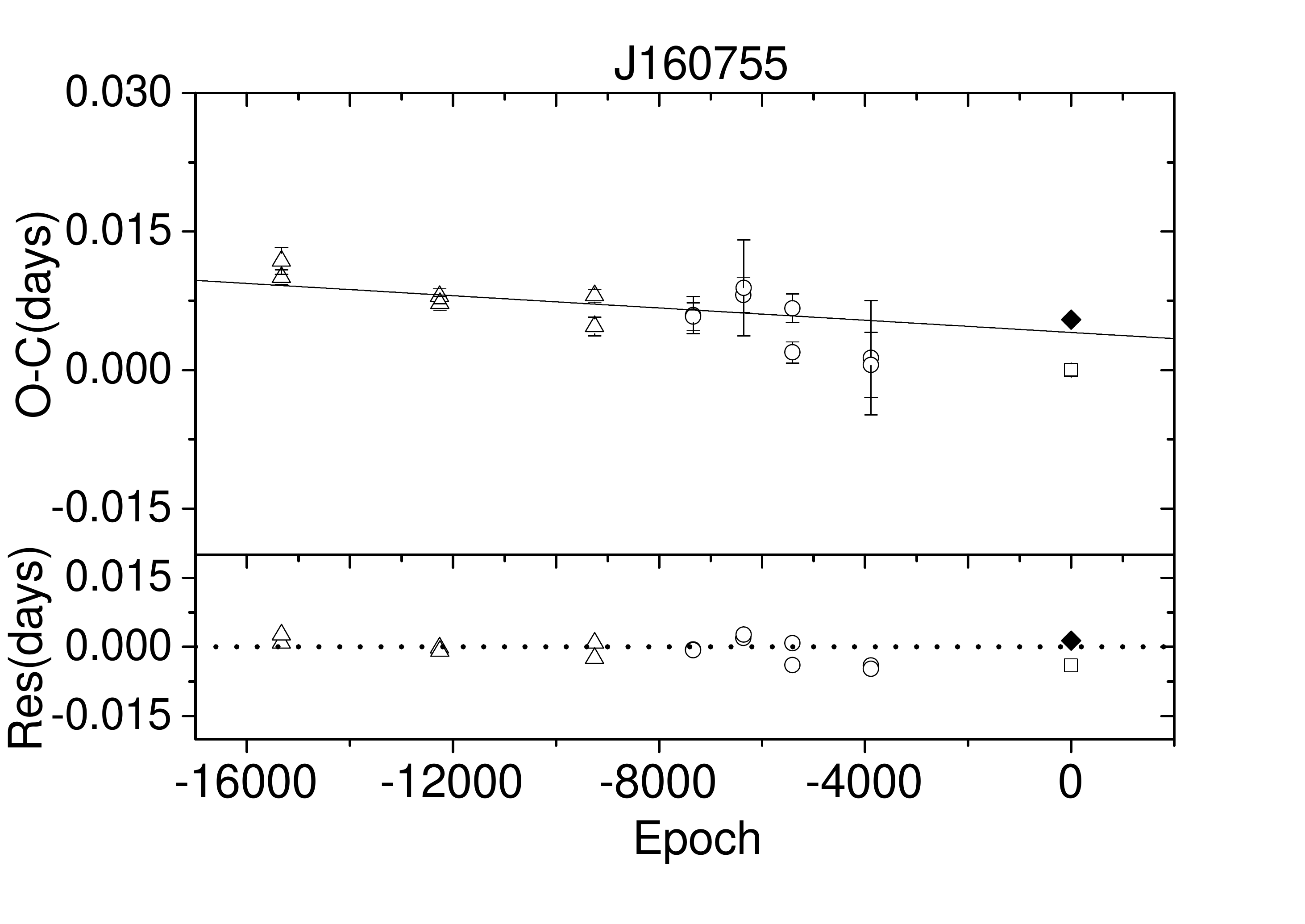}}
\subfigure{\label{fig:subfig:c}
\includegraphics[width=0.32\linewidth]{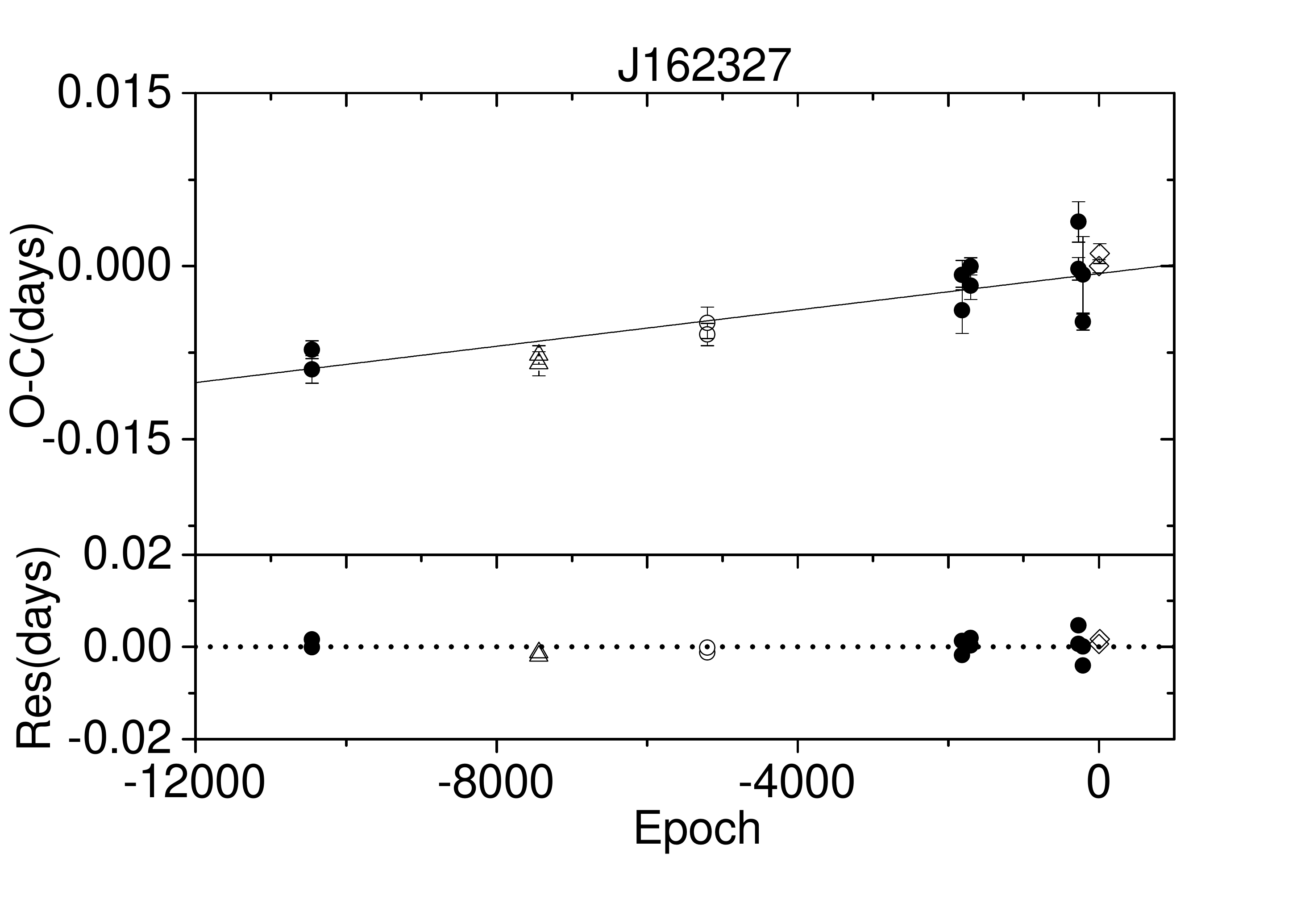}}
\subfigure{\label{fig:subfig:a}
\includegraphics[width=0.32\linewidth]{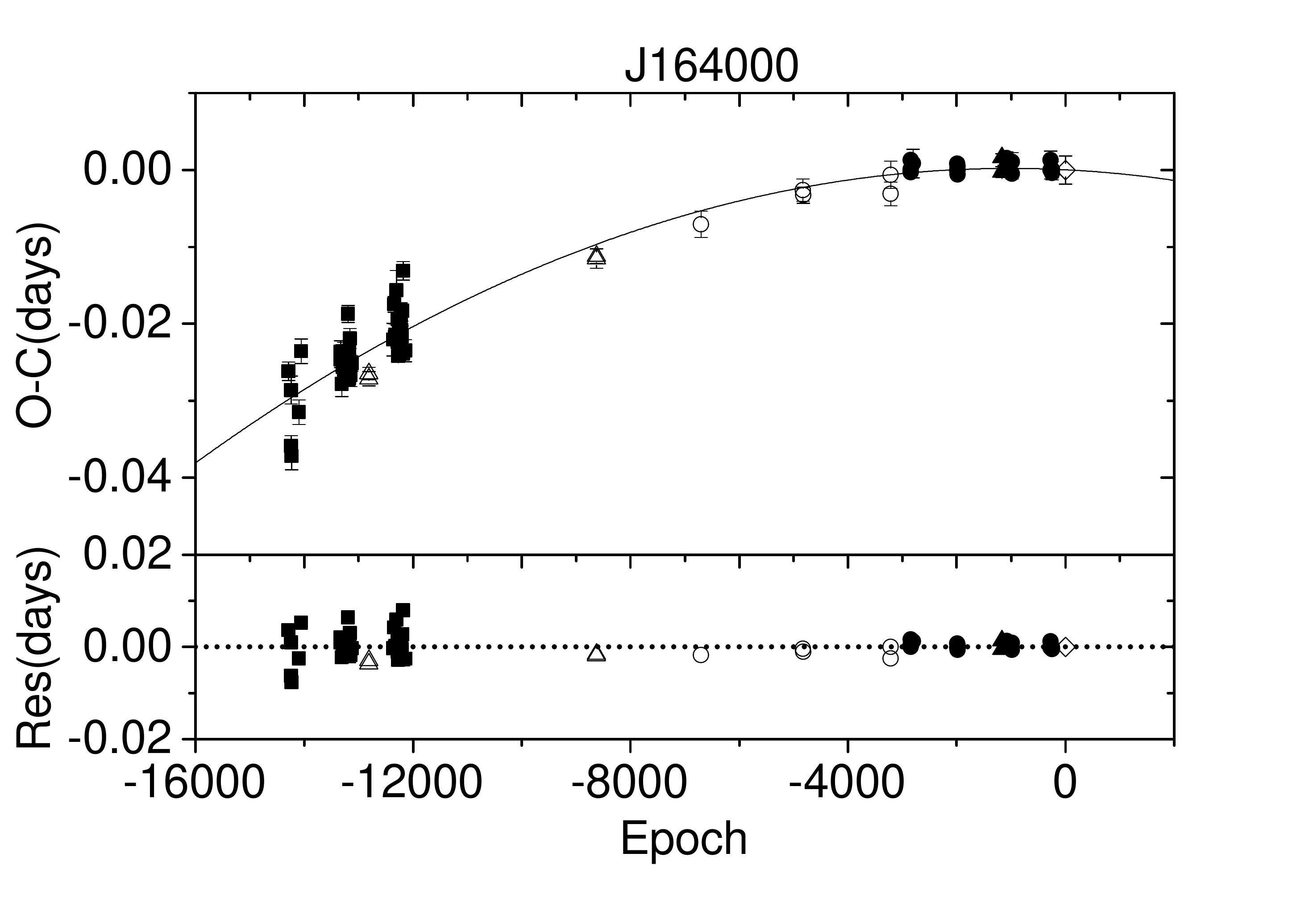}}
\subfigure{\label{fig:subfig:b}
\includegraphics[width=0.32\linewidth]{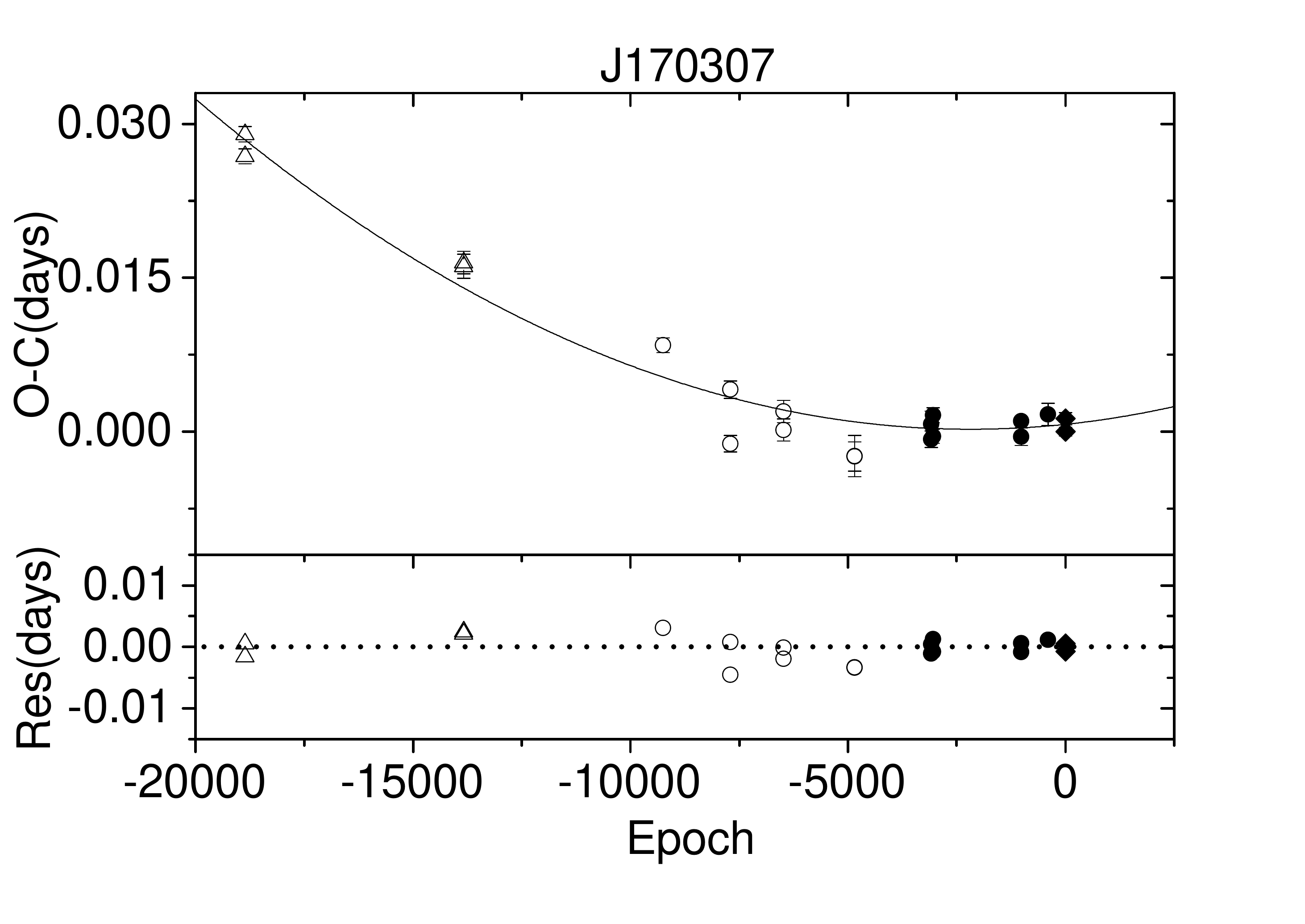}}
\subfigure{\label{fig:subfig:c}
\includegraphics[width=0.32\linewidth]{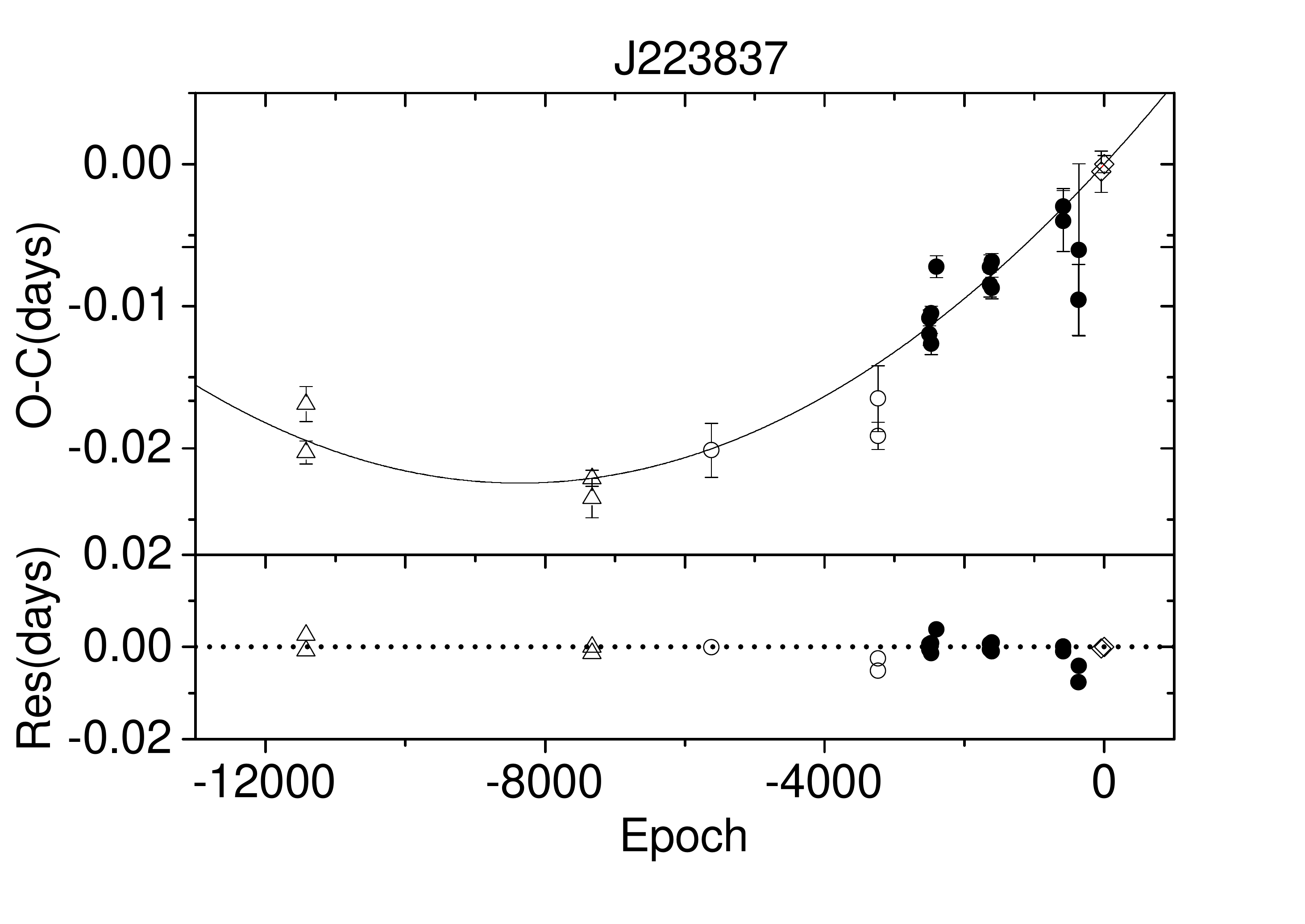}}
\subfigure{\label{fig:subfig:a}
\includegraphics[width=0.32\linewidth]{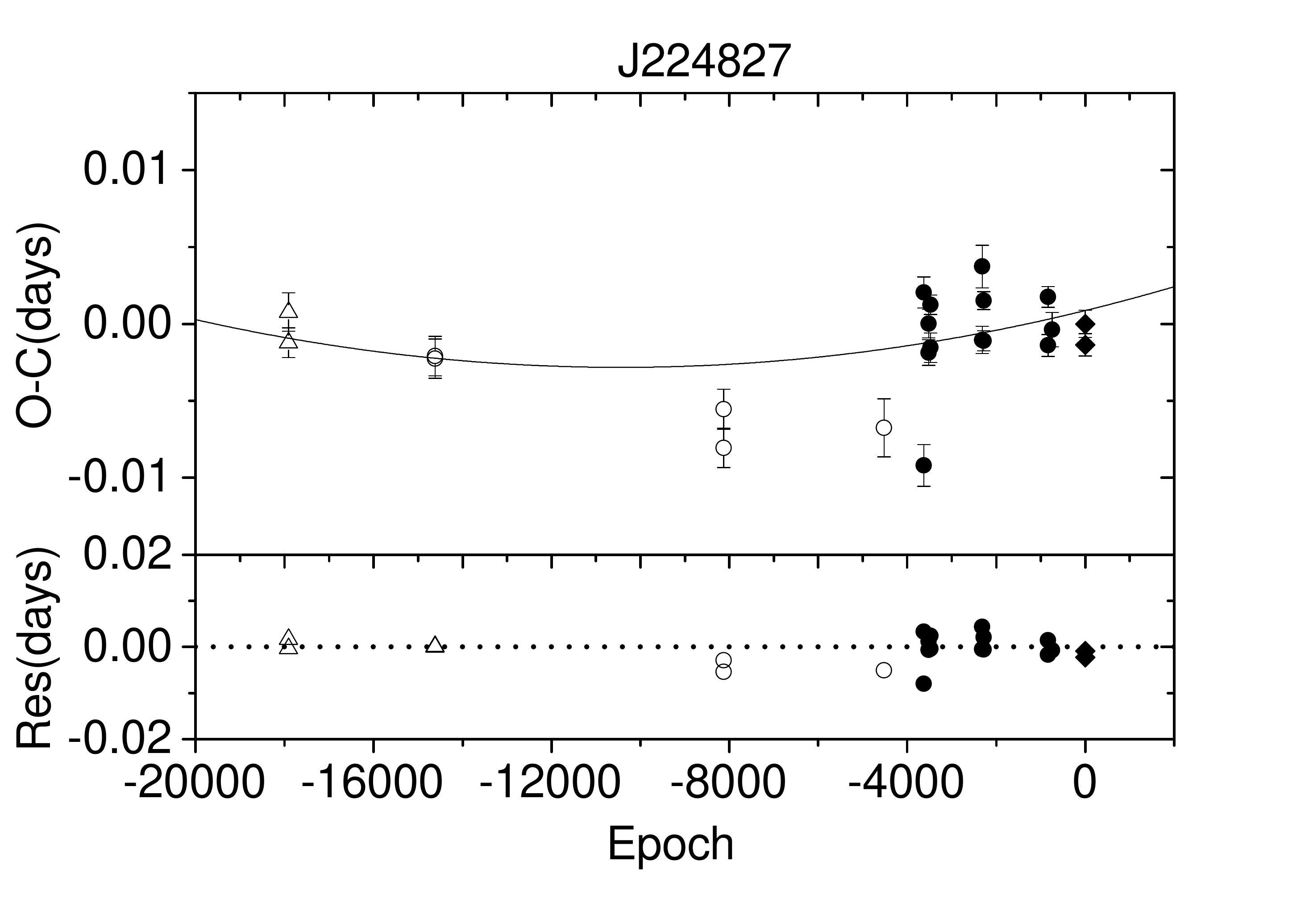}}
\subfigure{\label{fig:subfig:b}
\includegraphics[width=0.32\linewidth]{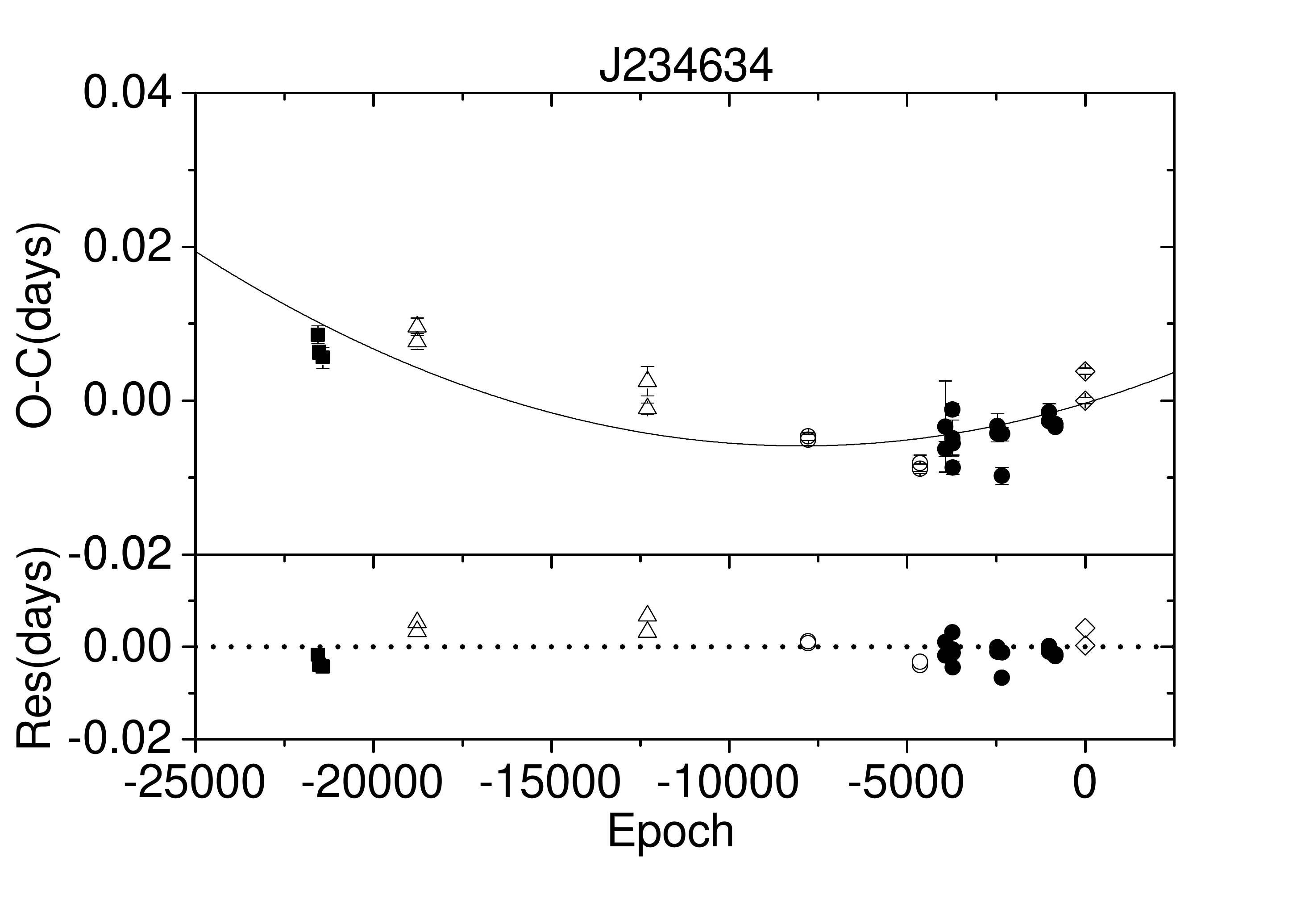}}
\caption{This figure depicts the O$-$C values. The residuals are shown at the bottom of the figure. ZTF $\bullet$; ASAS-SN $\circ$; TESS $\blacktriangle$; CSS $\bigtriangleup$; WASP $\blacksquare$; WHOT $\square$; OAN-SPM $\blacklozenge$; NEXT $\lozenge$; NAOs85cm $\bigstar$ }
\centering
\end{figure*}

\begin{table*}
\begin{center}
\caption{The corrected values based on O-C analysis.}
\label{table8}
\begin{tabular}{lrrrrcccccccccccc} 
\hline
Target &\multicolumn{2}{c}{Lin fit*}&\multicolumn{2}{c}{Qua fit*}& $\Delta$ $T_{0}$      & $\Delta P$                  & $dP/dt$                    & $dM_{1}/dt$          &$\tau$    &$\tau_{th}$\\  
       &  AIC    &BIC               &  AIC    &BIC               &($\times 10^{-4} days$)& ($days$)                    & ($d yr^{-1}$)              &($M_{\odot} yr^{-1}$ )&$yr$      &$yr$\\ 
\hline           
J133031&  -122.7 &-123.6            &  -119.9 &-121.3            & -14.20(6.50)          & $-1.45(0.33)\times 10^{-6}$ & -                          &   -& -& -\\                                 
J154254&  -122.4 &-123.4            &  -125.2 &-126.6            & -21.30(16.40)         & $-4.10(0.53)\times 10^{-10}$& $-1.11(0.17)\times 10^{-2}$& $-1.31\times 10^{-3}$&$1.01\times 10^{3}$&$1.16\times 10^{7}$\\ 
J155009&  -102.0 &-103.3            &  -114.8 &-116.8            &  2.22(8.41)           & $ 3.93(0.91)\times 10^{-10}$& $9.28(1.36)\times 10^{-3}$ & $9.52\times 10^{-4}$ &-&-\\ 
J155106&  -124.9 &-126.0            &  -142.3 &-144.0            &  14.00(6.48)          & $-2.67(0.24)\times 10^{-10}$& $-7.56(0.66)\times 10^{-3}$& $-8.88\times 10^{-4}$&$1.56\times 10^{3}$&$1.02\times 10^{7}$\\
J160755&  -79.2  &-80.8             &  -78.0  &-80.4             &  40.70(7.15)          & $-3.32(0.91)\times 10^{-7}$ & -                          &  -                   & -&-\\
J162327&  -82.8  &-84.7             &  -81.3  &-83.7             & -6.42(6.81)           & $7.87(1.37)\times 10^{-7}$  & -                          &  -                   & -&-\\
J164000&  -410.3 &-410.5            &  -426.2 &-426.4            &  0.33(3.65)           & $-1.71(0.18)\times 10^{-10}$& $-6.51(4.74)\times 10^{-4}$& $-8.09\times 10^{-5}$&$1.71\times 10^{4}$&$1.00\times 10^{7}$\\
J170307&  -90.8  &-92.2             &  -102.3 &-104.4            &  7.11(5.82)           & $1.01(0.12)\times 10^{-10}$ & $1.11(0.57)\times 10^{-3}$ & $1.58 \times 10^{-4}$& -&-\\
J223837&  -105.9 &-107.2            &  -110.6 &-112.2            &  -0.04(8.73)          & $3.21(0.41)\times 10^{-10}$ & $8.83(0.77)\times 10^{-3}$ & $9.40 \times 10^{-4}$& -&-\\
J224827&  -105.5 &-106.8            &  -105.7 &-107.7            &  8.67(10.00)          & $3.39(2.42)\times 10^{-11}$ & $1.61(0.99)\times 10^{-3}$ & $1.78 \times 10^{-4}$& -&-\\
J234634&  -124.3 &-125.4            &  -129.1 &-130.8            & -3.06(8.53)           & $8.72(1.53)\times 10^{-11}$ & $3.50(0.77)\times 10^{-3}$ & $3.91 \times 10^{-4}$& -&-\\
\hline
\end{tabular}
\begin{tablenotes}
	\footnotesize
	\item[1] *Lin fit: linear fitting; *Qua fit: quadratic fitting.  
\end{tablenotes}	
\end{center}
\end{table*}

\section{SPECTRA ANALYSIS}
Most W UMa-type eclipsing binaries may be chromospherically active binaries, because the spectral types of the two components are later than F \citep{77}. An important indicator of chromospheric activity is the presence of H$\alpha$ (6563\AA) emission line \citep{76}. H$\alpha$ emission line can be applied as the optical and near-infrared diagnostics of stellar chromospheric activity in late type rotating stars \citep{34,35,36}. In order to determine whether J133031, J154254, J155009, J160755, J162327 exist chromospherical activity, we analyzed their spectra. Due to the low $SNR$ of J155009' spectra (shown in Table 3), we didn't use them for the following analysis.

We performed flux normalization and decosmic ray processing on the spectra. In order to analyze H$\alpha$ emission line, the spectra subtraction technique performed by \citet{25} was adopted. The template spectra was generated using the iSpec \footnote{\href{https://iscottmark.gitee.io/iSpec/}{https://iscottmark.gitee.io/iSpec/}} software package according to temperatures and log $g$ (shown in Table 3). According to the redshift of the object spectra, we shifted the template spectra to keep the corresponding spectral line wavelengths are consistent. The subtracted spectra were obtained by subtracting the inactive synthesized spectra from the object spectra. The object, synthesized and subtracted spectra are shown in Fig. 5. We identified the H$\alpha$  emission line from the subtracted spectra of Fig. 5. The equivalent width (EW) of the H$\alpha$ emission line was obtained using the splot command in the iraf package and shown in Table 3. The existence of H$\alpha$ emission line indicates that J133031, J154254, J160755, J162327 exist chromospheric activity.

\begin{figure*}
\centering
\subfigure{\label{fig:subfig:a}
\includegraphics[width=0.48\linewidth]{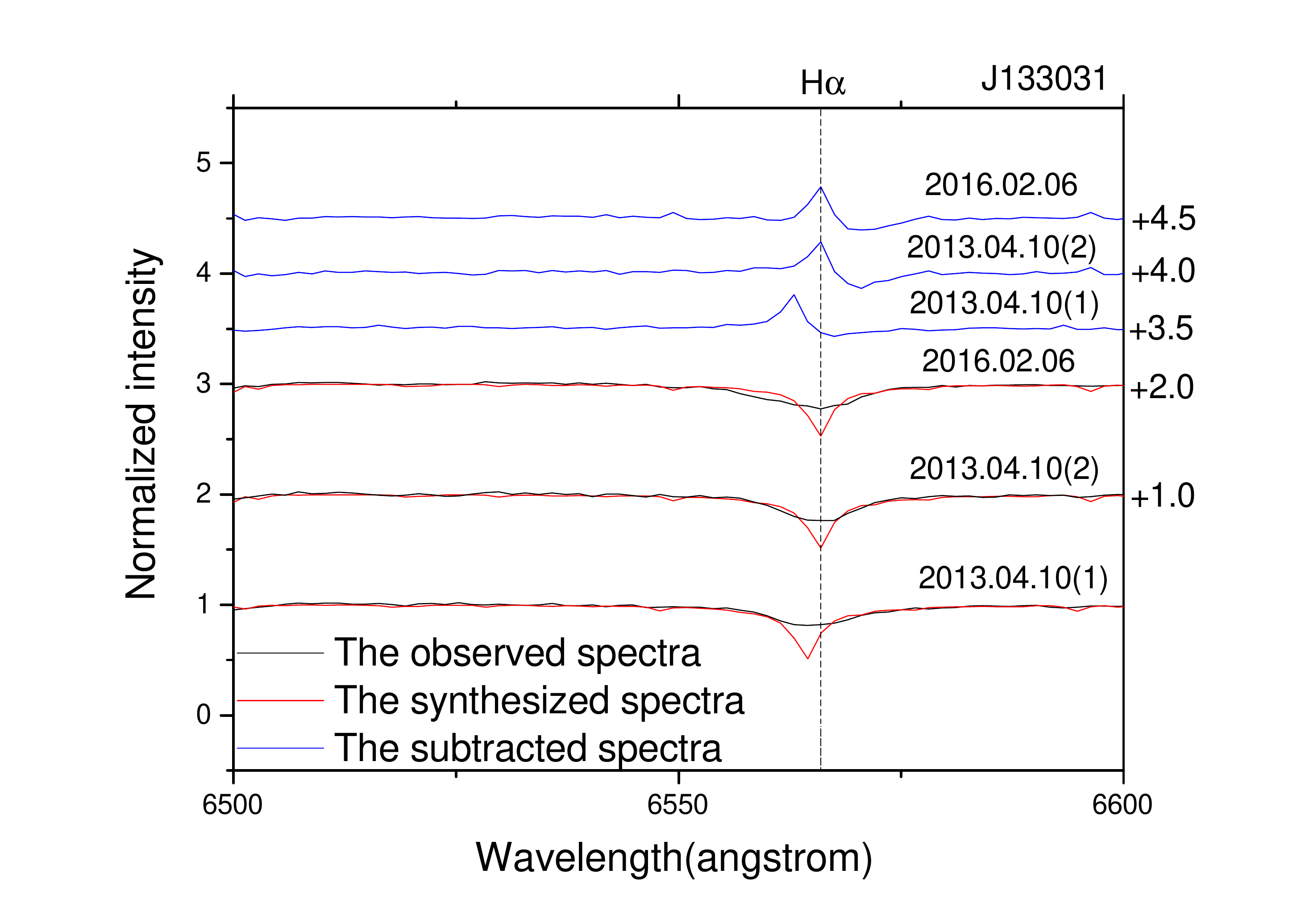}}
\subfigure{\label{fig:subfig:b}
\includegraphics[width=0.48\linewidth]{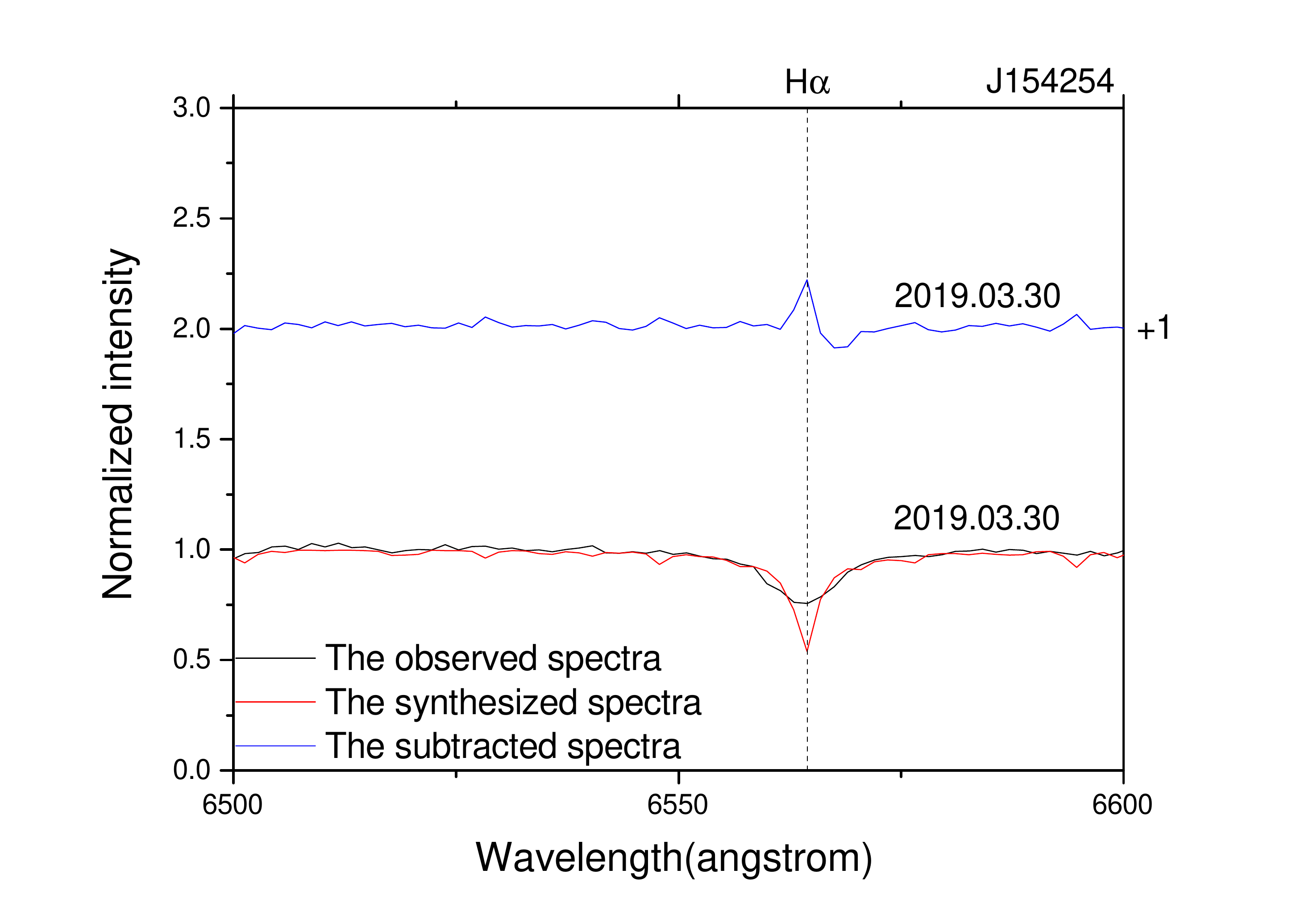}}
\subfigure{\label{fig:subfig:a}
\includegraphics[width=0.48\linewidth]{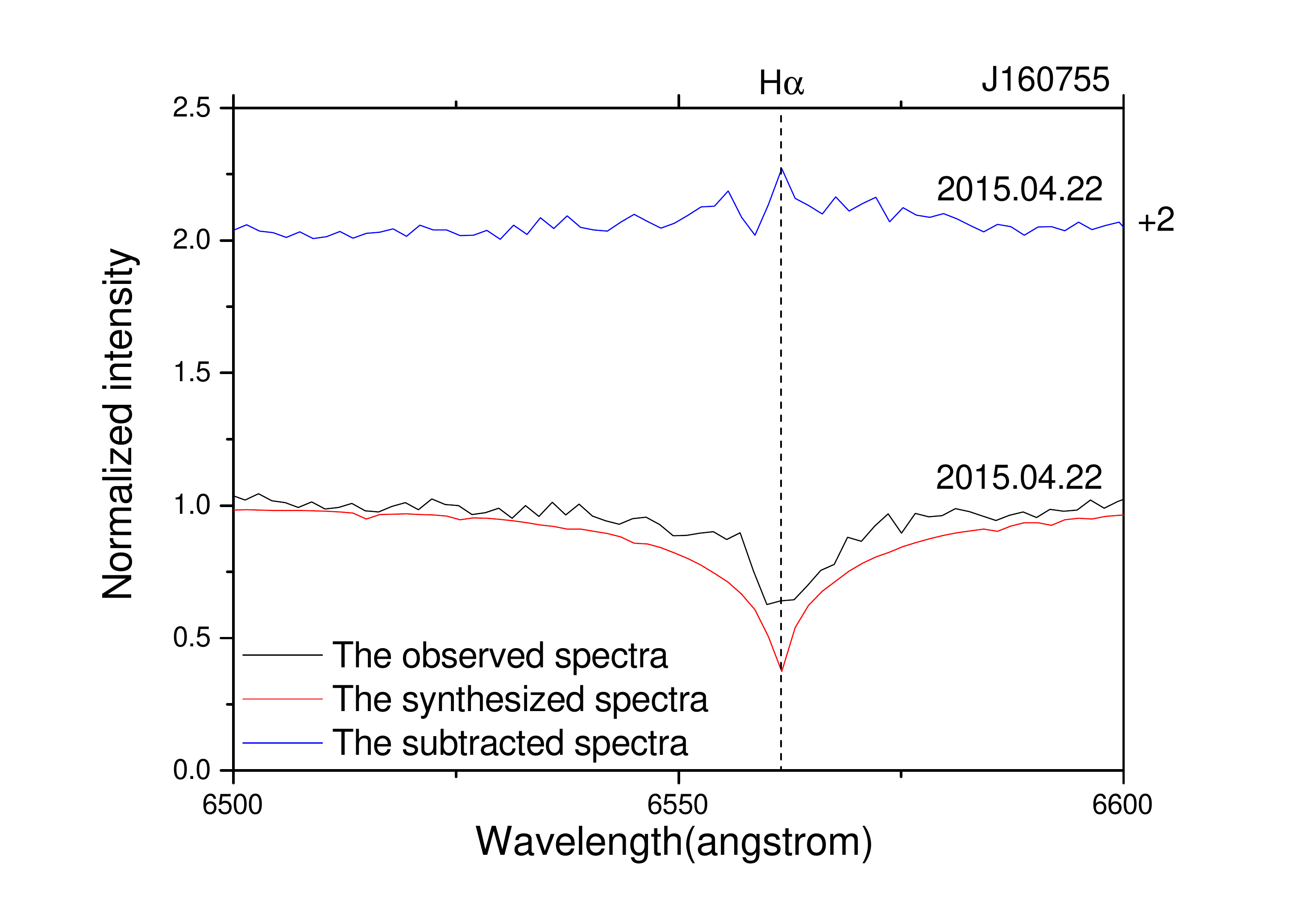}}
\subfigure{\label{fig:subfig:b}
\includegraphics[width=0.48\linewidth]{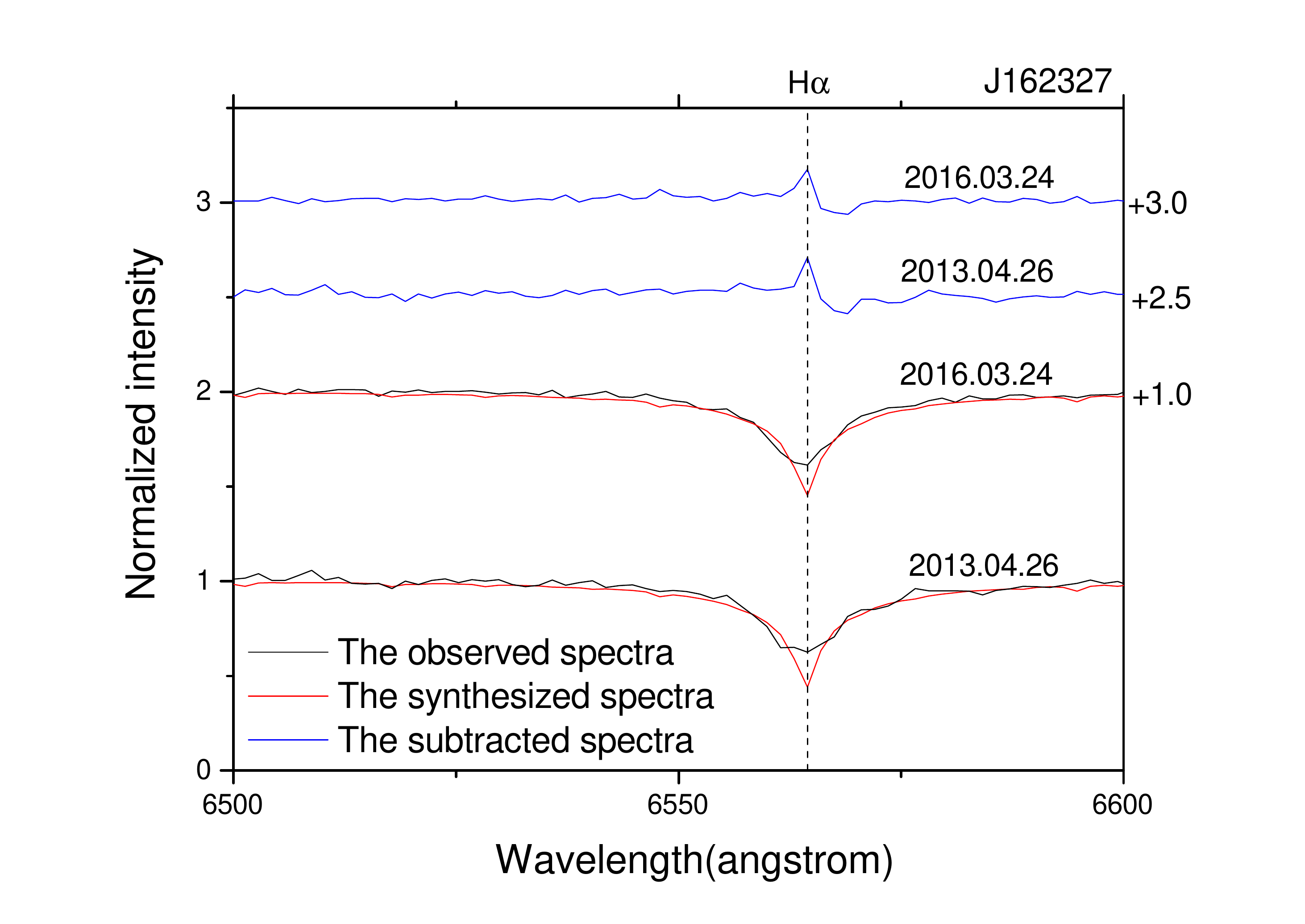}}
\caption{Normalized observed, synthesized and subtracted spectra. The H$\alpha$ emission line is marked with dotted lines.}
\centering
\end{figure*}

\section{DISCUSSTIONS AND CONCLUSIONS} 
Muti-band photometry and spectra analysis were carried out to get the eleven contact binaries’ physical and absolute parameters. Applying the W-D program, the photometric solutions were obtained. Due to the obvious O'Connell effect existing in the light curves of J133031, J154254, J155009, J170307, J224827 and J234634, we used the dark spot model to fit the distortion in the light curves. The mass ratios of the eleven contact binaries are all less than 0.1 and reliable because these systems are totally eclipsing binary systems \citep{17,18,92}. There are two W-subtype binary systems, they are J133031 and J154254, while the rest binary systems are A-subtype binary systems. Among the eleven systems, J164000 has the largest temperature difference between the two components, which is 563K. \citet{48} proposed a conclusion that contact binaries with low-mass ratios can present deep or medium or shallow degree of contact. \citet{47} defined that the contact binaries whose degree of contact exceeds 50$\%$ as deep contact binaries.We defined medium and shallow contact configuration as that the fill-out factors are between 25$\%$ and 50$\%$, and less than 25$\%$. According to the photometric solutions, we knew that the fill-out factors of eleven targets are between 18.9$\%$ and 94.3$\%$, which showed that our contact binaries have shallow, medium and deep contact systems. Therefore, it proved the conclusion proposed by \citet{48}. Table 9 listed the physical parameters obtained by us and those determined by \citet{12}. We compared these parameters and found that the mass ratios are similar, the fill-out degree and the inclination are quite different, and the errors of \cite{12} are generally large. Our photometric accuracy is better, the quality of sky survey data is relatively poor, and multi-band photometry are used by us. Therefore, to a certain extent, we believed that our results are more reliable.

\begin{table*}
\begin{center}
\caption{The basic geometric parameters we obtained and from paper.}
\label{table9}
\begin{tabular}{cccccccccccc} 
\hline
Target  & \multicolumn{2}{c}{T$_{1}$} & \multicolumn{2}{c}{T$_{2}$} & \multicolumn{2}{c}{$i$}  & \multicolumn{2}{c}{$q$}   & \multicolumn{2}{c}{$f$}\\
        & Ours     & Sun et al.       & Ours      & Sun et al.      & Ours       & Sun et al.  & Ours        & Sun et al.  & Ours        & Sun et al. \\
\hline
J133031 & 5860    & 5930     & 6019(12)  & 5961(71)    & 88.3(2)    & 79.38(210)  & 0.0980(4)   & 0.09(9)  & 0.560(27)   & 0.34(29)\\ 
J154254 & 5885    & 5837     & 6067(16)  & 5938(49)    & 84.4(3)    & 83.21(308)  & 0.0867(2)   & 0.07(9)  & 0.943(12)   &0.97(38)\\
J155009 & 6619    & 6388     & 6199(16)  & 5556(39)    & 85.7(2)    & 76.64(180)  & 0.0815(7)   & 0.07(9)  & 0.189(71)   &0.70(27)\\   
J155106 & 7147    & 7553     & 6819(21)  & 6962(94)    & 74.8(5)    & 71.70(160)  & 0.0885(15)  & 0.07(9)  & 0.693(124)  &0.86(48)\\
J160755 & 7987    & 7697     & 7476(62)  & 7287(69)    & 74.9(6)    & 72.59(105)  & 0.0991(23)  & 0.09(9)  & 0.361(36)   &0.48(31)\\ 
J162327 & 6914    & 7186     & 6380(18)  & 5845(58)    & 82.5(4)    & 78.67(131)  & 0.0968(8)   & 0.09(9)  & 0.274(64)   &0.52(22)\\  
J164000 & 7137    & 6740     & 6574(18)  & 5805(58)    & 83.5(2)    & 75.14(181)  & 0.0951(7)   & 0.09(9)  & 0.287(60)   &0.20(49)\\ 
J170307 & 5433    & 5451     & 5237(15)  & 5516(45)    & 75.7(4)    & 74.74(162)  & 0.0930(12)  & 0.07(9)  & 0.698(86)   &0.08(22)\\  
J223837 & 6919    & 7171     & 6756(25)  & 6686(55)    & 74.0(3)    & 75.41(125)  & 0.0933(11)  & 0.09(9)  & 0.449(78)   &-0.12(32)\\
J224827 & 6077    & 5930     & 5541(27)  & 5785(75)    & 83.2(6)    & 78.01(250)  & 0.0791(4)   & 0.07(9)  & 0.916(22)   &0.16(45)\\
J234634 & 5851    & 5816     & 5746(13)  & 5957(59)    & 74.6(3)    & 77.77(145)  & 0.0855(6)   & 0.07(9)  & 0.373(56)   &-0.08(39)\\
\hline 
\end{tabular}
\end{center}
\end{table*}

According to the three-dimensional correlations provided by \citep{37}, the absolute parameters of the two components of the eleven systems were determined. According to the Kepler's third law, 
\begin{equation}
\begin{aligned}
&M_{1}+M_{2} = 0.0134A^{3}/P^{2}\\
\end{aligned}
\end{equation}
the semi-major axis was obtained. The absolute parameters calculated by the above steps and errors on basis of error transfer formulas are shown in Table 10.

\begin{table*}
\begin{center}
\renewcommand\tabcolsep{2.0pt} 
\caption{The absolute parameters of eleven targets.}
\label{table10}
\begin{tabular}{llccccccccccc} 
\hline
Target  & M$_{1}$       & M$_{2}$       & R$_{1}$       & R$_{2}$       & L$_{1}$       & L$_{2}$       & \textbf{$A$}  \\   
        & (M$_{\odot})$ & (M$_{\odot})$ & (R$_{\odot})$ & (R$_{\odot})$ & (L$_{\odot})$ & (L$_{\odot})$ & (R$_{\odot})$ \\
\hline
J133031 & 1.162(0.254)  & 0.114(0.026)  & 1.237(0.121)  & 0.459(0.049)  & 2.018(0.605)  & 0.271(0.079)  & 2.059(0.015)  \\
J154254 & 1.317(0.281)  & 0.114(0.025)  & 1.464(0.139)  & 0.514(0.054)  & 3.215(0.947)  & 0.389(0.112)  & 2.379(0.017)  \\
J155009 & 1.599(0.321)  & 0.130(0.028)  & 1.882(0.170)  & 0.644(0.065)  & 6.429(1.835)  & 0.737(0.205)  & 3.015(0.023)  \\
J155106 & 1.384(0.290)  & 0.122(0.028)  & 1.559(0.148)  & 0.552(0.059)  & 3.806(1.128)  & 0.468(0.135)  & 2.537(0.019)  \\             
J160755 & 1.310(0.275)  & 0.130(0.030)  & 1.445(0.138)  & 0.537(0.058)  & 3.053(0.909)  & 0.414(0.119)  & 2.394(0.017)  \\
J162327 & 1.612(0.311)  & 0.156(0.032)  & 1.888(0.163)  & 0.695(0.067)  & 6.337(1.725)  & 0.873(0.224)  & 3.097(0.023)  \\
J164000 & 1.402(0.287)  & 0.133(0.028)  & 1.580(0.144)  & 0.577(0.058)  & 3.912(1.109)  & 0.512(0.142)  & 2.596(0.019)  \\
J170307 & 1.134(0.253)  & 0.105(0.024)  & 1.204(0.120)  & 0.436(0.048)  & 1.874(0.572)  & 0.271(0.072)  & 1.985(0.014)  \\
J223837 & 1.541(0.306)  & 0.144(0.030)  & 1.784(0.159)  & 0.646(0.064)  & 5.463(1.534)  & 0.704(0.192)  & 2.916(0.022)  \\
J224827 & 1.233(0.274)  & 0.098(0.022)  & 1.350(0.134)  & 0.456(0.050)  & 2.613(0.802)  & 0.292(0.087)  & 2.171(0.016)  \\
J234634 & 1.141(0.258)  & 0.098(0.023)  & 1.218(0.123)  & 0.425(0.047)  & 1.953(0.607)  & 0.233(0.071)  & 1.984(0.014)  \\  
\hline
\end{tabular}	
\end{center}
\end{table*}

The periods of J155109, J170307, J223837, J224827 and J234634 are slowly increasing, and of J154254, J155106 and J164000 are slowly decreasing. There are many explanations about the changes in period. The long-term increase in period can be explained by materials transfer from the less massive component to the more massive component. The long-term decrease in period can be explained by three reasons: materials transfer from the more massive component to the less massive component, the angular momentum loss or their combined effect. If the changes in period is caused by the materials transfer, the materials transfer rate can be calculated using the following equation,
\begin{equation}
\begin{aligned}
\frac{dM_{1}}{dt}=\frac{M_{1}M_{2}}{3P(M_{1}-M_{2})}\times\frac{dP}{dt}\\
\end{aligned}
\end{equation}
The results are shown in Table 8. To further determine the cause of the long-term period decrease, we calculated the timescale $\tau$ and the thermal timescale $\tau_{th}$ of the materials transfer rate. Using the following equation,
\begin{equation}
\begin{aligned}
\tau = \dfrac{M_{1}}{\frac{dM_{1}}{dt}}\\
\end{aligned}
\end{equation}
the timescale $\tau$ of the materials transfer rate was calculated and shown in Table 8. Assuming constant angular momentum and the total mass and using the following equation,
\begin{equation}
\begin{aligned}
\tau_{th} = \dfrac{GM_{1}^{2}}{R_{1}L_{1}}\\
\end{aligned}
\end{equation}
the thermal timescale $\tau_{th}$ was calculated and also shown in Table 8. The timescale $\tau$ and the thermal timescale $\tau_{th}$ of the materials transfer rate are quite different, meaning that the long-term orbital period decrease may be caused by the angular momentum loss.    

Chromospheric activity and starspot activity, triggered by magnetic fields and maintained by a magnetic dynamo, are the common phenomena in contact binaries \citep{50}. We noticed that the chromospheric activity indicators of H$\alpha$ line show emissions. According to Table 3, we found the equivalent widths of the H$\alpha$ emission line are all greater 0.75\AA. Comparing the criteria for the chromospheric activity derived by \citet{51}, J133031, J154254, J160755, J162327 were determined to exhibit chromospheric activity.

To further understand the evolutionary status of the eleven contact binaries, their mass-luminosity (left) and mass-radius distributions (right) are plotted, shown in Fig. 6. The filled symbols denote the primary components and the open symbols denote the secondary components. The zero-age main sequence (ZAMS) and terminal-age main sequence (TAMS) lines \citep{65} are represented by solid and dashed lines, respectively. The eleven targets are very close in the two figures, meaning that the evolution states of them are similar. The primary component is located between ZAMS and TAMS, indicating that the primary component is still in the main sequence evolution stage. The secondary components are above TAMS, indicating that they are over-luminous, and the reason is that energy transfer from the primary has raised their temperature to the level of that of the primary and increased their luminosity to a value much larger than what would be expected for their mass and size. \citet{4} proposed that the contact binary instability (Darwin's instability) occurs when $J_o \geqslant 3J_s$. To determine whether the eleven contact binaries are in stable state, $J_{s}/J_{o}$ is calculated according to the equation \citep{7},
\begin{equation}
\begin{aligned}
\dfrac{J_{s}}{J_{o}}=\dfrac{1+q}{q}[(k_{1}r_{1})^{2}+q((k_{1}r_{1})^{2})]\\
\end{aligned}
\end{equation}
where $q$ is the mass ratio, $r_{1}$,$r_{2}$ is the relative radius, and $k_{1}$,$k_{2}$ is the dimensionless rotation radius. The value of the dimensionless rotation radius $k_{1}$, $k_{2}$ depends on the internal structure of the star, so using the accurate dimensionless rotation radius is important to correctly calculate the value of $J_{s}/J_{o}$. Assuming $k_{1}^{2}=k_{2}^{2}=0.06$, as \citet{5} and \citet{7}, $J_{s}/J_{o}$ of the eleven contact binaries were calculated and the results are listed in Table 10. $J_{s}/J_{o}$ of of the eleven contact binaries are all less than $1/3$, so all of them should be in stable state. In fact, the value of $k_{1}$, $k_{2}$ can be further precise. The secondary component of contacat binaries approaching merger is a very low mass star, so the totally convective n=1.5 polytrope would be appropriate and $k^{2}=0.205$ \citep{8}. It is a good assumption for a very low mass star (M $_{2}$ < 0.3 M$_{\odot}$). We reestimated $k_{1}$ for primary component of different masses, using the tabulated results from \citet{28}, which considers the effects of tidal and rotational distortions on a star in binary systems. We calculated $k_{1}$ using the following linear relationships, $k_{1}=-0.250M+0.539$ for stars with mass in 0.5-1.4 $M_{\odot}$, $k_{1}=0.014M+0.152$ for stars with mass more than 1.4 $M_{\odot}$. After re-determining $k_{1}$, $k_{2}$, we calculated $J_{s}/J_{o}$ again and the results are also listed in Table 10.

We calculated the instability parameters using a series of derivations from \citet{3}, adopting the stability criterion \textbf{$dJ_{T}/dJ_{A}= 0$}. There is a linear relationship between $f$ and $q_{inst}$ for stars of different masses. After obtaining the values for $f = 1$ and $f = 0$ using the Equation (14), (15) in \citet{3}, the instability mass ratio for any given fill-out factors can be obtained. The instability seperation can be obtained by the Equations (7), (8), (10) in \citet{3}. After getting the instability seperation, the instability period can be calculated by the Kepler's third law. These instability parameters calculated by the above steps are shown in Table 11.

In the spectral class range from F0 to M0 (most contact binaries are this range), the earlier spectral class of systems may become unstable at extremely low mass ratios while the later class of systems may become unstable at mass ratios close to 0.25 \citep{3}. The spectral types of the eleven contact binaries are near the F type, and their mass ratios are all less than 0.1, and they are extremely low mass ratios, so they can be recognized as pre-merger candidates. Especially, the mass ratio, semi-major axis and period of J234634 exceeded the instability value calculated by the theoretical equations, so we argued that J234634 is on the verge of a merger. The eleven contact binaries have important research significance, therefore it is necessary to monitor them continuously in the future.

\begin{table*}
\begin{center}
\caption{The instability parameters of the eleven targets.}
\label{table11}
\begin{tabular}{cccccccccc} 
\hline
Target  &$J_{S}/J_{O}$   & $J_{S}/J_{O}$   &$q_{inst}$& $A_{inst}$ ($R_{\odot}$)& $P_{inst} (d)$\\  
        &$(k_{1}=k_{2})$ & $(k_{1}\neq k_{2})$ &      &                         &       \\
\hline
J133031 & 0.24           & 0.26            & 0.0847   & 1.943                   & 0.2774\\
J154254 & 0.29           & 0.22            & 0.0679   & 2.097                   & 0.2939\\
J155009 & 0.29           & 0.16            & 0.0418   & 2.418                   & 0.3310\\
J155106 & 0.28           & 0.18            & 0.0595   & 2.096                   & 0.2861\\
J160755 & 0.24           & 0.18            & 0.0651   & 2.007                   & 0.2744\\
J162327 & 0.24           & 0.13            & 0.0416   & 2.329                   & 0.3095\\     
J164000 & 0.25           & 0.13            & 0.0558   & 1.937                   & 0.2519\\
J170307 & 0.26           & 0.29            & 0.0903   & 1.959                   & 0.2852\\
J223837 & 0.26           & 0.14            & 0.0463   & 2.213                   & 0.2935\\
J224827 & 0.32           & 0.29            & 0.0782   & 2.138                   & 0.3136\\
J234634 & 0.28           & 0.31            & 0.0857   & 2.026                   & 0.2999\\
\hline
\end{tabular}
\end{center}
\end{table*}

\begin{figure*}
\centering
\subfigure{\label{fig:subfig:a}
\includegraphics[width=0.48\linewidth]{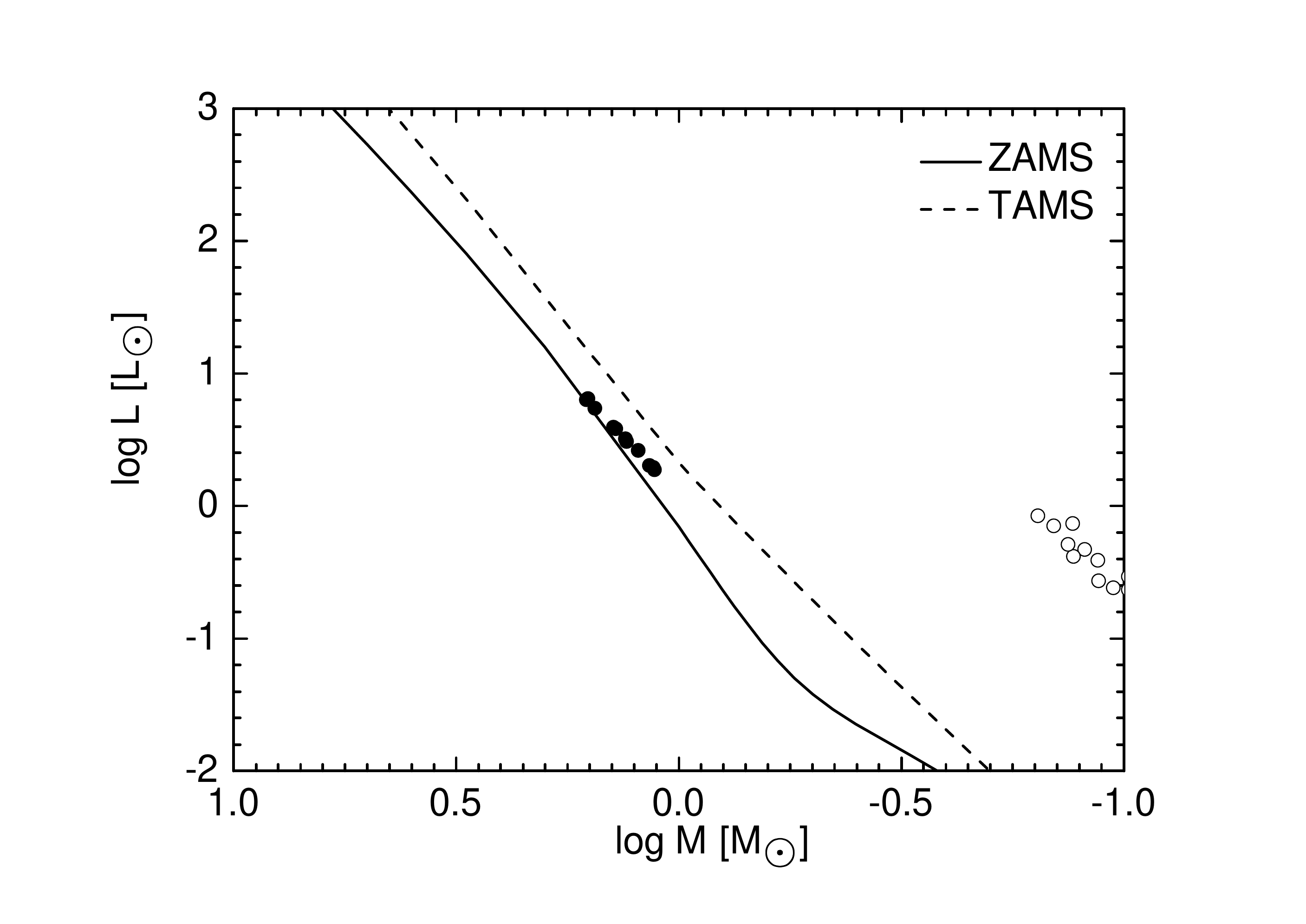}}
\subfigure{\label{fig:subfig:b}
\includegraphics[width=0.48\linewidth]{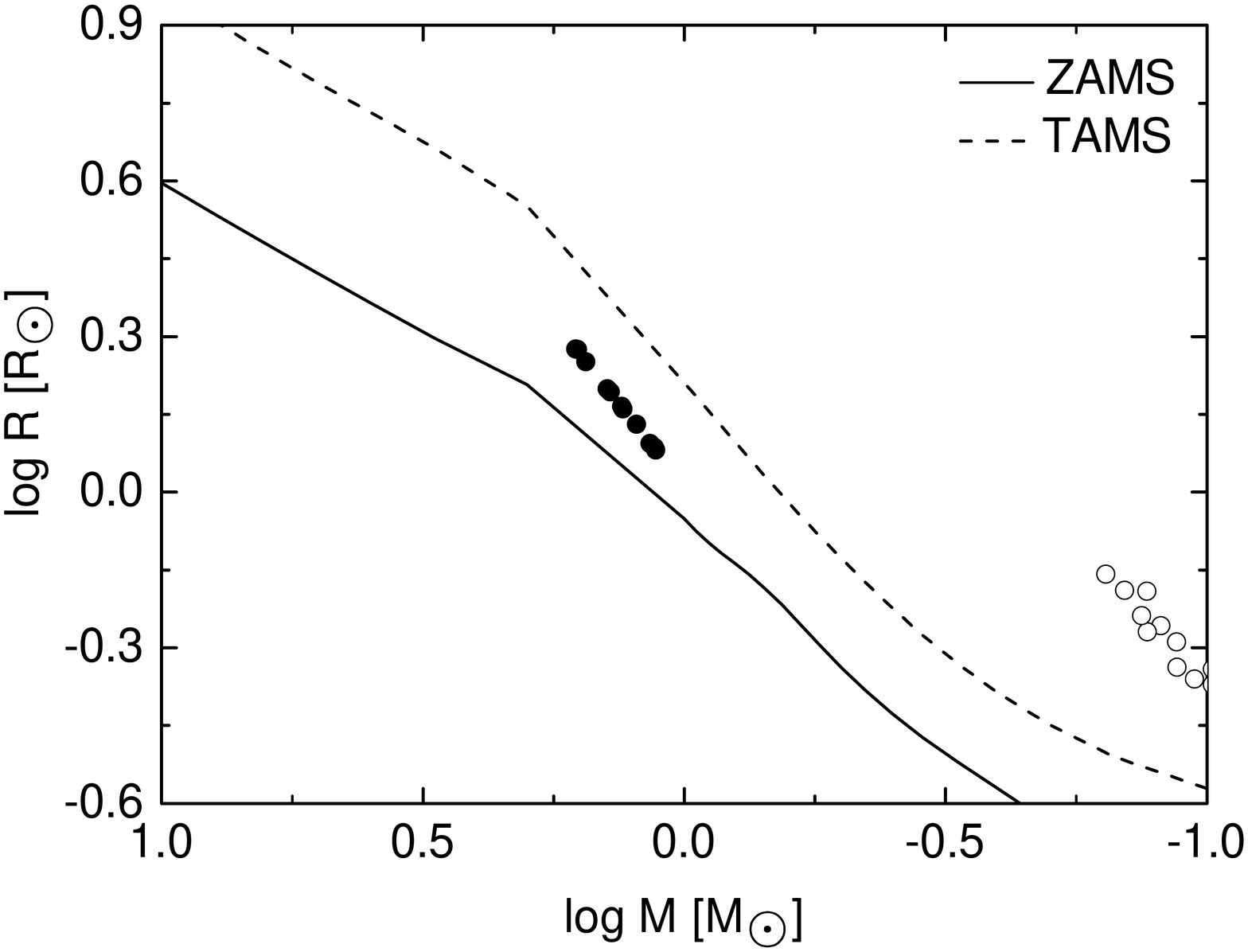}}
\caption{The mass – luminosity diagram (left) and the mass – radius diagram (right).}
\centering
\end{figure*}

\section*{Acknowledgements}
Great thanks to the referee for very helpful comments and suggestions that improved our manuscript a lot. This work is supported by National Natural Science Foundation of China (NSFC) (No. 12273018), and the Joint Research Fund in Astronomy (No. U1931103) under cooperative agreement between NSFC and Chinese Academy of Sciences (CAS), and by the Natural Science Foundation of Shandong Province (Nos. ZR2014AQ019), and by the Qilu Young Researcher Project of Shandong University, and by Young Scholars Program of Shandong University, Weihai (Nos. 20820171006) and by the Cultivation Project for LAMOST Scientific Payoff and Research Achievement of CAMS-CAS, and by the Chinese Academy of Science Interdisciplinary Innovation Team. The calculations in this work were carried out at Supercomputing Center of Shandong University, Weihai.

We acknowledge the support of the staff of NEXT, WHOT, OAN-SPM, NAOs85cm. This work of NAOs85cm was partially supported by the Open Project Program of the Key Laboratory of Optical Astronomy, the National Astronomical Observatories, and the Chinese Academy of Sciences.

The spectral data were provided by Guoshoujing Telescope (the Large Sky Area Multi-Object Fiber Spectroscopic Telescope LAMOST), which is a National Major Scientific Project built by the Chinese Academy of Sciences. Funding for the project has been provided by the National Development and Reform Commission. LAMOST is operated and managed by the National Astronomical Observatories, Chinese Academy of Sciences.

This work includes data collected by the TESS mission. Funding for the TESS mission is provided by NASA Science Mission Directorate. We acknowledge the TESS team for its support of this work.

We thank Las Cumbres Observatory and its staff for their continued support of ASAS-SN. ASAS-SN is funded in part by the Gordon and Betty Moore Foundation through grants GBMF5490 and GBMF10501 to the Ohio State University, and also funded in part by the Alfred P. Sloan Foundation grant G-2021-14192.

This paper makes use of data from ZTF. ZTF is supported by the National Science Foundation under grant no. AST-1440341 and a collaboration including Caltech, IPAC, the Weizmann Institute for Science, the Oskar Klein Center at Stockholm University, the University of Maryland, the University of Washington, Deutsches Elektronen-Synchrotron and Humboldt University, Los Alamos National Laboratories, the TANGO Consortium of Taiwan, the University of Wisconsin at Milwaukee, and Lawrence Berkeley National Laboratories. Operations are conducted by COO, IPAC, and UW.

This work has made use of data from the European Space Agency (ESA) mission {\it Gaia} (\url{https://www.cosmos.esa.int/gaia}), processed by the {\it Gaia} Data Processing and Analysis Consortium (DPAC,\url{https://www.cosmos.esa.int/web/gaia/dpac/consortium}). Funding for the DPAC has been provided by national institutions, in particular the institutions participating in the {\it Gaia} Multilateral Agreement.

This paper makes use of data from CSS. The CSS is funded by NASA under grant NNG05GF22G issued through the Science Mission Directorate Near-Earth Objects Observations Program. The Catalina Real-Time Transient Survey is supported by the U.S. National Science Foundation (NSF) under grants AST0909182 and AST-1313422.

This publication makes use of data products from the AA VSO Photometric All Sky Survey (APASS). Funded by the Robert Martin Ayers Sciences Fund and the National Science Foundation.

This publication makes use of data products from the Two Micron All Sky Survey, which is a joint project of the University of Massachusetts and the Infrared Processing and Analysis Center/California Institute of Technology, funded by the National Aeronautics and Space Administration and the National Science Foundation.

This paper makes use of data from from the DR1 of the WASP data \citet{81} as provided by the WASP consortium, and the computing and storage facilities at the CERIT Scientific Cloud, reg. no. CZ.1.05/3.2.00/08.0144 which is operated by Masaryk University, Czech Republic.

\section*{Data Availability}
The TESS data are publicly available at \href{http://archive.stsci.edu/tess/bulk_downloads.html}{http://archive.stsci.edu/tess/bulk downloads.html}. 
The ASAS-SN data are publicly available at \href{https://asas-sn.osu.edu/variables/lookup}{https://asas-sn.osu.edu/variables/lookup}.  
The SuperWASP data are publicly available at \href{https://wasp.cerit-sc.cz/search}{https://wasp.cerit-sc.cz/search}. 
The CSS data are publicly available at \href{https://catalina.lpl.arizona.edu/search/node}{https://catalina.lpl.arizona.edu/search/node}.
The ZTF data are publicly available at \href{https://www.ztf.caltech.edu/ztf-public-releases.html}{https://www.ztf.caltech.edu/ztf-public-releases.html}. 

\bibliography{bib}




\appendix
\boldmath{
\section{Our photometric data}
\begin{table*}
\begin{center}
\caption{Our photometric data.}
\begin{tabular}{cccccccccc} 
\hline
Target   &HJD-V          &$\Delta$MAG-V     &HJD-R          &$\Delta$MAG-R   &HJD-I          &$\Delta$MAG-I\\ 
\hline
J133031  &2459322.98762  &-0.211            &2459322.98847  &-0.199          &2459322.98900  &-0.205       \\         
		 &2459322.98980  &-0.228            &2459322.99065  &-0.211          &2459322.99118  &-0.206       \\
		 &2459322.99199  &-0.248            &2459322.99284  &-0.219          &2459322.99337  &-0.225       \\
		 &2459322.99418  &-0.258            &2459322.99502  &-0.249          &2459322.99555  &-0.209       \\
		 &2459322.99650  &-0.271            &2459322.99741  &-0.250          &2459322.99794  &-0.234       \\
\hline
\end{tabular}
\begin{tablenotes}
\footnotesize
\item[1] This table is available in its entirety in machine-readable form in the online version of this article. 
\end{tablenotes}
\end{center}
\end{table*} 
}

\bsp	
\label{lastpage}
\end{document}